%% file: paper.tex
\pdfoutput=1
\documentclass[runningheads]{llncs}

\newif\ifanonymous
\anonymousfalse

\usepackage[T1]{fontenc}
\usepackage[utf8]{inputenc}
\usepackage{amsmath}
\usepackage{amssymb}
\usepackage{mathtools}
\usepackage{stmaryrd}
\usepackage[Euler]{upgreek}
\usepackage{graphicx}
\usepackage{booktabs}
\usepackage{array}
\usepackage{float}
\usepackage{xcolor}
\usepackage{colortbl}
\usepackage{algpseudocodex}
\usepackage{tikz}
\usepackage{pgfplots}
\usepackage{pifont}
\pgfplotsset{compat=1.16}
\usepgfplotslibrary{groupplots}
\usetikzlibrary{calc}
\usetikzlibrary{arrows,arrows.meta,positioning}
\usepackage[most]{tcolorbox}
\usepackage[resetlabels]{multibib}
\usepackage[hidelinks]{hyperref}
\usepackage[capitalise,nameinlink]{cleveref}
\newcites{code}{Code References}
\makeatletter
\let\mb@citecode\citecode
\renewcommand\citecode[1]{{\def\@cite##1##2{[{code~##1\if@tempswa , ##2\fi}]}%
  \mb@citecode{#1}}}
\let\mb@bibliographycode\bibliographycode
\renewcommand\bibliographycode[1]{{\let\mb@thebibliography\thebibliography
  \def\thebibliography##1{\mb@thebibliography{code~##1}%
    \def\@biblabel####1{code~####1.}}%
  \mb@bibliographycode{#1}}}
\makeatother

\newcommand{\todo}[1]{}

\input{isabelle_permalinks}

\renewcommand{\checkmark}{\ding{51}}
\newcommand{\xmark}{\ding{55}}

\input{macros}

\begin{document}

\title{IsaGraphLib: Formally Verified Efficient Graph Algorithms in Isabelle/HOL}

  \author{Mohammad~Abdulaziz\inst{1} 
  \and
    Thomas~Ammer\inst{1} 
    \and
    Peter~Lammich\inst{2} 
    \and
    Adem~Rimpapa\inst{1}
    }
  \authorrunning{M. Abdulaziz et al.}
  \institute{King's College London, London, United Kingdom\\
    \email{\{mohammad.abdulaziz,thomas.ammer,adem.rimpapa\}@kcl.ac.uk} 
    \and
  University of Twente, Enschede, The Netherlands\\
    \email{p.lammich@utwente.nl}
    }

\maketitle

\begin{abstract}
  We present a library of formally verified graph algorithms in Isabelle/HOL.
  The library covers the main problems of combinatorial optimisation on graphs: reachability, shortest paths, minimum spanning trees, matching, maximum flow, and minimum-cost flow.
  The algorithms in the library use most of the standard algorithm design techniques: graph search, greedy algorithms on matroids, dynamic programming, augmenting paths, blossom contraction, blocking flows, push-relabel, scaling, the primal-dual method, randomisation, and certificate checking.
  The library has adaptors between different types of graphs (directed, undirected, multigraphs), which allow results to be reused.
  Each algorithm is verified independently of its data structures, and the main algorithms are refined to efficient imperative implementations, from which we generate code in Standard ML, Scala, and Go.
  We developed parts of the library with large language models (LLMs), improved the quality of their proofs with a method based on statistics of our own proofs, and labelled which parts were developed by humans, by LLMs, or collaboratively.
  We perform a detailed experimental analysis of representative algorithms in our library, comparing it with state-of-the-art mainstream libraries, showing that our implementations are practical.
\end{abstract}

\input{paper_intro}
\input{paper_methodology}

\input{paper_empirical}

\input{paper_ai_development}
\input{paper_conclusion}

\clearpage
\section*{Data Availability}

Our formal development, the style guideline that we gave to the LLM when it restyled its proofs, the two scripts that compute the style statistics of \cref{tab:hk-style} from the theory sources (\cref{app:style-statistics}), and the scripts of our experiments are archived at \url{https://doi.org/10.5281/zenodo.23805452}.
The library is developed further at \url{https://github.com/mabdula/Isabelle-Graph-Library}.
The links to the formal development in this paper point to the version of the library in this repository that has the tag \texttt{tacas27-artefact}:\\
\url{https://github.com/mabdula/Isabelle-Graph-Library/tree/tacas27-artefact}.

\bibliographystyle{splncs04}
\bibliography{claude_added,thomas_refs,adem_refs,refs,long_paper}

\input{paper_appendix}

\end{document}

%% file: isabelle_permalinks.tex
{\catcode`\#=12 \gdef\ghhash{#}}

{\catcode`\_=12
 \gdef\ghrepo{https://github.com/mabdula/Isabelle-Graph-Library}%
 \gdef\ghcommit{3518b79900ff56129575022d4fb8e7ba463ffa5a}%
 \gdef\afprepo{https://github.com/isabelle-prover/mirror-afp-2025-2}%
 \gdef\afpcommit{fa30b560ab754bfbb7a721ee45a921d4b3c04358}%
}

{\catcode`\_=12 \catcode`\&=12
 \gdef\ghurl#1#2#3{%
   \ghrepo/blob/\ghcommit/#1%
   \ghhash L#2\ifnum#2=#3 \else -L#3\fi}%
 \gdef\afpurl#1#2#3{%
   \afprepo/blob/\afpcommit/thys/#1%
   \ghhash L#2\ifnum#2=#3 \else -L#3\fi}%
}

\newcommand{\isalogo}{%
  \raisebox{-0.5ex}{\includegraphics[height=2.4ex]{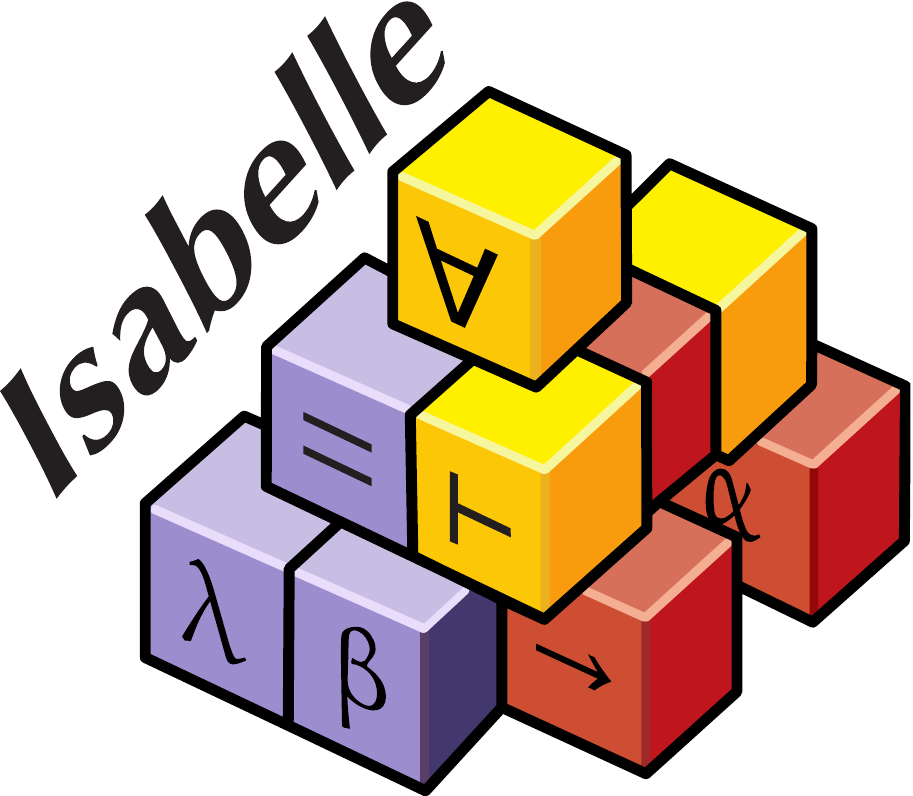}}}

\newcommand{\isalink}[3]{\href{\ghurl{#1}{#2}{#3}}{\isalogo}}
\newcommand{\afplink}[3]{\href{\afpurl{#1}{#2}{#3}}{\isalogo}}

\newcommand{\isainl}[4]{\href{\ghurl{#1}{#2}{#3}}{\textcolor{black}{\texttt{#4}}}}

%% file: macros.tex
\providecommand{\inst}{\ensuremath{\tvsal}}
\newcommand{\act}{\ensuremath{\pi}}

\makeatletter
\newcommand{\oset}[2]{%
  {\mathop{#2}\limits^{\vbox to -1\ex@{\kern-\tw@\ex@
   \hbox{\scriptsize #1}\vss}}}}
\makeatother

\renewcommand{\vec}[1]{\ensuremath{\oset{$\rightarrow$}{\ensuremath{#1}}}}
\newcommand{\as}{\ensuremath{\vec{{\act}}}}

\newcommand{\satpreas}[2]{\ensuremath{sat_precond_as(s, \as)}}

\newcommand{\tvsal}{{\pitchfork}}

\providecommand{\path}{\ensuremath{\gamma}}

\renewcommand{\path}{\ensuremath{\gamma}}

\providecommand{\graph}{}
\providecommand{\cal}{}
\renewcommand{\cal}{}
\renewcommand{\graph}{{\cal G}}

\renewcommand{\min}{\textsf{\upshape min}}

\newcommand{\edges}{\ensuremath{E}}

\newcommand{\settype}{\mbox{\upshape \textsf{set}}}

\newcommand{\Omit}[1]{}

\newcommand{\snapshotsymbol}{|\kern-.7ex\raise.08ex\hbox{\scalebox{0.7}{$\bullet$}}}

\newcommand{\ssgraph}[1]{\graph_\ss}

\newcommand{\vertexgen}{\ensuremath{u}}
\newcommand{\vertexa}{{\ensuremath{\vertexgen_1}}}
\newcommand{\vertexb}{{\ensuremath{\vertexgen_2}}}

\renewcommand{\prod}{\ensuremath{{{{{\mathlarger{\mathlarger {{\mathlarger {\Pi}}}}}}}}}}

\tikzset{dots/.style args={#1per #2}{line cap=round,dash pattern=on 0 off #2/#1}}
\providecommand{\moham}[1]{\fbox{{\bf \@Mohammad: }#1}}

\makeatletter
\newcommand*{\my@test@it}{it}
\newcommand*{\mathulifit}[1]{%
  \ifx\f@shape\my@test@it
    \underline{\mathit{#1}}
  \else
    \mathit{#1}
  \fi
}
\makeatother

\tikzset{
    robot1/.pic={
        \draw node at (0,0) {\includegraphics[height=2cm]{assets/robot1.png}};
    };
}
\tikzset{
    robot2/.pic={
        \draw node at (0,0) {\includegraphics[height=2cm]{assets/robot2.png}};
    };
}
\tikzset{
    block/.pic={
        \draw node at (0,0) {\includegraphics[height=1cm]{assets/block.png}};
    };
}
\tikzset{
    blockinv/.pic={
        \draw node at (0,0) {\includegraphics[height=1cm]{assets/block_inv.png}};
    };
}
\tikzset{
    r1b1l/.pic={
        \draw pic at (0,0) {robot1}
            pic at (-1.25,0.5) {block};
    },
    r1b1r/.pic={
        \draw pic at (0,0) {robot1}
            pic at (1.25,0.5) {block};
    },
    r2b1l/.pic={
        \draw pic at (0,0) {robot2}
            pic at (-1.25,0.5) {block};
    },
    r2b1r/.pic={
        \draw pic at (0,0) {robot2}
            pic at (1.25,0.5) {block};
    },
    r1b2l/.pic={
        \draw pic at (0,0) {robot1}
            pic at (-1.25,0.5) {blockinv};
    },
    r1b2r/.pic={
        \draw pic at (0,0) {robot1}
            pic at (1.25,0.5) {blockinv};
    },
    r2b2l/.pic={
        \draw pic at (0,0) {robot2}
            pic at (-1.25,0.5) {blockinv};
    },
    r2b2r/.pic={
        \draw pic at (0,0) {robot2}
            pic at (1.25,0.5) {blockinv};
    };
}
\tikzset{
    dooropen/.pic={
        \draw[color=gray] (0,0) -- ++(90:0.5) -- ++(300:0.5)
        (0,2) -- ++(270:0.5) -- ++(60:0.5);
    },
    doorclosed/.pic={
        \draw[color=gray] (0,0) -- (0,2);
    },
    doorhalfopen/.pic={
        \draw[color=gray] (0,0) ++(90:0.5) -- ++(15:0.5)
        (0,2) ++(270:0.5) -- ++(345:0.5);
        \draw[color=gray] (0,0) -- (0,2);
        \draw[color=gray] (0,0) ++(90:0.5) -- ++(300:0.5)
        (0,2) ++(270:0.5) -- ++(60:0.5);
    };
}
\tikzset{
    global scale/.style={
        scale=#1,
        every node/.style={scale=#1}
    }
}

\tikzset{>={Stealth}}

\tikzset{
    room width/.initial=3,
    room height/.initial=1.5,
    pics/room/.style 2 args={
        code={
            \draw[color=gray] (0,0) rectangle
                (\pgfkeysvalueof{/tikz/room width},\pgfkeysvalueof{/tikz/room height});
            \draw pic[global scale=0.5*\pgfkeysvalueof{/tikz/room height}]
                at (0.5*\pgfkeysvalueof{/tikz/room width},0) {#1};
            \draw node[anchor=north]
                at (0.5*\pgfkeysvalueof{/tikz/room width},0) {#2};
        }
    }
}

\tikzset{
    action box/.style={draw=#1!60,solid,fill=#1!20},
    invalid action box/.style={draw=none,pattern=north west lines,pattern color=#1!20}
}

\tikzset{
    to*/.style={
        shorten >=.25em,#1-to,
        to path={-- node[inner sep=0pt,at end,sloped] {${}^*$} (\tikztotarget) \tikztonodes}
    },
    to*/.default=
}

\makeatletter
\pgfarrowsdeclare{to*}{to*}
{
  \pgfutil@tempdima=-0.84pt%
  \advance\pgfutil@tempdima by-1.3\pgflinewidth%
  \pgfutil@tempdimb=0.21pt%
  \advance\pgfutil@tempdimb by.625\pgflinewidth%
  \advance\pgfutil@tempdimb by2.5pt%
  \pgfarrowsleftextend{+\pgfutil@tempdima}
 textual matterfarrowsrightextend{+\pgfutil@tempdimb}
}
{
  \pgfutil@tempdima=0.28pt%
  \advance\pgfutil@tempdima by.3\pgflinewidth%
  \pgfsetlinewidth{0.8\pgflinewidth}
  \pgfsetdash{}{+0pt}
  \pgfsetroundcap
  \pgfsetroundjoin
  \pgfpathmoveto{\pgfqpoint{-3\pgfutil@tempdima}{4\pgfutil@tempdima}}
  \pgfpathcurveto
  {\pgfqpoint{-2.75\pgfutil@tempdima}{2.5\pgfutil@tempdima}}
  {\pgfqpoint{0pt}{0.25\pgfutil@tempdima}}
  {\pgfqpoint{0.75\pgfutil@tempdima}{0pt}}
  \pgfpathcurveto
  {\pgfqpoint{0pt}{-0.25\pgfutil@tempdima}}
  {\pgfqpoint{-2.75\pgfutil@tempdima}{-2.5\pgfutil@tempdima}}
  {\pgfqpoint{-3\pgfutil@tempdima}{-4\pgfutil@tempdima}}
  \pgfusepathqstroke
  \begingroup
    \pgftransformxshift{2.5pt}
    \pgftransformyshift{2pt}
    \pgftransformscale{.7}
    \pgfuseplotmark{asterisk}
  \endgroup
}

\newcommand{\splitatcommas}[1]{%
  \begingroup
  \begingroup\lccode`~=`, \lowercase{\endgroup
    \edef~{\mathchar\the\mathcode`, \penalty0 \noexpand\hspace{0pt plus 1em}}%
  }\mathcode`,="8000 #1%
  \endgroup
}

\renewcommand{\paragraph}[1]{\noindent{\bf #1}}

\makeatletter
\newtcolorbox{pseudocodebox}{enhanced, colback=black!4, frame hidden,
  boxrule=0pt, arc=1pt, left=3pt, right=3pt, top=3pt, bottom=3pt,
  before skip=6pt, after skip=6pt,
  before upper={\footnotesize\parindent\z@
    \begin{list}{}{\topsep\z@\partopsep\z@\parsep\z@\itemsep\z@
      \labelwidth\z@\labelsep\z@\itemindent\z@\listparindent\z@
      \leftmargin-1em\rightmargin\z@}\item[]\relax},
  after upper={\end{list}}}
\makeatother

\newcommand{\uedge}[2]{\ensuremath{\{#1,#2\}}}
\newcommand{\typevar}{\ensuremath{\alpha}}
\newcommand{\hastype}{\mathrel{::}}
\newcommand{\bigO}[1]{\ensuremath{O(#1)}}
\newcommand{\numverts}{\ensuremath{n}}
\newcommand{\numedges}{\ensuremath{m}}
\newcommand{\vdeg}{\ensuremath{d}}
\newcommand{\rankingratio}{\ensuremath{1-1/e}}

%% file: paper_intro.tex
\section{Introduction}
\label{sec:intro}

Algorithms for combinatorial optimisation problems on graphs have influenced both the theory and the practice of computing.
Theoretically, polynomial running time as the notion of efficient computation was proposed by Edmonds in his paper on matching in general graphs~\cite{edmond1965blossom}.
Practically, matching algorithms are used in kidney exchange programmes~\cite{EUKidneyMatchingReview} and in allocating online advertisements~\cite{AdWords2007}, minimum-cost flow algorithms compute the cheapest way to transport goods~\cite{optCritMinCostFlows}, and shortest path algorithms are used in route planning~\cite{bellmanFordI}, AI planning~\cite{bonetPlanningHeuristicSearch2001}, and model checking~\cite{principlesMC}.
Ensuring the correctness of implementations of these algorithms is difficult without formal verification, for two reasons.
First, their correctness arguments are mathematically involved, with many concepts and complicated constructions.
Second, practical implementations employ low-level optimisations, e.g.\ storing graphs in the compressed sparse row (CSR) format and renumbering vertices to improve cache locality, whose correctness is hard to ensure within a large and complicated algorithm.
We assume that the reader is familiar with common graph algorithms, otherwise we refer to standard literature~\cite{KorteVygenOptimisation,schrijverBook,ahujaNetworkFlowsTheory1993,VaziraniApprox}.

\begin{table}[!t]
  \caption{The verified algorithms, grouped by the problems they solve.
    A citation marks an algorithm whose formalisation was published before; the formalisations of the other algorithms are new.
    \emph{Proved}: termination (T), correctness (C), optimality (O), and approximation or competitive ratio (A).
    The Isabelle logo links to the main theorem in the formal development.
    \emph{Lines}: the number of lines of the theories specific to the algorithm.
    \Cref{tab:algorithms} in \cref{app:library} gives the details of the verification of every algorithm, and \cref{tab:ai-algorithms} in \cref{app:ai} shows which of them have abstract (i.e.\ non-executable), functional, and imperative versions. The library has about 300,000 lines of Isabelle/HOL.
    \todo{Check this list against the algorithms that have efficient implementations.}}
  \label{tab:library}
  \centering
  \scriptsize
  \setlength{\tabcolsep}{2pt}
  \newcommand{\algproblem}[1]{\multicolumn{4}{@{}l}{\emph{#1}}\tabularnewline}
  \begin{minipage}[t]{0.495\textwidth}
    \begin{tabular}[t]{@{}>{\raggedright\arraybackslash}p{30mm}lrr@{}}
      \toprule
      Algorithm & Proved & & Lines\tabularnewline
      \midrule
      \algproblem{Reachability and cycle detection}
      Depth-first search~\cite{fdsBook} & T, C & \isalink{Set_Graphs/Graph_Algorithms/DFS.thy}{739}{762} & 877\tabularnewline
      Breadth-first search~\cite{fdsBook} & T, C & \isalink{Set_Graphs/Graph_Algorithms/BFS_2.thy}{1504}{1524} & 1,805\tabularnewline
      Cycle detection & T, C & \isalink{Set_Graphs/Graph_Algorithms/DFS_Cycles.thy}{1185}{1200} & 3,305\tabularnewline
      Directed cycle detection & T, C & \isalink{Set_Graphs/Graph_Algorithms/DFS_DirCycle.thy}{404}{406}\,\isalink{Set_Graphs/Graph_Algorithms/DFS_DirCycle.thy}{676}{678} & 697\tabularnewline
      \midrule
      \algproblem{Shortest paths}
      BFS with distances & T, C & \isalink{Set_Graphs/Graph_Algorithms/BFS_3.thy}{2901}{2917} & 2,901\tabularnewline
      Bellman--Ford & C, O & \isalink{Set_Graphs/Graph_Algorithms/Bellman_Ford.thy}{1999}{2003} & 2,837\tabularnewline
      Dijkstra & T, C, O & \isalink{Set_Graphs/Graph_Algorithms/Dijkstra.thy}{1645}{1647} & 2,865\tabularnewline
      \midrule
      \algproblem{Maximum flow}
      Dinic & T, C, O & \isalink{Dinic/Dinic.thy}{426}{428} & 1,455\tabularnewline
      Blocking flow & T, C & \isalink{Dinic/Compute_Blocking_Simple.thy}{587}{597} & 713\tabularnewline
      DFS with backtracking & T, C & \isalink{Dinic/DFS_Collect_Backtrack.thy}{577}{583} & 746\tabularnewline
      Push-relabel & T, C, O & \isalink{Push_Relabel/Push_Relabel_Generic.thy}{1971}{1971} & 4,897\tabularnewline
      \midrule
      \algproblem{Minimum-cost flow}
      Successive shortest paths~\cite{ScalingIsabelle} & T, C, O & \isalink{Mincost_Flow_Algorithms/Successive_Shortest_Path.thy}{1056}{1060} & 1,158\tabularnewline
      Capacity scaling~\cite{ScalingIsabelle} & T, C, O & \isalink{Mincost_Flow_Algorithms/Scaling.thy}{1624}{1629} & 1,785\tabularnewline
      Orlin~\cite{ScalingIsabelle} & T, C, O & \isalink{Mincost_Flow_Algorithms/Orlins.thy}{2837}{2844} & 12,080\tabularnewline
      Network simplex & T, C, O & \isalink{Mincost_Flow_Algorithms/Network_Simplex_Final.thy}{479}{482} & 7,582\tabularnewline
      Acyclic flows & T, C & \isalink{Mincost_Flow_Algorithms/Acyclic_Flow.thy}{5449}{5453} & 10,691\tabularnewline
      \bottomrule
    \end{tabular}
  \end{minipage}\hfill
  \begin{minipage}[t]{0.495\textwidth}
    \begin{tabular}[t]{@{}>{\raggedright\arraybackslash}p{30mm}lrr@{}}
      \toprule
      Algorithm & Proved & & Lines\tabularnewline
      \midrule
      \algproblem{Spanning trees and matroids}
      Best-in-greedy~\cite{MatroidIsabelle} & T, C, A, O & \isalink{Matroids_Greedy/BestInGreedy_stablesort.thy}{629}{632} & 5,750\tabularnewline
      Kruskal~\cite{MatroidIsabelle} & T, C, O & \isalink{Spanning_Trees/Kruskal_Functional.thy}{1097}{1098} & 1,559\tabularnewline
      Greedoid greedy~\cite{MatroidIsabelle} & T, C, O & \isalink{Greedoids/Greedoids_Optimisation.thy}{1595}{1598} & 2,217\tabularnewline
      Jarn\'{\i}k--Prim~\cite{MatroidIsabelle} & T, C, O & \isalink{Spanning_Trees/Jarnik_Prim.thy}{497}{499} & 1,277\tabularnewline
      Matroid intersection~\cite{MatroidIsabelle} & T, C, O & \isalink{Matroid_Intersection/Matroid_Intersection_Path_Loop.thy}{336}{338} & 2,646\tabularnewline
      Weighted matroid intersection & T, C, O & \isalink{Matroid_Intersection/Matroid_Weighted_Intersection_Path_Loop.thy}{1144}{1147} & 5,274\tabularnewline
      Matroid partitioning & C, O & \isalink{Matroid_Intersection/Matroid_Partitioning.thy}{899}{900} & 1,215\tabularnewline
      \midrule
      \algproblem{Matching}
      Edmonds' blossom~\cite{BlossomIsabelleJAR} & T, C, O & \isalink{Edmonds_Blossom_Executable/Find_Max_Matching.thy}{321}{323} & 10,460\tabularnewline
      Weighted blossom & C, O & \isalink{Edmonds_Weighted_Blossom/Naive_Weighted_Blossom.thy}{3704}{3709} & 5,801\tabularnewline
      Edmonds--Gallai decomposition & T, C & \isalink{Edmonds_Gallai/Edmonds_Gallai_Computation.thy}{574}{580} & 5,007\tabularnewline
      Hopcroft--Karp & T, C, O & \isalink{Hopcroft_Karp/Hopcroft_Karp_Top_Loop.thy}{395}{396} & 2,381\tabularnewline
      Hungarian method~\cite{abdulazizFormalPrimalDualAlgorithm2026b} & T, C, O & \isalink{Hungarian_Method/Hungarian_Method_Top_Loop.thy}{571}{573} & 4,123\tabularnewline
      Primal-dual (Egerv\'ary)~\cite{abdulazizFormalPrimalDualAlgorithm2026b} & T, C, O & \isalink{Hungarian_Method/Naive_Primal_Dual.thy}{1239}{1242} & 1,863\tabularnewline
      \midrule
      \algproblem{Online matching}
      RANKING~\cite{RankingIsabelle} & A & \isalink{RANKING/Ranking_Probabilistic.thy}{1967}{1967} & 6,973\tabularnewline
      RANKING, primal-dual~\cite{abdulazizFormalPrimalDualAlgorithm2026b} & A & \isalink{Primal_Dual_Bipartite_Matching/Ranking_Order_Probabilistic.thy}{1298}{1312} & 4,861\tabularnewline
      AdWords~\cite{abdulazizFormalPrimalDualAlgorithm2026b} & C, A & \isalink{Primal_Dual_Bipartite_Matching/AdWords.thy}{1352}{1354} & 1,347\tabularnewline
      \bottomrule
    \end{tabular}
  \end{minipage}
\end{table}

In this paper, we present practical formally verified (implementations of) algorithms in Isabelle/HOL~\cite{IsabelleHOLRef} for most of the main combinatorial optimisation problems on graphs (\cref{tab:library}).
In particular: 1.\ we present a library of these implementations, from which we generate code in Standard ML, Scala, and Go~\cite{haftmannCodeGenerationHigherOrder2010,isabelleGo},
2.\ we describe the methodology underlying the library (\cref{sec:methodology}), which extends that of Abdulaziz~\cite[Chapter~24]{fdsBook} to substantially more involved algorithms, 
3.\ we refine our algorithms to imperative implementations without a custom mechanism for non-determinism, which, as far as we are aware, is a first (\cref{sec:imperative}), using a separation logic framework for Imperative HOL which we extended, e.g.\ with monadic folds and while-loops, 
4.\ we report on developing parts of the library with large language models (LLMs) (\cref{sec:ai}): we label parts of the library as developed by humans, by an LLM, or collaboratively, and we restyle LLM-generated proofs using statistics of our own proofs, and 
5.\ we evaluate the implementations of representative algorithms experimentally against LEMON~\cite{dezsoLEMON2011}, SciPy~\cite{scipyManual}, NetworkX~\cite{networkxManual}, JGraphT~\cite{michailJGraphTJavaLibrary2020}, and Sage~\cite{sageMatroid} (\cref{sec:evaluation}).

Throughout the paper, the Isabelle logo~\isalogo{} and the Isabelle names in typewriter font link to the corresponding lines of our formal development (see the data availability statement).

%% file: paper_methodology.tex
\section{Methodology}
\label{sec:methodology}

Our library uses Abdulaziz's~\cite[Chapter~24]{fdsBook} methodology for verifying graph algorithms:
the abstract data type of graphs (specified
as an Isabelle/HOL locale) and the mathematical representation of graphs as sets of
edges (related to the abstract data type by an abstraction function)
are both taken from that chapter. However, that chapter applies the methodology
to depth-first search (DFS) and breadth-first search (BFS), while we apply it to a far larger set of
algorithms~\cite{ScalingIsabelle,MatroidIsabelle,BlossomIsabelleJAR,abdulazizFormalPrimalDualAlgorithm2026b}
whose verification is substantially more involved: our formalisation of minimum-cost flow
algorithms~\isalink{Mincost_Flow_Algorithms/Orlins.thy}{2837}{2844}~\cite{ScalingIsabelle}, including Orlin's algorithm, is 27,000 lines of
Isabelle/HOL, that of Edmonds' blossom algorithm~\isalink{Edmonds_Blossom_Executable/Find_Max_Matching.thy}{321}{323}~\cite{BlossomIsabelleJAR} is about 10,000
lines, while the formalisations of DFS~\isalink{Set_Graphs/Graph_Algorithms/DFS.thy}{739}{741} and BFS~\isalink{Set_Graphs/Graph_Algorithms/BFS_2.thy}{1517}{1519} accompanying the chapter are about 800 and 1,500 lines, respectively.

\subsection{Mathematical Models of Graphs and Reasoning Principles}
\label{sec:graphs}
An initial motivation for our library was that the Isabelle community has produced a large number of graph libraries, covering both theory and algorithms.
Each of those libraries came with its own
definitions of graphs and lemmas about them: in the Archive of Formal Proofs (AFP) alone, a graph is
represented as a record of a vertex set, a set of arcs, and two functions
mapping every arc to its tail and head~\cite{noschinskiGraphLib}, as a set
of vertices together with a set of edges, each of which is a set of
vertices~\cite{Undirected_Graph_Theory-AFP}, as a multiset of such
edges~\cite{Hypergraph_Basics-AFP}, as a set of unordered vertex pairs~\cite{Kruskal-AFP}, as a value of an abstract type of undirected graphs
whose edges are given as pairs of vertices~\cite{Prim_Dijkstra_Simple-AFP}, and
as a function mapping every pair of vertices to a
capacity~\cite{Flow_Networks-AFP}. Every result in a given representation has to be proved again for every other representation.
As we designed our library, we were aiming for results concerning graphs to be proved once and then reused by all the algorithms that need them.

\begin{figure}[t]
  \centering
  \begin{tikzpicture}[
      every node/.style={font=\scriptsize, align=center},
      box/.style={draw, rounded corners, inner sep=3pt, text width=30mm},
      rep/.style={box, very thick},
      ext/.style={box, fill=black!8},
      adt/.style={box},
      impl/.style={box, dashed},
      lbl/.style={font=\tiny, fill=white, inner sep=1pt},
      import/.style={-{Stealth}},
      abs/.style={-{Stealth}, dotted, thick},
      adapt/.style={{Stealth}-{Stealth}, densely dashed}]
    \node[ext, text width=48mm] (gt) at (0,0) {\textbf{Graph library from the AFP}~\afplink{Graph_Theory/Digraph.thy}{27}{29}\\
      \texttt{Graph\_Theory}~\cite{noschinskiGraphLib}, locales\\
      \texttt{wf\_digraph}, \texttt{nomulti\_digraph}};
    \node[rep]  (multi) at (-4.4,-1.9) {\textbf{Multigraphs}~\isalink{Set_Graphs/Directed_Graphs/Multigraph.thy}{125}{129}\\
      \texttt{multigraph} $\edges$ \texttt{fst} \texttt{snd}};
    \node[rep]  (dir)   at (0,-1.9)    {\textbf{Directed set graphs}\\
      $\graph \hastype (\typevar \times \typevar)\,\settype$~\isalink{Set_Graphs/Directed_Graphs/Pair_Graph.thy}{17}{17}};
    \node[rep]  (und)   at (4.4,-1.9)  {\textbf{Undirected set graphs}~\isalink{Set_Graphs/Undirected_Graphs/Undirected_Set_Graphs.thy}{763}{765}\\
      $\graph \hastype \typevar\,\settype\,\settype$};
    \node[adt]  (madt)  at (-4.4,-3.8) {\textbf{Multigraph ADTs}$^\dagger$~\isalink{Set_Graphs/Directed_Graphs/Edge_Iterators.thy}{17}{37}\\
      \texttt{outgoing\_edge\_iterator},\\ \texttt{Multi\_Graph\_Specs}};
    \node[adt]  (spec)  at (0,-3.8)    {\textbf{Directed ADTs}~\isalink{Set_Graphs/Directed_Graphs/Pair_Graph_Specs.thy}{243}{251}\\
      \texttt{Pair\_Graph\_Specs},\\ \texttt{Adj\_Map\_Specs2}};
    \node[adt]  (uspec) at (4.4,-3.8)  {\textbf{Undirected ADT}~\isalink{Set_Graphs/Undirected_Graphs/Pair_Graph_U_Specs.thy}{498}{502}\\
      \texttt{Pair\_Graph\_U\_Specs}};
    \node[impl, text width=22mm] (csr) at (-4.4,-5.5) {\textbf{Arrays (CSR)}$^\dagger$~\isalink{Set_Graphs/Directed_Graphs/CSR_Graph_Spec.thy}{347}{348}\\
      \texttt{csr\_graph}};
    \node[impl, text width=22mm] (fun) at (-1.2,-5.5) {\textbf{Functional}\\
      RBT~\isalink{Set_Graphs/Directed_Graphs/Pair_Graph_RBT.thy}{54}{55}, AVL~\isalink{Set_Graphs/Directed_Graphs/Pair_Graph_AVL.thy}{68}{72}};
    \node[impl, text width=29mm] (imp) at (2.5,-5.5) {\textbf{Imperative}~\isalink{Set_Graphs/Directed_Graphs/Refinements/Pair_Graph_Imperative.thy}{13}{19}\\
      \texttt{Pair\_Graph\_Imperative}};
    \draw[adapt]  (gt)    -- node[lbl, align=center] {adaptors, e.g.\ walks~\isalink{Set_Graphs/Directed_Graphs/Adaptors/Pair_Graph_Library_Adaptor.thy}{175}{177}} (dir);
    \draw[adapt]  (dir)   -- node[lbl, above=1pt, align=center] {adaptors,\\ e.g.\ walks\\\isalink{Set_Graphs/Undirected_Graphs/Pair_Graph_Berge_Adaptor.thy}{95}{96}} (und);
    \draw[abs]    (multi) -- node[lbl, above=1pt] {\texttt{make\_pair}} (dir);
    \draw[abs]    (madt)  -- node[lbl] {\texttt{out\_abstract}} (multi);
    \draw[abs]    (spec)  -- node[lbl] {\texttt{digraph\_abs}} (dir);
    \draw[abs]    (uspec) -- node[lbl] {\texttt{ugraph\_abs}} (und);
    \draw[import] (uspec) -- (spec);
    \draw[import] (csr)   -- (madt);
    \draw[import] (fun)   -- (spec);
    \draw[import] (fun)   -- (uspec);
    \draw[import] (imp)   -- (spec);
  \end{tikzpicture}
  \caption{Representations of graphs in the library and the connections
    between them. Thick boxes are graph representations, the grey box is
    an external library from the AFP, plain boxes are ADTs, and
    dashed boxes are their implementations. Dashed double arrows are adaptors;
    dotted arrows are abstraction functions; solid arrows are locale imports
    or interpretations. Daggers mark ADTs and implementations
    that are only in development branches of the library.
    \Cref{tab:graph-adts} in \cref{app:library} describes the abstract data
    types.}
  \label{fig:graph-locales}
\end{figure}
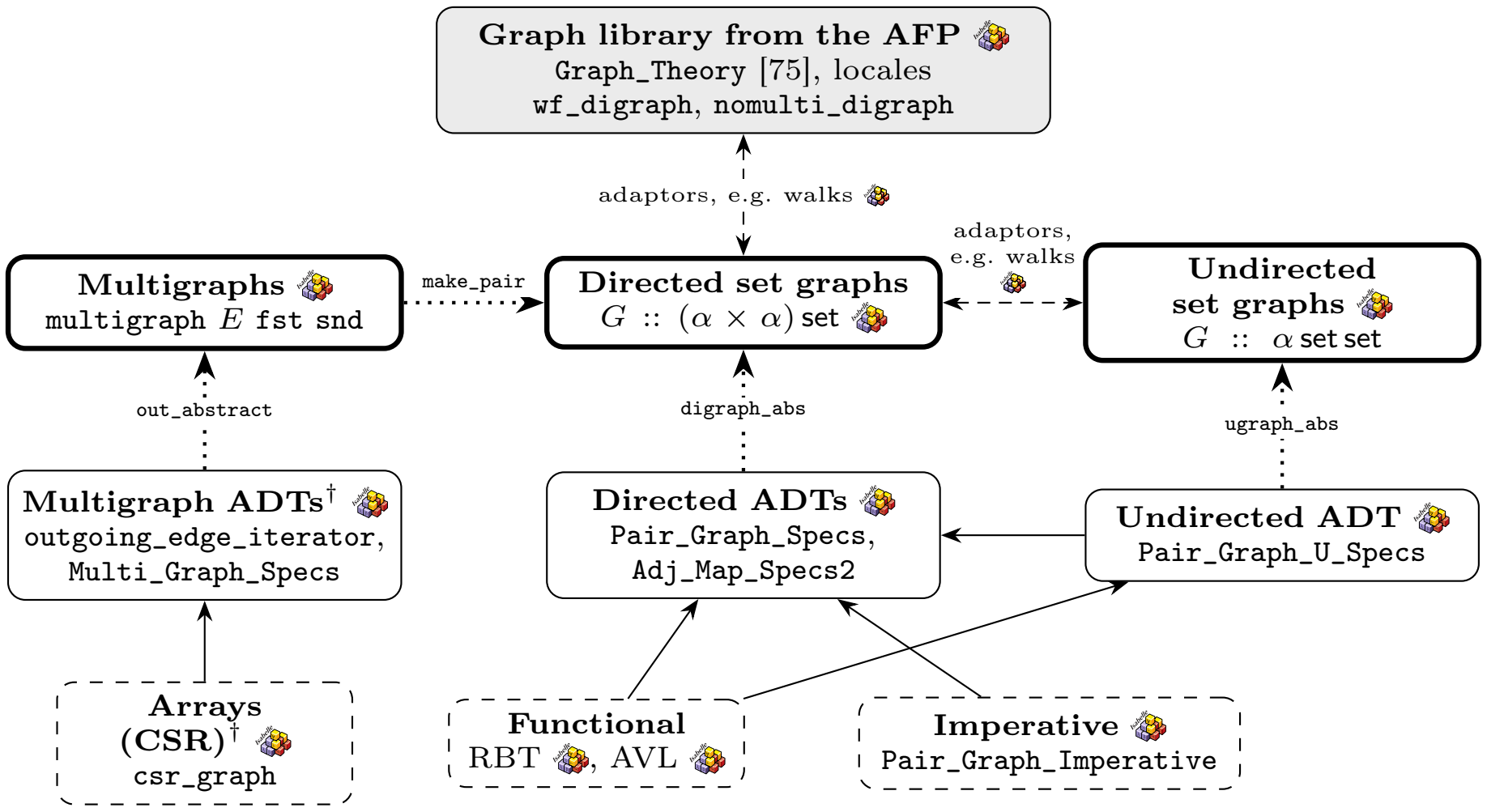

Our library has two main representations, shown in \cref{fig:graph-locales}
along with the connections between them: directed graphs, i.e.\ sets of pairs of
vertices~\cite[Chapter~24]{fdsBook}, and undirected graphs, i.e.\ sets of
two-element sets of vertices~\cite{BlossomIsabelleJAR}. Each of them has its own
notions of walks,
\isainl{Set_Graphs/Directed_Graphs/Vwalk.thy}{13}{16}{vwalk}
and
\isainl{Set_Graphs/Directed_Graphs/Awalk_Defs.thy}{29}{30}{awalk}
for directed graphs and
\isainl{Set_Graphs/Undirected_Graphs/Paths.thy}{105}{105}{walk\_betw}
for undirected graphs. The two
representations are connected by \emph{adaptors}, which are theories relating
the notions of one representation to those of the other, e.g.\ they relate paths, given as lists of vertices, of two distinct representations.
For example,
\isainl{Set_Graphs/Undirected_Graphs/Pair_Graph_Berge_Adaptor.thy}{95}{96}{walk\_betw\_iff\_vwalk\_bet}
states that a walk in an undirected graph $\graph$ is a walk in the directed graph
that has both orientations of every edge of $\graph$, and vice versa.
Thus, results proved for one representation can be used for the other.
Multigraphs, whose edges have an abstract type, are an example of this reuse: a path in a
multigraph, \isainl{Set_Graphs/Directed_Graphs/Multigraph.thy}{54}{55}{multigraph\_path},
is defined as a walk, \texttt{awalk}, over its edges mapped to vertex pairs
by \texttt{make\_pair}, so all lemmas about walks in directed graphs apply to
multigraphs without being proved anew.
Finally, adaptors to Noschinski's library~\cite{noschinskiGraphLib}, which was developed independently of ours,
prove that our definitions of arcs, reachability, walks, trails, paths, cycles
and decompositions of walks into cycles, (induced) subgraphs,
connectivity, and strongly connected components of directed graphs agree with
those of that library.

Some theorems about graphs are proved by induction on the graph's edge set. For example,
\isainl{Set_Graphs/Undirected_Graphs/Connected_Components.thy}{2124}{2128}{component\_has\_path\_distinct}
states that if every vertex of a finite graph $\graph$ has degree at most two,
then every connected component of $\graph$ is the vertex set of a path
without repeated vertices, and it is proved by induction on $\graph$. The
induction step has to know how the connected components change when an edge
$\uedge{\vertexa}{\vertexb}$ is added to $\graph$, which is given by a case
analysis
\isainl{Set_Graphs/Undirected_Graphs/Connected_Components.thy}{1528}{1542}{in\_insert\_connected\_componentE}
of Abdulaziz and Mehlhorn~\cite{BlossomIsabelleJAR}: every connected component
of the extended graph is a connected component of $\graph$, the new component
$\uedge{\vertexa}{\vertexb}$, the component of one endpoint extended by the
other endpoint, or the union of the components of both endpoints.

The library has further reasoning principles of this kind for other concepts,
none of which has been described before:
(1)~the same case analysis for walks~\isalink{Set_Graphs/Undirected_Graphs/Paths.thy}{316}{323},
for reachability~\isalink{Set_Graphs/Undirected_Graphs/Paths.thy}{2010}{2016},
for walks in directed graphs~\isalink{Set_Graphs/Directed_Graphs/Vwalk.thy}{1327}{1335},
and for the union of two graphs~\isalink{Set_Graphs/Undirected_Graphs/Paths.thy}{454}{461};
(2)~a lemma that splits reachability in the extended graph into reachability in
$\graph$ to and from the endpoints of the new
edge~\isalink{Set_Graphs/Undirected_Graphs/Paths.thy}{2034}{2038};
(3)~induction principles that build a walk one edge at a time, for walks in
undirected graphs~\isalink{Set_Graphs/Undirected_Graphs/Paths.thy}{197}{201},
for walks in directed graphs given as lists of vertices, extended at the
start~\isalink{Set_Graphs/Directed_Graphs/Vwalk.thy}{416}{420}
or at the end~\isalink{Set_Graphs/Directed_Graphs/Vwalk.thy}{484}{488},
or as lists of edges~\isalink{Set_Graphs/Directed_Graphs/Awalk.thy}{1314}{1320},
and for paths in multigraphs~\isalink{Set_Graphs/Directed_Graphs/Multigraph.thy}{194}{199};
(4)~induction on reachability in directed graphs, forwards from the
source~\isalink{Set_Graphs/Directed_Graphs/Pair_Graph.thy}{99}{103}
or backwards from the
target~\isalink{Set_Graphs/Directed_Graphs/Pair_Graph.thy}{106}{110};
and (5)~an induction principle that reduces a property of all walks between two
vertices to the walks without repeated vertices by removing closed
subwalks~\isalink{Set_Graphs/Directed_Graphs/Vwalk.thy}{1096}{1102}.

For proving algorithms correct, we need intricate graph-theoretic results, such as, for minimum-cost flows, a criterion for optimality and the preservation thereof~\isalink{Flow_Theory/Cost_Optimality.thy}{667}{669}\,\isalink{Mincost_Flow_Algorithms/Path_Aug_Opt.thy}{2520}{2525}, or Berge's lemma~\isalink{Basic_Matching/Berge.thy}{744}{748} for matchings.
Although in this paper we clearly focus on the algorithmic side, note that our infrastructure allows formal proofs of deep graph-theoretic results that we do not use for proving algorithms correct, for example, the Tutte--Berge formula~\isalink{Berge_Formula/Berge_Formula.thy}{1482}{1488}, or the Edmonds--Gallai decomposition~\isalink{Edmonds_Gallai/Edmonds_Gallai.thy}{1431}{1451}~\cite{gallaiKritischeGraphenII1963,gallaiMaximaleSystemeUnabhangiger1964,edmond1965blossom}.

\subsection{Implementing Graphs}
\label{sec:graph-impl}

The algorithms do not operate on the mathematical representations of graphs
directly, but on abstract data types (ADTs) of
graphs~\cite[Chapter~24]{fdsBook}. An ADT is a locale (\cref{sec:locales}) that
fixes the operations of a data structure and an invariant on its values, and assumes the specification of these operations in terms of an abstraction
function, which maps every value satisfying the invariant to one of the
mathematical representations. For example,
\isainl{Set_Graphs/Directed_Graphs/Pair_Graph_Specs.thy}{243}{251}{Pair\_Graph\_Specs}
combines a map ADT with an ADT of sets from which an element can be
chosen, maps every vertex to its out-neighbourhood, and is related to
directed graphs by
\isainl{Set_Graphs/Directed_Graphs/Pair_Graph_Specs.thy}{111}{111}{digraph\_abs}.
Accordingly, an algorithm is defined in terms of the ADT's operations, and
its proof uses the lemmas about the mathematical representation, which apply to
the abstraction of the graph. \Cref{tab:graph-adts} in \cref{app:library}
describes the graph ADTs of the library. The four ADTs in the main branch represent a graph as a map from
every vertex to the set or list of its neighbours, which suits algorithms that
look up and modify neighbourhoods, e.g.\ DFS and BFS.
The ADTs in development branches represent multigraphs whose edges at a vertex
are visited with a cursor, which suits algorithms that resume the scan of the
edges of a vertex where an earlier scan stopped, e.g.\ Dijkstra's algorithm.

An implementation of an ADT is an interpretation of its locale with concrete
operations, for which the locale assumptions have to be proved. The
interpretation yields the definitions and theorems of every algorithm that is
defined in terms of the ADT for these operations. The functional
implementations use the red-black trees (RBTs) and AVL trees of the Isabelle/HOL
distribution~\cite{fdsBook}, both for the map from vertices to their
neighbourhoods and for neighbourhoods themselves, e.g.\ the implementation
of \texttt{Pair\_Graph\_Specs} with RBTs~\isalink{Set_Graphs/Directed_Graphs/Pair_Graph_RBT.thy}{54}{55}.
The imperative implementations use the array maps or hash maps of the
separation logic framework for Imperative HOL~\cite{lammichSepRef} for the map,
and duplicate-free lists for the
neighbourhoods~\isalink{Set_Graphs/Graph_Algorithms/Refinements/DFS_Imperative.thy}{239}{240}\,\isalink{Set_Graphs/Graph_Algorithms/Refinements/DFS_Imperative.thy}{254}{255}.
In development branches, the CSR
representation~\isalink{Set_Graphs/Directed_Graphs/CSR_Graph_Spec.thy}{347}{348}
stores the edges in one array sorted by their sources, together with an array
of the positions at which the edges of every vertex begin and an array of
cursors. The edges leaving a vertex are thus contiguous in memory, which makes
scanning them cache-efficient, and advancing or resetting the cursor of a
vertex takes constant time. Note that since the first two arrays are never
modified, the CSR representation suits algorithms whose graph does not change.

\subsection{Modelling Algorithms}
\label{sec:locales}

The main mechanism we use to structure the library is Isabelle/HOL's
\emph{locales}~\cite{LocalesBallarin}.
A locale is a named context that fixes a number of constants, the parameters of the locale, and states assumptions about them; definitions in the locale can refer to these constants, and theorems proved in the locale can use these assumptions.
A locale can extend other locales with further constants and assumptions, and can later be
instantiated by giving concrete values for its constants and proving its assumptions for these values. This yields instances of all the definitions and theorems of the locale for these values. We usually split locales modelling algorithms into a \emph{specification locale}, which only fixes constants and their types, and a \emph{proof locale}, which extends the former by the assumptions.
For example, the specification locale \isainl{Set_Graphs/Directed_Graphs/Multigraph.thy}{11}{15}{multigraph\_spec}
fixes an edge set $\edges$, functions \texttt{fst} and \texttt{snd}
returning an edge's endpoints, and \texttt{create\_edge}
to construct an edge from two vertices. The proof locale
\isainl{Set_Graphs/Directed_Graphs/Multigraph.thy}{125}{129}{multigraph} assumes that $\edges$ is finite, that \texttt{fst} and \texttt{snd} return the vertices given to
\texttt{create\_edge}, etc.
We split locales in this way because a specification locale has no assumptions and therefore admits a \emph{global interpretation} for arbitrary values of its constants. Such a global interpretation is what Isabelle/HOL's code generator needs to produce executable code from the locale definitions.

We model the algorithms following the methodology of
Abdulaziz~\cite[Chapter~24]{fdsBook}, which was also used for the minimum-cost
flow algorithms~\cite{ScalingIsabelle}, the greedy and matroid intersection
algorithms~\cite{MatroidIsabelle}, and the Hungarian
method~\cite{abdulazizFormalPrimalDualAlgorithm2026b}. The variables that
change during a loop of an algorithm form its \emph{program state}, which is
modelled as a record with one field for each variable. The loop
is modelled as a recursive function that maps a state to the state in which the
loop stops. For instance, depth-first search,
\isainl{Set_Graphs/Graph_Algorithms/DFS.thy}{66}{84}{DFS},
is a function on records of type
\isainl{Set_Graphs/Graph_Algorithms/DFS.thy}{19}{19}{DFS\_state},
whose fields are a stack, a set of visited vertices, and a return value.
Termination is not proved when such a function is defined: instead,
Isabelle/HOL's function package~\cite{functionPackageIsabelle} provides a domain
predicate, e.g.\ \texttt{DFS\_dom}, which holds for the states on which the
function terminates, and we first prove the function partially correct for all
states in its domain and only then prove that the states that satisfy the
invariants are in the domain.

Locales are used in two ways when modelling algorithms. First, the data
structures used by the algorithm are parameters of its locale, specified by
the ADTs of \cref{sec:graph-impl}. Thus, the algorithm is defined and verified
once and can be executed with every implementation of these ADTs.
Second, a sub-procedure that the algorithm calls is a parameter of its locale, specified by assumptions that state what the sub-procedure computes, so that the
algorithm can be verified before the sub-procedure is defined. For instance,
the locale of the main loop of Edmonds' blossom
algorithm~\isalink{Edmonds_Blossom_Executable/Find_Max_Matching.thy}{32}{41}
fixes a function that searches for an augmenting path and assumes that this
function finds such a path whenever one exists and that every path it returns
is augmenting.
This is stepwise refinement~\cite{BlossomIsabelleJAR}: the assumed
sub-procedures are defined and verified in later steps, until no assumption is
left that is not satisfied by a definition.

\Cref{tab:library} states, for every algorithm in the library, which
properties of it are proved and the size of the theories that verify it, and
\cref{tab:algorithms} in \cref{app:library} also states whether its loops are
modelled with the methodology of \cref{sec:proofs}, the number of invariants of
its loops, and its main theorem.

\todo{Revise the line numbers and formalisation sizes, as they are AI
generated: the \emph{Lines} and \emph{Inv.} columns of \cref{tab:algorithms}
and the \emph{Lines} column of \cref{tab:library},
and the sizes in the opening paragraph of \cref{sec:methodology}.}

\subsection{Proofs about Algorithms}
\label{sec:proofs}

The induction principle that the function package derives for a function modelling a loop contains the function's entire control flow, making it hard to use in
interactive proofs~\cite[Chapter~24]{fdsBook}. Accordingly, for every execution
path of the loop, we define a predicate stating that the path is taken and
a function computing the state at the end of the path, e.g.\ \isainl{Set_Graphs/Graph_Algorithms/DFS.thy}{89}{99}{DFS\_call\_1\_conds}
and
\isainl{Set_Graphs/Graph_Algorithms/DFS.thy}{110}{117}{DFS\_upd1}
for the path that pushes an unvisited vertex onto the
stack. From these definitions we prove three lemmas that characterise the
function only in terms of its execution paths: (1)~a case analysis,
\isainl{Set_Graphs/Graph_Algorithms/DFS.thy}{196}{201}{DFS\_cases},
stating that one of the paths is taken; (2)~simplification rules,
\isainl{Set_Graphs/Graph_Algorithms/DFS.thy}{213}{218}{DFS\_simps},
stating the result of the function for each path; and (3)~a computation
induction principle,
\isainl{Set_Graphs/Graph_Algorithms/DFS.thy}{226}{231}{DFS\_induct},
whose induction step has one hypothesis for every recursive path, only
mentioning the predicate and the update function of that path. For every
path predicate, we also prove introduction and elimination rules, e.g.\ the
elimination rule
\isainl{Set_Graphs/Graph_Algorithms/DFS.thy}{101}{106}{DFS\_call\_1\_conds},
which turns the predicate into the facts about the state that it implies.

Invariants are predicates on the state, and, instead of one comprehensive loop
invariant, we state many separate invariants, e.g.\ sixteen for the search for
alternating paths in Edmonds' blossom algorithm~\cite{BlossomIsabelleJAR}, each
of which is proved to be preserved assuming whatever other invariants are
needed. Every invariant is first proved to be preserved by every recursive path,
e.g.\ \isainl{Set_Graphs/Graph_Algorithms/DFS.thy}{315}{317}{invar\_1\_holds\_1},
which is where the mathematical reasoning about the algorithm takes place and
where the reasoning principles of \cref{sec:graphs} are used. It is then
lifted to the whole function by the computation induction principle, e.g.\ \isainl{Set_Graphs/Graph_Algorithms/DFS.thy}{329}{331}{invar\_1\_holds}.
These two kinds of lemmas are combined by Isabelle/HOL's classical reasoner and
simplifier: we collect the introduction and elimination rules of the path
predicates and of the invariants, and the preservation lemmas, in named
collections of theorems~\isalink{Set_Graphs/Graph_Algorithms/DFS.thy}{23}{30}. They are given to the classical reasoner as introduction and elimination
rules, so that e.g.\ the proof of \texttt{invar\_1\_holds} consists of the
computation induction, the case analysis, and a single call of \texttt{auto}.
Finally, termination is proved by a measure that decreases on every recursive
path of a state that satisfies some of the invariants, and the correctness
theorems, e.g.\ \isainl{Set_Graphs/Graph_Algorithms/DFS.thy}{739}{741}{DFS\_correct\_1},
follow from the invariants and the predicates of the paths on which the loop
returns.

The loops of breadth-first
search~\isalink{Set_Graphs/Graph_Algorithms/BFS_2.thy}{246}{251},
the best-in-greedy algorithm, which yields Kruskal's
algorithm~\isalink{Matroids_Greedy/BestInGreedy_base.thy}{181}{187},
successive shortest
paths~\isalink{Mincost_Flow_Algorithms/Successive_Shortest_Path.thy}{258}{262},
capacity
scaling~\isalink{Mincost_Flow_Algorithms/Scaling.thy}{1379}{1384},
the two inner loops of Orlin's
algorithm~\isalink{Mincost_Flow_Algorithms/Send_Flow.thy}{421}{426}\,\isalink{Mincost_Flow_Algorithms/Maintain_Forest.thy}{331}{335},
the Hungarian
method~\isalink{Hungarian_Method/Hungarian_Method_Top_Loop.thy}{175}{180},
Dinic's algorithm~\isalink{Dinic/Dinic.thy}{99}{103},
and matroid
intersection~\isalink{Matroid_Intersection/Matroid_Intersection_Path_Loop.thy}{140}{145}
are verified in the same way, where each link points to the computation
induction principle of the loop. However, Edmonds' blossom algorithm, which was
the first algorithm in the library, uses path predicates only in its search for
alternating paths and reasons about its other loops with the induction
principles that the function package derives.

\subsection{Imperative Implementations}
\label{sec:imperative}
So far, we have described functional algorithms as locales parameterised over
their data structures and operations.
However, imperative implementations are much faster than functional ones: on
the BFS benchmarks of \cref{sec:evaluation}, the imperative export to SML is
about two orders of magnitude faster than the functional one on the smallest
grids, and about four orders of magnitude faster on the largest grids and on
the road networks (\cref{tab:bfs-square,tab:bfs-road}).
One approach to verifying imperative code in Isabelle/HOL is Lammich's
Isabelle Refinement Framework~\cite{lammichMonads} with its tool
Sepref~\cite{LammichRefinement}.
The Refinement Framework, however, requires the abstract algorithm to be stated in its
non-determinism error monad.
Accordingly, using it would require restating every algorithm of the library in
that monad, and would give up a property of our methodology: an algorithm is a
recursive function in a locale whose correctness is proved by induction.
We favour this approach because reading or extending the library needs no vocabulary beyond what an
average Isabelle/HOL user already knows.
The problem we had to solve was thus to refine our algorithms to imperative
implementations without changing the locale-based methodology.
Note that the monad models non-determinism, e.g.\ choosing an arbitrary element
from a set, and failure.
We solve this problem with an intermediate functional implementation: we first
instantiate the functional locale with a functional data structure, e.g.\ a
list, which fixes the choices that the abstract algorithm leaves open, and we
then refine this deterministic functional implementation to an imperative one.

We use the separation logic framework for partial correctness in Imperative HOL (SL-Framework for short), which we originally developed for total correctness and extended to partial correctness to implement a WebAssembly interpreter~\cite{WattTLM23}.
For this project, we improved its automation, added concepts such as monadic folds and while-loops, and proved that Hoare triples are admissible for fixed-point induction over recursive partial functions in the heap monad. We also reuse its library of imperative data structures.

Imperative HOL shallowly embeds imperative programs into a heap-error monad, from which Isabelle's code generator produces actual imperative programs~\cite{Bulwahn+08}.
The SL-Framework adds a separation logic and a verification condition generator. Its central concept is the Hoare triple $\langle P\rangle\,c\,\langle\lambda x.~Q~x\rangle$, with separation logic assertions $P$ and $Q$. It expresses partial correctness: if $P$ holds for the initial heap and $c$ returns without errors, then $Q~x$ holds for the result $x$ and the final heap. An error means non-termination or a runtime failure, e.g.\ a failed bounds check.

Since we mainly verify an imperative implementation w.r.t.\ the functional one, all our assertions are \emph{refinement assertions}.
Such an assertion $A~a~c$ relates an abstract value $a$ to a concrete one $c$ that may refer to the heap.
For example, \isainl{Imperative/Separation_Logic_Imperative_HOL_Partial/Examples/Array_Map_Impl.thy}{20}{21}{is\_iam m a} states that \texttt{a} points to an array on the heap that implements the map \texttt{m} from natural numbers to values, the indices being the keys.

To refine a graph algorithm, we set up a locale that inherits from the algorithm's functional locale and fixes imperative versions of its data structures and subroutines, assuming that they are correct.
Thus, the functional locale refines abstract concepts such as sets and graphs to functional implementations, and the imperative locale refines these further to imperative ones.

In DFS, for instance, the open choice is the next successor to process, which the set ADT provides by a choice operation~\isalink{Data_Structures/Set_Choose.thy}{11}{15}, and we implement vertex sets by distinct lists, which we refine differently depending on their use.
Successor sets remain purely functional lists, as DFS only picks elements from them and iterates over them.
The visited set, however, needs an efficient membership test, so we refine its list back to a set and then use any suitable imperative set implementation, e.g.\ bit-sets or hash-sets.
We note that fixing an implementation, e.g.\ of sets using lists, is only a workaround. 
\todo{@Mohammad: maybe explain here why not allowing different implementations in the functional locale, but only having, e.g., one vset. It also restricts your choices for the purely functional implementation, but probably reduces boilerplate?}

Depending on the complexity of the algorithm and the intended flexibility, the imperative locale either partially instantiates the functional locale with functional data structures, or leaves all instantiations to the last step\todo{@Mohammad: what's the advantage of the latter, as, e.g., done in BFS? Its disadvantage is clearly bigger locale-parameter lists and more boiler-plate heavy instantiations. }.
In both cases, it is parameterised over the imperative implementations, which we choose when instantiating it. For DFS, for example, we generate one implementation with natural numbers as vertices and array-based maps and sets~\isalink{Set_Graphs/Graph_Algorithms/Refinements/DFS_Imperative.thy}{245}{247}, and one based on hash maps and hash sets~\isalink{Set_Graphs/Graph_Algorithms/Refinements/DFS_Imperative.thy}{260}{262}.

We define a refinement assertion for the state, typically as the product of the refinements of the state components. We then implement refinements of the functions of the functional version, i.e.\ some helpers and the main loop, and prove that they indeed refine their functional counterparts: on related arguments, they return related results.
The loop is implemented as a partial function~\cite{Krauss06,functionPackageIsabelle} in Imperative HOL's heap monad, and we prove by fixed-point induction that it refines the tail-recursive partial function\todo{this should be explained above, and a name for "tail-recursive partial function" should be introduced} of the functional version. For DFS, we prove~\isalink{Set_Graphs/Graph_Algorithms/Refinements/DFS_Imperative.thy}{101}{101} that, on implementations of a graph and a state, the imperative DFS leaves the graph unchanged and returns an implementation of the functional DFS's result state.
The proof is largely canonical: inside the induction, we unfold one step of the functional implementation and split it into its conditional branches; for each branch, we unfold the imperative implementation, eliminate its other branches, and prove the refinement with the verification condition generator. This step can be largely automated\todo{@Mohammad: But I see that you decided to keep it very-low level, single stepping over bind-rules in many proofs, presenting the automated version as a commented-out alternative. Explain why.}.

Finally, we combine both refinements into a Hoare triple stating that the imperative algorithm satisfies the abstract specification, again in a canonical and partially automated way. For example, from the result of the imperative DFS, we obtain a walk to the target or the absence of one~\isalink{Set_Graphs/Graph_Algorithms/Refinements/DFS_Imperative.thy}{219}{224}.

\subsection{Other Methodological Points}
\label{sec:other-methodology}

\paragraph{Algorithms on abstract structures.}
Some algorithms are verified on abstract combinatorial structures and then
instantiated to graphs, so that one correctness proof serves several
algorithms.
For instance, Kruskal's algorithm is an interpretation of the locale of the
best-in-greedy algorithm on matroids, with the forests of the input graph as the
independent
sets~\isalink{Spanning_Trees/Kruskal_Functional.thy}{235}{254},
and maximum cardinality bipartite matching is computed by matroid intersection,
where the independent sets of the two matroids are the sets of edges that share
no vertex on the left and on the right side,
respectively~\isalink{Matroid_Intersection/Applications/Max_Bipartite_Matching_Matroid.thy}{1545}{1552}.

\paragraph{Verified algorithms as building blocks.}
Verified algorithms are also used as concrete sub-procedures of other
algorithms, in contrast to the sub-procedures that are locale parameters
(\cref{sec:locales}).
For instance, the independence oracle of Kruskal's algorithm runs the verified
DFS to decide whether an edge closes a cycle in the current
forest~\isalink{Spanning_Trees/Kruskal_Functional.thy}{218}{222},
and the proof of the oracle uses the correctness theorems of DFS.

\paragraph{Verified reductions.}
We verify reductions between problems, so that an algorithm for one problem
solves another.
For instance, we prove that every minimum-cost flow problem with capacities can
be transformed into an uncapacitated one, such that flows of the two problems
correspond, costs are preserved, and optimal flows are mapped to optimal
flows~\isalink{Flow_Theory/Hitchcock_Reduction.thy}{72}{103}.
Accordingly, Orlin's algorithm, which is verified for uncapacitated
problems~\cite{ScalingIsabelle}, also solves problems with
capacities~\isalink{Mincost_Flow_Algorithms/Use_Orlins_Algorithm/Usage_Capacitated.thy}{691}{693}.

\paragraph{Shared optimality theorems.}
We prove the characterisations of optimal solutions separately from the
algorithms and share them between algorithms, so that the proof of an algorithm
only needs to show the condition of the characterisation.
For instance, Berge's lemma~\cite{BergesLemma}, i.e.\ that a matching has
maximum cardinality if and only if there is no augmenting
path~\isalink{Basic_Matching/Berge.thy}{744}{748},
is used in the proofs of the blossom algorithm, the Hungarian method, and
RANKING, and the characterisation of minimum-cost flows by the absence of
augmenting cycles of negative
cost~\isalink{Flow_Theory/Cost_Optimality.thy}{667}{669}
is used in the proofs of successive shortest paths, capacity scaling, and
Orlin's algorithm.

\paragraph{Running-time analysis.}
We analyse running times in a locale that extends the locale of the algorithm
with the costs of the ADT operations as parameters. It defines a function
with the same recursion as the algorithm, which returns the time along with
the result.
For Orlin's algorithm, we prove in such a
locale~\isalink{Mincost_Flow_Algorithms/Orlins_Time.thy}{13}{18}
that the time is bounded by a sum of products of the operation costs and terms
in the number of vertices, and that the result is that of the
algorithm~\isalink{Mincost_Flow_Algorithms/Orlins_Time.thy}{1789}{1798}.
Accordingly, the algorithm and its correctness proof are not changed by the
analysis.

\paragraph{Certification.}
\todo{Name the hybrid solvers of \cref{sec:evaluation}: neither the text,
nor the figures, nor the appendix says which they are, and Overleaf has
dropped ``with verified certificate checking and cleanup'' from that
sentence. Also check whether the verified checker is in the library at
49f0962.}
In addition to verified algorithms, we use verified code to check, and if
necessary to clean up, the output of an unverified solver, which is the
approach of LEDA's certifying algorithms~\cite{LEDAbook}, except that the
checker itself is verified~\cite{matchingCertIsabelle}.

%% file: paper_empirical.tex
\section{Empirical Evaluation}
\label{sec:evaluation}

The library contains many verified implementations of graph algorithms and, given the space limitations, we focus on evaluating those for the main problems: 1.\ depth-first (DFS) and breadth-first search (BFS) for reachability, 2.\ Dijkstra's algorithm for single-source shortest paths, 3.\ Kruskal's algorithm~\cite{fc0df122-3305-33a8-bac5-7a6fc3666dfb} for minimum spanning forests, 4.\ push-relabel (PR) for maximum flow~\cite{goldbergNewApproachMaximumflow1988}, 5.\ the Hungarian method (HM)~\cite{Jacobi,HungarianMethodAssignment,kuhnVariantsHungarianMethod1956} for minimum-weight perfect bipartite matching (minPBM), 6.\ (weighted) bipartite ($b$-)matching, with matroid intersection (MI)~\cite{Lawler1975,FRANK1981328}, and 7.\ hybrid solvers for minPBM and minimum-cost flows. 
Our focus is on running time: we compare the code exported from our verified implementations against unverified implementations in widely used graph libraries, such as SciPy (Cython), Sage and NetworkX~\cite{scipyManual,sageMatroid,networkxManual} (Python), JGraphT~\cite{michailJGraphTJavaLibrary2020} (Java), LEMON~\cite{dezsoLEMON2011} (C++), which is widely recognised as one of the fastest graph libraries, and against reference solvers recommended for the DIMACS challenges, from which most of the problem families are also taken.


We export code to SML, Scala, and Go using Isabelle's code generator~\cite{haftmannCodeGenerationHigherOrder2010} and its Go extension~\cite{isabelleGo}, which we extended to export Imperative HOL to Go.
The exceptions are the hybrid solvers, whose verified part we export only to SML.
We export to 32-bit integers (SML default) and compile SML with \texttt{mlton} with high inlining.
We conduct the measurements on an AMD EPYC 7343 (Milan, Zen 3) with a timeout of 1200\,s.
The runs are scheduled to share the RAM but not the cache hierarchy.
Because of the graphs' sizes, we only take between 3 and 10 instances per size, run each instance 3 times, take the median for each instance, and average over all instances.

If a library can combine building data structures or graph representations with input file parsing, we count the respective time as parsing, which is excluded from the measurement.
If the data representation has to be built separately from parsing, as in the case of our verified code, it is included in the measurement.
Except for Kruskal's algorithm, all our imperative implementations scan graphs by neighbourhoods, and we use the cache-efficient CSR representation.
Despite the cost of CSR buildup being an additional $\mathcal{O}(n+m)$, we generally use it since our implementations of simpler algorithms, which themselves are $\mathcal{O}(n+m)$ or close, are intended as repeatedly executed sub-procedures of more complex algorithms, which amortises the cost.
Note the convention that $n$ denotes the number of vertices and $m$ denotes the number of edges in a graph.

We present plots (running time in seconds against the number of vertices $n$ on a log-log scale) and tables, with the latter in the appendix.
Our code exports' plots usually have the same slope, with the vertical shift accounting for the constant factor from language differences.
SML beats Scala, which in turn beats Go, and imperative implementations beat functional ones by orders of magnitude (OoM).
Imperative SML, which is always our best-performing variant and which we focus on, is plotted as~\ref{leg-search:sml_csr}.
Generally, the gap to the top-performing reference implementations (`the libraries') is much smaller for simple algorithms, probably because there is less room for heuristics and low-level optimisations.
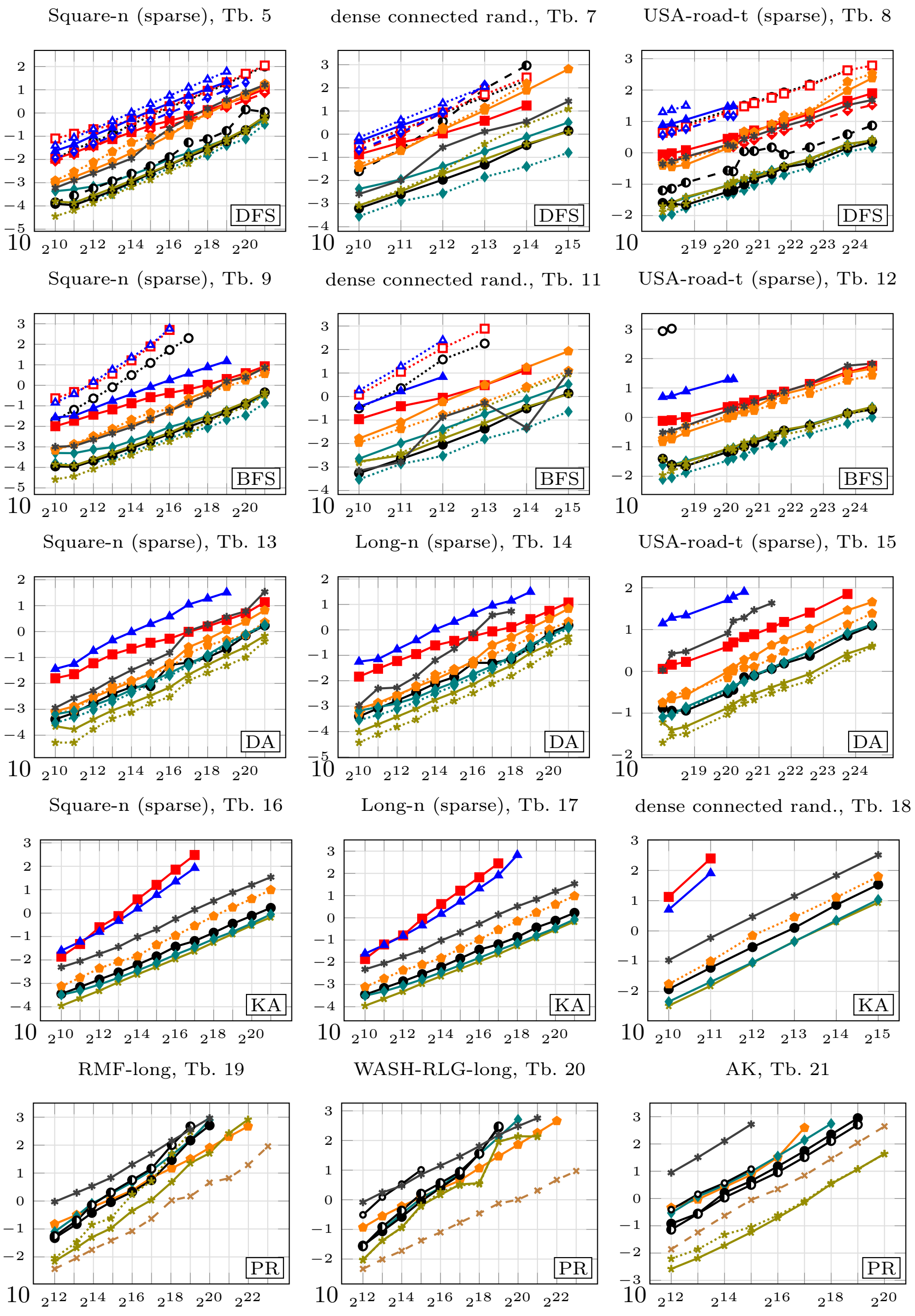
\begin{figure}[p]
\begin{minipage}[t][\textheight][t]{\textwidth}
\centering
\input{measurements/tex_figures/search}\par\nointerlineskip
\input{measurements/tex_figures/dijkstra}\par\nointerlineskip
\input{measurements/tex_figures/kruskal}\par\nointerlineskip
\input{measurements/tex_figures/push_relabel}
\caption{\ref{leg-search:sml_csr}~SML CSR, \ref{leg-search:sml_lists}~SML adjacency lists, \ref{leg-search:sml_rbt}~SML RBT, \ref{leg-search:scala_csr}~Scala CSR, \ref{leg-search:scala_lists}~Scala  adjacency lists, \ref{leg-search:scala_rbt}~Scala RBT, \ref{leg-search:go_csr}~Go CSR, \ref{leg-search:go_lists}~Go adjacency lists, \ref{leg-search:go_rbt}~Go RBT, \ref{leg-search:lemon_csr}~LEMON CSR, \ref{leg-search:lemon_linked}~LEMON linked, \ref{leg-search:jgrapht_csr}~JGraphT CSR, \ref{leg-search:jgrapht_hashgraph}~JGraphT hash, \ref{leg-search:scipy_csr}~SciPy CSR, \ref{leg-search:scipy_prebuilt}~SciPy prebuilt, \ref{leg-search:networkx}~NetworkX. PR: \ref{leg-push_relabel:sml_hl}~SML HL, \ref{leg-push_relabel:sml_fifo}~SML FIFO, \ref{leg-push_relabel:sml_afp}~SML AFP, \ref{leg-push_relabel:jgrapht}~JGraphT, \ref{leg-push_relabel:scipy_prebuilt}~SciPy, \ref{leg-push_relabel:hipr}~GC.}
\label{fig:dfs}
\label{fig:bfs}
\label{fig:dijkstra}
\label{fig:push-relabel}
\label{fig:kruskal}
\end{minipage}
\end{figure}

\paragraph{DFS and BFS.}
We let the solvers explore all vertices reachable from a random source on three families of graphs.
\Cref{fig:bfs} shows a performance difference of OoMs between imperative (with CSR) and purely functional (with RBTs) code exports.
For DFS, we also include a semi-imperative export with an adjacency list representation, whose performance is in between but closer to CSR-based (RBT-based) on sparse (dense) graphs.
For the libraries, we measure both with a native graph representation that can be built during parsing and with the more efficient CSRs.
SciPy's native representation is CSR, and here we have measurements with and without its buildup:
the cost of buildup is clearly visible on dense graphs, where it is also asymptotically highest.
Since the CSRs are only used once, CSR-based implementations of DFS and BFS generally perform worse in comparison to those where the representation is built during parsing.
For sparse graphs, our imperative SML code with CSRs is on par with the libraries, and is within an OoM for the dense family.
If the libraries are also charged with the cost of building CSRs, it seems that we can beat them by a small margin.

\paragraph{Dijkstra's (DA) and Kruskal's Algorithms (KA).}
\Cref{fig:dijkstra} shows that our SML is still within an OoM of the best library or narrows to it at scale.
For DA, the difference is bigger for DFS and BFS, which we attribute to the higher implementation complexity of the algorithm, in particular to the priority queue. 
For KA, where we reuse an implementation of a union-find data structure by Stevens and Ghidini~\cite{stevens_unionfind}, our SML is also well within an OoM of the best library's performance.

\paragraph{Push-Relabel (PR).}
Apart from the libraries, we evaluate our PR exports with highest-label (HL) and FIFO selection heuristics, an export of the older formalisation~\cite{LammichFlows} (AFP) with FIFO, and Goldberg and Cherkassky's implementation~\cite{cherkasskyImplementingPushRelabel1997} (GC), which has been tuned for decades.
SciPy does not offer PR, so we use its implementation of Dinic's algorithm instead.

The algorithm has ample room for heuristics, e.g.\ all implementations use the gap heuristic~\cite{ahujaNetworkFlowsTheory1993,cherkasskyImplementingPushRelabel1997}, and \cref{fig:push-relabel} seemingly shows the effect:
on RMF and WASH-RLG~\cite{johnsonNetworkFlowsMatching1993}, JGraphT and NetworkX, both much slower for the simpler algorithms from above, beat (or start to beat at scale) both of our SML exports, which hit the timeout.
In contrast to our implementation (and to LEMON), they offer an effective heuristic: BFS-based global relabelling (GL).
Our two code exports offer HL and FIFO, of which HL is known to usually perform a bit better~\cite{ahujaNetworkFlowsTheory1993,cherkasskyImplementingPushRelabel1997}.
GL seems to be good enough to compensate for JGraphT's FIFO heuristic and the language difference.
On RMF and WASH-RLG, we only beat the code export from the older formalisation~\cite{LammichFlows}.
The latter does not even work at larger sizes because of an overflow during the buildup of its memory-consuming data structures (adjacency matrices).
Our code performs asymptotically and in terms of OoM far worse than GC.
By design, the AK family affects the gap heuristic~\cite{cherkasskyImplementingPushRelabel1997}, which interacts with GL~\cite{cherkasskyImplementingPushRelabel1997}, and GL seems to become much less effective, which might reduce LEMON's disadvantage of lacking it.
LEMON's plot is parallel to those of our SML exports and two OoMs below them, which indicates a large constant factor.

We plan to add GL to our PR, leading to the combination of gap heuristic, GL and HL, which is known for its strong performance~\cite{cherkasskyImplementingPushRelabel1997}.
We expect a significant speedup since GL is seemingly able to make NetworkX, which is pure Python, beat compiled SML.


\paragraph{Hungarian method (HM).}
Our HM for minPBM is optimised for sparse graphs, and we borrow a few, but not all, practical optimisations from Jonker and Volgenant's (JV) high-performing version~\cite{jonkerShortestAugmentingPath1987}.
We compare against SciPy's JV for dense (LSA) and sparse adjacency matrices (MWFBM), JGraphT's dense HM (KM) and sparse HM (MWBM), Goldberg and Kennedy's CSA solver~\cite{goldbergEfficientCostScaling1995}, and Google's OR-Tools~\cite{ortoolsManual} (OR).
The latter two implement cost-scaling algorithms.
We use the families \textit{low-cost (LC)}, \textit{two-cost (TC)}, and \textit{geometric (GEO)}~\cite{goldbergEfficientCostScaling1995}.

In \cref{fig:hungarian}, our HM in SML is within two OoMs of the best three solvers, and for TC, even within one, but the effect does not seem to scale for LC.
We think that the smaller gap for TC is due to 1.\ more productive primal-dual adjustments because there are only two cost values, and 2.\ more scaling phases because one of the two costs is large.
Our measurements for PR and HM confirm the higher difficulty of producing verified implementations of complex and heuristic-heavy algorithms, especially ones that are somewhat competitive with respected libraries that have often been tuned for decades.

\input{measurements/tex_figures/hungarian}

\paragraph{Matroid Intersection (MI).}
Matroids are used to model $b$-matching, in which every vertex $v$ can be matched up to $b(v)$ times.
The abstract algorithms obtain information about the concrete problem by \textit{oracles}.
One iteration of our MI takes $\mathcal{O}(m^2)$ oracle calls.
We offer constant-time oracles to model bipartite matching as MI.
Sage also offers unweighted and weighted MI in pure Python, though the weighted one uses a different algorithm.
The pure algorithms also take $\mathcal{O}(m^2)$ oracle calls per iteration, and internal glue code adds a factor of $\mathcal{O}(m)$.
An oracle built from Sage's primitives takes $\mathcal{O}(m)$ time, and our own customised one takes constant time, resulting in $\mathcal{O}(m^4)$ and $\mathcal{O}(m^3)$ per iteration, respectively.
Bipartite matching as MI takes $\mathcal{O}(B \cdot \iota)$ time, where $\iota$ is the time of one iteration and $B = \sum \min \lbrace \deg(v), b(v) \rbrace$.

We benchmark the resulting solvers for (un-)weighted ($b$-)matching on LC, TC, and GEO.
\Cref{fig:mi} shows the results (see \cref{fig:mi-appendix} in the appendix for more).
The figures mostly seem to confirm the asymptotics.
TC tends to be solved a bit faster than LC, which we attribute to the simpler weight structure.
Unexpectedly, Sage instantiated with the faster Python oracle improves relatively for the dense GEO family, perhaps due to the weight variance.
Usually, SML beats the fast Python implementation by OoMs, but for weighted matching on GEO, they are on par and SML times out for the biggest instance.
However, for weighted $b$-matching (see \cref{fig:mi-appendix} in the appendix), SML changes little and Sage falls behind, which must be the effect of allowing multiple matching edges per vertex.

\paragraph{Hybrid Solvers.}
We offer certificate checking for minimum-cost flows (MCF) and weighted bipartite matching (WBM).
In combination with the verified network simplex and HM, respectively, we build hybrid solvers that first let an unverified program (`oracle') solve the problem and generate a certificate.
If the verified checker rejects the certificate, we use our verified network simplex or HM, respectively, to solve the problem.
In contrast to HM, the network simplex can be warm-started, i.e.\ it is initialised with the unverified solver's rejected solution.
Both checkers examine a simple property of every edge, in $\mathcal{O}(n+m)$ time.

For example, by extending implementations of different algorithms from LEMON with certificate generation, we obtain four oracles for MCFs from which the user can select.
The \textit{cost ratio of verification (CV)} is, for the fastest oracle, the ratio between the total CPU time (which includes certificate generation and checking) and the time of the oracle's initialisation and main loop.
\Cref{tab:mcf-summary} shows the results.
On \texttt{netgen\_deg}, where for fixed $n$ the density grows up to $n^2$, the CV increases up to $3.69$.
Since both oracle and checker have to deal with bigger graphs, a roughly constant CV would be expected, or even a decreasing one since no oracle's running time is linear.
These are worst-case running times, however, and we believe the CV increases because, as the graph grows, for a fixed increase in the checker's time, the oracle has the same or better chances to route the flow.
For growth in $n$ and moderate or no growth in density (\texttt{netgen\_lo\_sr} and \texttt{road\_paths}), the CV decreases as expected.
We benchmark on 18 families, and $3.69$ is the highest CV that we measured for any family and size.

\Cref{tab:mcf-summary} also gives the CV of the hybrid solver for minPBM that combines the CSA solver~\cite{goldbergEfficientCostScaling1995} with our verified certificate checker.
We use the same families as for HM, in which the graphs grow in both $n$ and $m$, and observe the same behaviour as for \texttt{netgen\_lo\_sr} and \texttt{road\_paths}: the CV decreases at scale.
For LC and TC, the certificate checker does the same amount of work, but since CSA does worse on TC, the CV is smaller.
On the dense GEO family, the CV is much higher since the greater density increases the checker's relative work.

Hybrid solvers perform almost by design well within an OoM of unverified ones but we cannot reduce the CV to $1$.
They are still practical, especially for dense instances, where a fully verified but slower main loop would hurt most.

\input{measurements/tex_tables/mcf_summary}

%% file: measurements/tex_figures/search.tex
\begin{tikzpicture}
\begin{groupplot}[group style={group size=3 by 2, horizontal sep=0.7cm, vertical sep=1.1cm}, xmode=log, ymode=log, log basis x=2, ytickten={-6,...,6}, log basis y=10, yticklabel={$\pgfmathparse{int(round(\tick))}\pgfmathresult$}, after end axis/.code={\node[anchor=north east, font=\footnotesize, inner sep=1pt] at (rel axis cs:0,0) {$10$};}, grid=major, grid style={gray!25}, tick label style={font=\tiny}, label style={font=\scriptsize}, every axis plot/.append style={thick, mark size=1.5pt, legend image code/.code={\draw[mark repeat=2, mark phase=2] plot coordinates {(0cm,0cm) (0.15cm,0cm) (0.3cm,0cm)};}}, width=0.405\textwidth, height=4cm, title style={font=\scriptsize, align=center}]
\nextgroupplot[extra description/.code={\node[anchor=south east, draw=black, fill=white, font=\scriptsize, inner sep=1.5pt] at ([xshift=-2pt,yshift=2pt]rel axis cs:1,0) {DFS};}, title={Square-n (sparse), Tb.~\ref{tab:dfs-square}}, xtick={1024,2025,4096,8190,16384,32761,65536,131044,262144,524176,1048576,2096704}, xticklabels={$2^{10}$,,$2^{12}$,,$2^{14}$,,$2^{16}$,,$2^{18}$,,$2^{20}$,}]
\addplot[teal, mark=diamond*, mark options={solid}] coordinates {(1024,0.000433) (2025,0.000519) (4096,0.000698) (8190,0.001093) (16384,0.002185) (32761,0.004355) (65536,0.010552) (131044,0.019159) (262144,0.037084) (524176,0.076141) (1048576,0.155529) (2096704,0.550976)};
\label{leg-search:scipy_csr}
\addplot[teal, densely dotted, mark=diamond*, mark options={solid}] coordinates {(1024,0.000129) (2025,0.000153) (4096,0.000218) (8190,0.000339) (16384,0.000842) (32761,0.001855) (65536,0.004301) (131044,0.008593) (262144,0.015142) (524176,0.040155) (1048576,0.076393) (2096704,0.330905)};
\label{leg-search:scipy_prebuilt}
\addplot[black, mark=*, mark options={solid}] coordinates {(1024,0.000123) (2025,0.000108) (4096,0.000217) (8190,0.000449) (16384,0.001002) (32761,0.00252) (65536,0.005811) (131044,0.01303) (262144,0.02765) (524176,0.061261) (1048576,0.184465) (2096704,0.728551)};
\label{leg-search:sml_csr}
\addplot[black, dashed, mark=halfcircle*, mark options={solid, rotate=90}] coordinates {(1024,0.000143) (2025,0.000279) (4096,0.000562) (8190,0.001146) (16384,0.00243) (32761,0.005092) (65536,0.012568) (131044,0.053676) (262144,0.07462) (524176,0.168571) (1048576,1.42325) (2096704,1.1119)};
\label{leg-search:sml_lists}
\addplot[black, densely dotted, mark=*, mark options={solid, fill=white}] coordinates {(1024,0.010764) (2025,0.026324) (4096,0.05928) (8190,0.117847) (16384,0.326318) (32761,0.721163) (65536,1.75033) (131044,3.69561) (262144,8.11216) (524176,17.8888) (1048576,43.8441) (2096704,100.514)};
\label{leg-search:sml_rbt}
\addplot[red, mark=square*, mark options={solid}] coordinates {(1024,0.010836) (2025,0.019528) (4096,0.040326) (8190,0.075298) (16384,0.150693) (32761,0.293712) (65536,0.458033) (131044,0.758089) (262144,1.32403) (524176,2.5008) (1048576,4.98455) (2096704,10.914)};
\label{leg-search:scala_csr}
\addplot[red, dashed, mark=halfsquare*, mark options={solid}] coordinates {(1024,0.008289) (2025,0.016221) (4096,0.03474) (8190,0.078916) (16384,0.151786) (32761,0.178881) (65536,0.281434) (131044,0.536453) (262144,0.900401) (524176,1.71572) (1048576,3.56248) (2096704,7.27122)};
\label{leg-search:scala_lists}
\addplot[red, densely dotted, mark=square*, mark options={solid, fill=white}] coordinates {(1024,0.080309) (2025,0.128234) (4096,0.200737) (8190,0.296534) (16384,0.501342) (32761,0.911866) (65536,1.95354) (131044,4.00042) (262144,9.03499) (524176,21.0925) (1048576,49.1451) (2096704,110.894)};
\label{leg-search:scala_rbt}
\addplot[blue, mark=triangle*, mark options={solid}] coordinates {(1024,0.02331) (2025,0.044035) (4096,0.107112) (8190,0.234924) (16384,0.548487) (32761,1.22651) (65536,2.63032) (131044,5.48898) (262144,10.9498) (524176,21.7153)};
\label{leg-search:go_csr}
\addplot[blue, dashed, mark=halfdiamond*, mark options={solid}] coordinates {(1024,0.011668) (2025,0.021622) (4096,0.047379) (8190,0.101508) (16384,0.217783) (32761,0.466368) (65536,1.01394) (131044,2.14779) (262144,4.51533) (524176,9.77885) (1048576,20.3689)};
\label{leg-search:go_lists}
\addplot[blue, densely dotted, mark=triangle*, mark options={solid, fill=white}] coordinates {(1024,0.039367) (2025,0.085393) (4096,0.201629) (8190,0.445591) (16384,1.039) (32761,2.45202) (65536,5.58371) (131044,12.3934) (262144,27.7242) (524176,60.4643)};
\label{leg-search:go_rbt}
\addplot[olive, mark=star, mark options={solid}] coordinates {(1024,0.000156) (2025,0.000137) (4096,0.000283) (8190,0.000581) (16384,0.001241) (32761,0.002742) (65536,0.005968) (131044,0.01473) (262144,0.033534) (524176,0.078097) (1048576,0.20271) (2096704,0.705079)};
\label{leg-search:lemon_csr}
\addplot[olive, densely dotted, mark=star, mark options={solid}] coordinates {(1024,3.6e-05) (2025,6.5e-05) (4096,0.000134) (8190,0.000281) (16384,0.000668) (32761,0.001331) (65536,0.003171) (131044,0.006723) (262144,0.020542) (524176,0.058352) (1048576,0.152284) (2096704,0.471881)};
\label{leg-search:lemon_linked}
\addplot[orange, mark=pentagon*, mark options={solid}] coordinates {(1024,0.000949) (2025,0.00176) (4096,0.003701) (8190,0.008004) (16384,0.019456) (32761,0.04067) (65536,0.107156) (131044,0.36988) (262144,0.93708) (524176,2.49028) (1048576,5.70855) (2096704,15.1672)};
\label{leg-search:jgrapht_csr}
\addplot[orange, densely dotted, mark=pentagon*, mark options={solid}] coordinates {(1024,0.001237) (2025,0.002953) (4096,0.007511) (8190,0.021487) (16384,0.054236) (32761,0.096296) (65536,0.141014) (131044,0.305228) (262144,0.9459) (524176,2.37042) (1048576,6.40339) (2096704,18.0295)};
\label{leg-search:jgrapht_hashgraph}
\addplot[darkgray, mark=asterisk, mark options={solid}] coordinates {(1024,0.000597) (2025,0.001225) (4096,0.002517) (8190,0.005283) (16384,0.010969) (32761,0.055246) (65536,0.213093) (131044,0.605122) (262144,1.5418) (524176,3.73831) (1048576,7.77751) (2096704,16.2702)};
\label{leg-search:networkx}
\nextgroupplot[extra description/.code={\node[anchor=south east, draw=black, fill=white, font=\scriptsize, inner sep=1.5pt] at ([xshift=-2pt,yshift=2pt]rel axis cs:1,0) {DFS};}, title={dense connected rand., Tb.~\ref{tab:dfs-dense}}, xtick={1024,2048,4096,8192,16384,32768}]
\addplot[teal, mark=diamond*, mark options={solid}] coordinates {(1024,0.004322) (2048,0.011024) (4096,0.039883) (8192,0.177571) (16384,0.749344) (32768,3.21126)};
\addplot[teal, densely dotted, mark=diamond*, mark options={solid}] coordinates {(1024,0.000285) (2048,0.001267) (4096,0.002779) (8192,0.014269) (16384,0.040254) (32768,0.158542)};
\addplot[black, mark=*, mark options={solid}] coordinates {(1024,0.000628) (2048,0.002619) (4096,0.010768) (8192,0.048297) (16384,0.335387) (32768,1.39641)};
\addplot[black, dashed, mark=halfcircle*, mark options={solid, rotate=90}] coordinates {(1024,0.025055) (2048,0.233687) (4096,3.55609) (8192,90.6161) (16384,922.722)};
\addplot[black, densely dotted, mark=*, mark options={solid, fill=white}] coordinates {(1024,0.398949) (2048,1.46568) (4096,8.15365) (8192,39.9148) (16384,209.788)};
\addplot[red, mark=square*, mark options={solid}] coordinates {(1024,0.134872) (2048,0.432) (4096,1.08063) (8192,3.78956) (16384,17.2858)};
\addplot[red, dashed, mark=halfsquare*, mark options={solid}] coordinates {(1024,0.179911) (2048,0.913481) (4096,8.86485) (8192,118.912)};
\addplot[red, densely dotted, mark=square*, mark options={solid, fill=white}] coordinates {(1024,0.364912) (2048,1.41556) (4096,8.61905) (8192,51.5952) (16384,280.584)};
\addplot[blue, mark=triangle*, mark options={solid}] coordinates {(1024,0.49632) (2048,2.36797) (4096,9.54764)};
\addplot[blue, dashed, mark=halfdiamond*, mark options={solid}] coordinates {(1024,0.205183) (2048,1.17002) (4096,8.50637) (8192,133.164)};
\addplot[blue, densely dotted, mark=triangle*, mark options={solid, fill=white}] coordinates {(1024,0.716212) (2048,3.96042) (4096,21.3535) (8192,110.225)};
\addplot[olive, mark=star, mark options={solid}] coordinates {(1024,0.000838) (2048,0.003348) (4096,0.019325) (8192,0.082497) (16384,0.375208) (32768,1.28167)};
\addplot[olive, densely dotted, mark=star, mark options={solid}] coordinates {(1024,0.000813) (2048,0.004043) (4096,0.023801) (8192,0.378823) (16384,2.69155) (32768,12.2792)};
\addplot[orange, mark=pentagon*, mark options={solid}] coordinates {(1024,0.035713) (2048,0.191303) (4096,1.57367) (8192,10.6339) (16384,77.6889) (32768,647.495)};
\addplot[orange, densely dotted, mark=pentagon*, mark options={solid}] coordinates {(1024,0.050246) (2048,0.195733) (4096,1.90142) (8192,14.0975) (16384,155.463)};
\addplot[darkgray, mark=asterisk, mark options={solid}] coordinates {(1024,0.00259) (2048,0.010084) (4096,0.268519) (8192,1.30413) (16384,3.54299) (32768,26.3185)};
\nextgroupplot[extra description/.code={\node[anchor=south east, draw=black, fill=white, font=\scriptsize, inner sep=1.5pt] at ([xshift=-2pt,yshift=2pt]rel axis cs:1,0) {DFS};}, title={USA-road-t (sparse), Tb.~\ref{tab:dfs-road}}, xtick={524288,1048576,2097152,4194304,8388608,16777216}]
\addplot[teal, mark=diamond*, mark options={solid}] coordinates {(264346,0.023861) (321270,0.026269) (435666,0.036113) (1070376,0.092663) (1207945,0.10295) (1524453,0.135922) (1890815,0.187726) (2758119,0.276089) (3598623,0.368924) (6262104,0.627112) (14081816,1.85949) (23947347,2.51004)};
\addplot[teal, densely dotted, mark=diamond*, mark options={solid}] coordinates {(264346,0.00953) (321270,0.011023) (435666,0.017463) (1070376,0.044564) (1207945,0.050744) (1524453,0.064099) (1890815,0.091853) (2758119,0.140107) (3598623,0.198456) (6262104,0.340742) (14081816,1.08726) (23947347,1.48921)};
\addplot[black, mark=*, mark options={solid}] coordinates {(264346,0.025324) (321270,0.023346) (435666,0.021857) (1070376,0.057409) (1207945,0.06429) (1524453,0.0985) (1890815,0.144545) (2758119,0.214131) (3598623,0.34647) (6262104,0.469297) (14081816,1.41941) (23947347,2.23979)};
\addplot[black, dashed, mark=halfcircle*, mark options={solid, rotate=90}] coordinates {(264346,0.062925) (321270,0.071914) (435666,0.112428) (1070376,0.27108) (1207945,0.254109) (1524453,1.11981) (1890815,1.13819) (2758119,1.48425) (3598623,0.882097) (6262104,1.49738) (14081816,3.87422) (23947347,7.39114)};
\addplot[black, densely dotted, mark=*, mark options={solid, fill=white}] coordinates {(264346,5.26958) (321270,6.07432) (435666,8.17704) (1070376,22.0799) (1207945,24.8354) (1524453,31.9299) (1890815,41.8493) (2758119,60.1804) (3598623,83.8972) (6262104,150.76) (14081816,424.54)};
\addplot[red, mark=square*, mark options={solid}] coordinates {(264346,0.886904) (321270,0.966676) (435666,1.21156) (1070376,2.77849) (1207945,3.00452) (1524453,3.97543) (1890815,5.1602) (2758119,7.36342) (3598623,9.39901) (6262104,17.314) (14081816,45.317) (23947347,78.787)};
\addplot[red, dashed, mark=halfsquare*, mark options={solid}] coordinates {(264346,0.585861) (321270,0.657377) (435666,0.758397) (1070376,1.55173) (1207945,1.52253) (1524453,1.9924) (1890815,2.50433) (2758119,3.77573) (3598623,4.99928) (6262104,8.53156) (14081816,21.7215) (23947347,33.2475)};
\addplot[red, densely dotted, mark=square*, mark options={solid, fill=white}] coordinates {(264346,4.32148) (321270,5.09335) (435666,6.94511) (1070376,19.3434) (1207945,22.1703) (1524453,30.0982) (1890815,38.0652) (2758119,56.7135) (3598623,77.0018) (6262104,138.94) (14081816,424.581) (23947347,605.167)};
\addplot[blue, mark=triangle*, mark options={solid}] coordinates {(264346,7.65988) (321270,8.44313) (435666,11.3981) (1070376,28.938) (1207945,30.4813)};
\addplot[blue, dashed, mark=halfdiamond*, mark options={solid}] coordinates {(264346,3.6512) (321270,4.41603) (435666,5.7663) (1070376,15.1868) (1207945,14.8093)};
\addplot[blue, densely dotted, mark=triangle*, mark options={solid, fill=white}] coordinates {(264346,19.7464) (321270,22.9931) (435666,31.67)};
\addplot[olive, mark=star, mark options={solid}] coordinates {(264346,0.020083) (321270,0.025689) (435666,0.039451) (1070376,0.088688) (1207945,0.110172) (1524453,0.144991) (1890815,0.16838) (2758119,0.257068) (3598623,0.376565) (6262104,0.65203) (14081816,1.9232) (23947347,2.43841)};
\addplot[olive, densely dotted, mark=star, mark options={solid}] coordinates {(264346,0.013273) (321270,0.018013) (435666,0.023691) (1070376,0.099073) (1207945,0.129505) (1524453,0.158012) (1890815,0.231609) (2758119,0.265872) (3598623,0.381959) (6262104,0.600649) (14081816,1.55186) (23947347,2.71582)};
\addplot[orange, mark=pentagon*, mark options={solid}] coordinates {(264346,0.391954) (321270,0.352612) (435666,0.419298) (1070376,1.39998) (1207945,1.69565) (1524453,4.34172) (1890815,4.4472) (2758119,8.16534) (3598623,11.0592) (6262104,20.3341) (14081816,95.5584) (23947347,239.342)};
\addplot[orange, densely dotted, mark=pentagon*, mark options={solid}] coordinates {(264346,0.455039) (321270,0.436719) (435666,0.656268) (1070376,1.53956) (1207945,1.84832) (1524453,4.73615) (1890815,4.58003) (2758119,7.75952) (3598623,15.542) (6262104,20.3652) (14081816,185.496) (23947347,331.059)};
\addplot[darkgray, mark=asterisk, mark options={solid}] coordinates {(264346,0.439592) (321270,0.526456) (435666,0.73854) (1070376,1.81221) (1207945,1.62185) (1524453,2.76377) (1890815,3.43079) (2758119,5.31617) (3598623,7.00821) (6262104,11.9285) (14081816,33.0744) (23947347,48.1849)};
\nextgroupplot[extra description/.code={\node[anchor=south east, draw=black, fill=white, font=\scriptsize, inner sep=1.5pt] at ([xshift=-2pt,yshift=2pt]rel axis cs:1,0) {BFS};}, title={Square-n (sparse), Tb.~\ref{tab:bfs-square}}, xtick={1024,2025,4096,8190,16384,32761,65536,131044,262144,524176,1048576,2096704}, xticklabels={$2^{10}$,,$2^{12}$,,$2^{14}$,,$2^{16}$,,$2^{18}$,,$2^{20}$,}]
\addplot[teal, mark=diamond*, mark options={solid}] coordinates {(1024,0.000504) (2025,0.000494) (4096,0.000753) (8190,0.000959) (16384,0.001933) (32761,0.003907) (65536,0.009212) (131044,0.016781) (262144,0.031229) (524176,0.058134) (1048576,0.117882) (2096704,0.346425)};
\addplot[teal, densely dotted, mark=diamond*, mark options={solid}] coordinates {(1024,0.000162) (2025,0.000134) (4096,0.000182) (8190,0.00025) (16384,0.000604) (32761,0.001159) (65536,0.002612) (131044,0.005448) (262144,0.008527) (524176,0.020919) (1048576,0.034451) (2096704,0.134344)};
\addplot[black, mark=*, mark options={solid}] coordinates {(1024,0.000112) (2025,0.000106) (4096,0.000208) (8190,0.000408) (16384,0.000876) (32761,0.00191) (65536,0.004247) (131044,0.00965) (262144,0.020088) (524176,0.048528) (1048576,0.125138) (2096704,0.451912)};
\addplot[black, densely dotted, mark=*, mark options={solid, fill=white}] coordinates {(1024,0.019475) (2025,0.064111) (4096,0.229841) (8190,0.808042) (16384,3.14675) (32761,12.3245) (65536,53.2233) (131044,199.625)};
\addplot[red, mark=square*, mark options={solid}] coordinates {(1024,0.010166) (2025,0.01884) (4096,0.037007) (8190,0.067091) (16384,0.133594) (32761,0.264442) (65536,0.421023) (131044,0.639362) (262144,1.05599) (524176,2.02865) (1048576,3.91478) (2096704,8.4614)};
\addplot[red, densely dotted, mark=square*, mark options={solid, fill=white}] coordinates {(1024,0.233507) (2025,0.472933) (4096,1.11984) (8190,3.6812) (16384,16.4637) (32761,78.3191) (65536,501.306)};
\addplot[blue, mark=triangle*, mark options={solid}] coordinates {(1024,0.026977) (2025,0.032887) (4096,0.076139) (8190,0.173811) (16384,0.374572) (32761,0.836289) (65536,1.78382) (131044,3.71) (262144,7.57583) (524176,15.0539)};
\addplot[blue, densely dotted, mark=triangle*, mark options={solid, fill=white}] coordinates {(1024,0.144763) (2025,0.37603) (4096,1.46533) (8190,5.76328) (16384,22.7487) (32761,92.4803) (65536,577.112)};
\addplot[olive, mark=star, mark options={solid}] coordinates {(1024,0.000139) (2025,0.000123) (4096,0.000251) (8190,0.000522) (16384,0.001062) (32761,0.002292) (65536,0.004885) (131044,0.012573) (262144,0.029435) (524176,0.06385) (1048576,0.152217) (2096704,0.44872)};
\addplot[olive, densely dotted, mark=star, mark options={solid}] coordinates {(1024,2.6e-05) (2025,3.7e-05) (4096,8.3e-05) (8190,0.000188) (16384,0.000394) (32761,0.000964) (65536,0.001816) (131044,0.00408) (262144,0.018464) (524176,0.039915) (1048576,0.094715) (2096704,0.330578)};
\addplot[orange, mark=pentagon*, mark options={solid}] coordinates {(1024,0.000764) (2025,0.001322) (4096,0.002823) (8190,0.00586) (16384,0.012416) (32761,0.022697) (65536,0.053635) (131044,0.225984) (262144,0.511463) (524176,1.32423) (1048576,3.10917) (2096704,6.61838)};
\addplot[orange, densely dotted, mark=pentagon*, mark options={solid}] coordinates {(1024,0.000673) (2025,0.00137) (4096,0.003347) (8190,0.007588) (16384,0.021057) (32761,0.045882) (65536,0.070596) (131044,0.126877) (262144,0.460153) (524176,0.854043) (1048576,1.7432) (2096704,3.64439)};
\addplot[darkgray, mark=asterisk, mark options={solid}] coordinates {(1024,0.001014) (2025,0.001095) (4096,0.002215) (8190,0.004527) (16384,0.009223) (32761,0.022653) (65536,0.055844) (131044,0.144426) (262144,0.35444) (524176,1.55469) (1048576,2.48709) (2096704,7.283)};
\nextgroupplot[extra description/.code={\node[anchor=south east, draw=black, fill=white, font=\scriptsize, inner sep=1.5pt] at ([xshift=-2pt,yshift=2pt]rel axis cs:1,0) {BFS};}, title={dense connected rand., Tb.~\ref{tab:bfs-dense}}, xtick={1024,2048,4096,8192,16384,32768}]
\addplot[black, mark=*, mark options={solid}] coordinates {(1024,0.00055) (2048,0.002181) (4096,0.008958) (8192,0.04284) (16384,0.310156) (32768,1.37969)};
\addplot[black, densely dotted, mark=*, mark options={solid, fill=white}] coordinates {(1024,0.315025) (2048,2.26925) (4096,37.9101) (8192,180.131)};
\addplot[red, mark=square*, mark options={solid}] coordinates {(1024,0.108766) (2048,0.386461) (4096,0.884368) (8192,3.03707) (16384,13.7454)};
\addplot[red, densely dotted, mark=square*, mark options={solid, fill=white}] coordinates {(1024,1.1982) (2048,11.3029) (4096,115.339) (8192,772.268)};
\addplot[blue, mark=triangle*, mark options={solid}] coordinates {(1024,0.366111) (2048,1.67767) (4096,6.84876)};
\addplot[blue, densely dotted, mark=triangle*, mark options={solid, fill=white}] coordinates {(1024,1.81344) (2048,19.1065) (4096,248.33)};
\addplot[olive, mark=star, mark options={solid}] coordinates {(1024,0.001705) (2048,0.003118) (4096,0.016119) (8192,0.074897) (16384,0.358908) (32768,1.25953)};
\addplot[olive, densely dotted, mark=star, mark options={solid}] coordinates {(1024,0.001541) (2048,0.003913) (4096,0.026232) (8192,0.358524) (16384,2.05644) (32768,9.21665)};
\addplot[orange, mark=pentagon*, mark options={solid}] coordinates {(1024,0.016336) (2048,0.079912) (4096,0.587272) (8192,3.09271) (16384,17.3586) (32768,85.2085)};
\addplot[orange, densely dotted, mark=pentagon*, mark options={solid}] coordinates {(1024,0.010923) (2048,0.041889) (4096,0.16873) (8192,0.57742) (16384,2.56884) (32768,12.276)};
\addplot[teal, mark=diamond*, mark options={solid}] coordinates {(1024,0.002265) (2048,0.010383) (4096,0.040505) (8192,0.180153) (16384,0.759569) (32768,3.33591)};
\addplot[teal, densely dotted, mark=diamond*, mark options={solid}] coordinates {(1024,0.000297) (2048,0.001357) (4096,0.002952) (8192,0.015737) (16384,0.045843) (32768,0.228041)};
\addplot[darkgray, mark=asterisk, mark options={solid}] coordinates {(1024,0.000692) (2048,0.001622) (4096,0.133846) (8192,0.52188) (16384,0.046553) (32768,10.8428)};
\nextgroupplot[extra description/.code={\node[anchor=south east, draw=black, fill=white, font=\scriptsize, inner sep=1.5pt] at ([xshift=-2pt,yshift=2pt]rel axis cs:1,0) {BFS};}, title={USA-road-t (sparse), Tb.~\ref{tab:bfs-road}}, xtick={524288,1048576,2097152,4194304,8388608,16777216}]
\addplot[teal, mark=diamond*, mark options={solid}] coordinates {(264346,0.023499) (321270,0.024431) (435666,0.032949) (1070376,0.081364) (1207945,0.092353) (1524453,0.120489) (1890815,0.173172) (2758119,0.247354) (3598623,0.307367) (6262104,0.574742) (14081816,1.40022) (23947347,2.22498)};
\addplot[teal, densely dotted, mark=diamond*, mark options={solid}] coordinates {(264346,0.007787) (321270,0.008878) (435666,0.013249) (1070376,0.033888) (1207945,0.04159) (1524453,0.050396) (1890815,0.081404) (2758119,0.112989) (3598623,0.144455) (6262104,0.284421) (14081816,0.620629) (23947347,1.01649)};
\addplot[black, mark=*, mark options={solid}] coordinates {(264346,0.039455) (321270,0.024137) (435666,0.022181) (1070376,0.067125) (1207945,0.074834) (1524453,0.101624) (1890815,0.137834) (2758119,0.210465) (3598623,0.357647) (6262104,0.528354) (14081816,1.36228) (23947347,1.84791)};
\addplot[black, densely dotted, mark=*, mark options={solid, fill=white}] coordinates {(264346,853.962) (321270,1037.34)};
\addplot[red, mark=square*, mark options={solid}] coordinates {(264346,0.760615) (321270,0.802649) (435666,1.01659) (1070376,2.23362) (1207945,2.38943) (1524453,3.18067) (1890815,3.84941) (2758119,5.79354) (3598623,7.50513) (6262104,13.6857) (14081816,34.2339) (23947347,56.7882)};
\addplot[red, densely dotted, mark=square*, mark options={solid, fill=white}] coordinates {};
\addplot[blue, mark=triangle*, mark options={solid}] coordinates {(264346,4.94719) (321270,5.33363) (435666,7.67564) (1070376,19.1005) (1207945,19.9292)};
\addplot[blue, densely dotted, mark=triangle*, mark options={solid, fill=white}] coordinates {};
\addplot[olive, mark=star, mark options={solid}] coordinates {(264346,0.03915) (321270,0.02335) (435666,0.030423) (1070376,0.086392) (1207945,0.095619) (1524453,0.139287) (1890815,0.164177) (2758119,0.288406) (3598623,0.376369) (6262104,0.539904) (14081816,1.43221) (23947347,2.257)};
\addplot[olive, densely dotted, mark=star, mark options={solid}] coordinates {(264346,0.010898) (321270,0.015345) (435666,0.021622) (1070376,0.086019) (1207945,0.074534) (1524453,0.094561) (1890815,0.18058) (2758119,0.226156) (3598623,0.261115) (6262104,0.388954) (14081816,1.23766) (23947347,1.94708)};
\addplot[orange, mark=pentagon*, mark options={solid}] coordinates {(264346,0.147222) (321270,0.181228) (435666,0.307969) (1070376,0.940401) (1207945,1.13625) (1524453,2.03715) (1890815,2.18342) (2758119,4.43004) (3598623,6.22303) (6262104,10.3373) (14081816,32.3088) (23947347,45.7717)};
\addplot[orange, densely dotted, mark=pentagon*, mark options={solid}] coordinates {(264346,0.188115) (321270,0.261188) (435666,0.507084) (1070376,1.05902) (1207945,1.20041) (1524453,1.49551) (1890815,1.68754) (2758119,2.68704) (3598623,3.93491) (6262104,6.40951) (14081816,17.9953) (23947347,26.7648)};
\addplot[darkgray, mark=asterisk, mark options={solid}] coordinates {(264346,0.307913) (321270,0.352533) (435666,0.514246) (1070376,1.71376) (1207945,1.99289) (1524453,2.35884) (1890815,3.43729) (2758119,4.97034) (3598623,7.41252) (6262104,13.9208) (14081816,57.3742) (23947347,67.1225)};
\end{groupplot}
\pgfresetboundingbox\path ([yshift=0.55cm]group c1r1.outer west |- group c1r1.north) rectangle ([yshift=-0.55cm]group c3r2.outer east |- group c3r2.south);
\end{tikzpicture}

%% file: measurements/tex_figures/dijkstra.tex
\begin{tikzpicture}
\begin{groupplot}[group style={group size=3 by 1, horizontal sep=0.7cm, vertical sep=1.1cm}, xmode=log, ymode=log, log basis x=2, ytickten={-6,...,6}, log basis y=10, yticklabel={$\pgfmathparse{int(round(\tick))}\pgfmathresult$}, after end axis/.code={\node[anchor=north east, font=\footnotesize, inner sep=1pt] at (rel axis cs:0,0) {$10$};}, grid=major, grid style={gray!25}, tick label style={font=\tiny}, label style={font=\scriptsize}, every axis plot/.append style={thick, mark size=1.5pt, legend image code/.code={\draw[mark repeat=2, mark phase=2] plot coordinates {(0cm,0cm) (0.15cm,0cm) (0.3cm,0cm)};}}, width=0.405\textwidth, height=4cm, title style={font=\scriptsize, align=center}]
\nextgroupplot[extra description/.code={\node[anchor=south east, draw=black, fill=white, font=\scriptsize, inner sep=1.5pt] at ([xshift=-2pt,yshift=2pt]rel axis cs:1,0) {DA};}, title={Square-n (sparse), Tb.~\ref{tab:dijkstra-square}}, xtick={1024,2025,4096,8190,16384,32761,65536,131044,262144,524176,1048576,2096704}, xticklabels={$2^{10}$,,$2^{12}$,,$2^{14}$,,$2^{16}$,,$2^{18}$,,$2^{20}$,}]
\addplot[black, mark=*, mark options={solid}] coordinates {(1024,0.000413) (2025,0.000724) (4096,0.001711) (8190,0.003896) (16384,0.007189) (32761,0.007777) (65536,0.052105) (131044,0.063312) (262144,0.099798) (524176,0.216182) (1048576,0.707234) (2096704,1.69668)};
\label{leg-dijkstra:sml}
\addplot[red, mark=square*, mark options={solid}] coordinates {(1024,0.015737) (2025,0.022662) (4096,0.05973) (8190,0.131507) (16384,0.219741) (32761,0.370172) (65536,0.530186) (131044,0.977544) (262144,1.58419) (524176,2.79746) (1048576,5.02305) (2096704,13.5817)};
\label{leg-dijkstra:scala}
\addplot[blue, mark=triangle*, mark options={solid}] coordinates {(1024,0.036117) (2025,0.056556) (4096,0.178313) (8190,0.459123) (16384,0.953104) (32761,1.96907) (65536,3.8451) (131044,11.0391) (262144,18.8001) (524176,32.1864)};
\label{leg-dijkstra:go}
\addplot[olive, mark=star, mark options={solid}] coordinates {(1024,0.000222) (2025,0.000169) (4096,0.000407) (8190,0.000903) (16384,0.001711) (32761,0.00337) (65536,0.006968) (131044,0.021836) (262144,0.051842) (524176,0.118021) (1048576,0.251657) (2096704,0.691399)};
\label{leg-dijkstra:lemon_csr}
\addplot[olive, densely dotted, mark=star, mark options={solid}] coordinates {(1024,5.2e-05) (2025,5.2e-05) (4096,0.00017) (8190,0.00044) (16384,0.00075) (32761,0.001732) (65536,0.002955) (131044,0.012894) (262144,0.025822) (524176,0.048297) (1048576,0.100703) (2096704,0.424663)};
\label{leg-dijkstra:lemon_linked}
\addplot[orange, mark=pentagon*, mark options={solid}] coordinates {(1024,0.000831) (2025,0.001312) (4096,0.002887) (8190,0.005823) (16384,0.011554) (32761,0.023507) (65536,0.059432) (131044,0.249485) (262144,0.505494) (524176,1.176) (1048576,2.45135) (2096704,6.65034)};
\label{leg-dijkstra:jgrapht_csr}
\addplot[orange, densely dotted, mark=pentagon*, mark options={solid}] coordinates {(1024,0.000738) (2025,0.000919) (4096,0.002996) (8190,0.00783) (16384,0.012675) (32761,0.023574) (65536,0.037976) (131044,0.144671) (262144,0.290216) (524176,0.468228) (1048576,0.753155) (2096704,2.28367)};
\label{leg-dijkstra:jgrapht_hashgraph}
\addplot[teal, mark=diamond*, mark options={solid}] coordinates {(1024,0.000646) (2025,0.000863) (4096,0.001418) (8190,0.002605) (16384,0.005632) (32761,0.011597) (65536,0.024942) (131044,0.055849) (262144,0.126524) (524176,0.342719) (1048576,0.777601) (2096704,1.85046)};
\label{leg-dijkstra:scipy_csr}
\addplot[teal, densely dotted, mark=diamond*, mark options={solid}] coordinates {(1024,0.000295) (2025,0.000493) (4096,0.000927) (8190,0.001956) (16384,0.00421) (32761,0.009185) (65536,0.019733) (131044,0.04553) (262144,0.109041) (524176,0.303608) (1048576,0.73558) (2096704,1.70471)};
\label{leg-dijkstra:scipy_prebuilt}
\addplot[darkgray, mark=asterisk, mark options={solid}] coordinates {(1024,0.001137) (2025,0.002651) (4096,0.005178) (8190,0.0137) (16384,0.032357) (32761,0.068346) (65536,0.150174) (131044,1.05899) (262144,1.88701) (524176,3.71151) (1048576,6.08481) (2096704,34.1381)};
\label{leg-dijkstra:networkx}
\nextgroupplot[extra description/.code={\node[anchor=south east, draw=black, fill=white, font=\scriptsize, inner sep=1.5pt] at ([xshift=-2pt,yshift=2pt]rel axis cs:1,0) {DA};}, title={Long-n (sparse), Tb.~\ref{tab:dijkstra-long}}, xtick={1024,2048,4096,8192,16384,32768,65536,131072,262144,524288,1048576,2097152}, xticklabels={$2^{10}$,,$2^{12}$,,$2^{14}$,,$2^{16}$,,$2^{18}$,,$2^{20}$,}]
\addplot[black, mark=*, mark options={solid}] coordinates {(1024,0.000396) (2048,0.000816) (4096,0.001662) (8192,0.003467) (16384,0.007793) (32768,0.013941) (65536,0.050594) (131072,0.050297) (262144,0.07035) (524288,0.201585) (1048576,0.749683) (2097152,1.50778)};
\addplot[red, mark=square*, mark options={solid}] coordinates {(1024,0.014509) (2048,0.02953) (4096,0.059503) (8192,0.111628) (16384,0.24558) (32768,0.368419) (65536,0.565057) (131072,0.869574) (262144,1.28216) (524288,2.64213) (1048576,5.58256) (2097152,11.9427)};
\addplot[blue, mark=triangle*, mark options={solid}] coordinates {(1024,0.056018) (2048,0.070706) (4096,0.16973) (8192,0.391117) (16384,1.04307) (32768,2.07676) (65536,4.31152) (131072,9.01471) (262144,14.067) (524288,31.312)};
\addplot[olive, mark=star, mark options={solid}] coordinates {(1024,9.8e-05) (2048,0.000194) (4096,0.000386) (8192,0.000789) (16384,0.001776) (32768,0.003428) (65536,0.007244) (131072,0.018229) (262144,0.039411) (524288,0.122659) (1048576,0.263508) (2097152,0.54819)};
\addplot[olive, densely dotted, mark=star, mark options={solid}] coordinates {(1024,3.7e-05) (2048,7.8e-05) (4096,0.000153) (8192,0.000298) (16384,0.000802) (32768,0.001615) (65536,0.003299) (131072,0.00774) (262144,0.015555) (524288,0.041629) (1048576,0.120663) (2097152,0.343827)};
\addplot[orange, mark=pentagon*, mark options={solid}] coordinates {(1024,0.000798) (2048,0.001348) (4096,0.002764) (8192,0.005718) (16384,0.011217) (32768,0.024706) (65536,0.059447) (131072,0.250611) (262144,0.506954) (524288,1.1655) (1048576,2.72033) (2097152,6.83522)};
\addplot[orange, densely dotted, mark=pentagon*, mark options={solid}] coordinates {(1024,0.000511) (2048,0.001215) (4096,0.002418) (8192,0.006243) (16384,0.017272) (32768,0.028816) (65536,0.044605) (131072,0.105626) (262144,0.21102) (524288,0.487678) (1048576,1.0487) (2097152,2.07132)};
\addplot[teal, mark=diamond*, mark options={solid}] coordinates {(1024,0.000667) (2048,0.000912) (4096,0.001281) (8192,0.002165) (16384,0.004491) (32768,0.008615) (65536,0.018474) (131072,0.037755) (262144,0.08811) (524288,0.255531) (1048576,0.594295) (2097152,1.30912)};
\addplot[teal, densely dotted, mark=diamond*, mark options={solid}] coordinates {(1024,0.000284) (2048,0.000445) (4096,0.000772) (8192,0.001462) (16384,0.003178) (32768,0.006337) (65536,0.013514) (131072,0.028867) (262144,0.071149) (524288,0.19238) (1048576,0.449317) (2097152,1.16799)};
\addplot[darkgray, mark=asterisk, mark options={solid}] coordinates {(1024,0.001065) (2048,0.004938) (4096,0.005368) (8192,0.014617) (16384,0.063573) (32768,0.181565) (65536,0.727469) (131072,3.8115) (262144,5.36319)};
\nextgroupplot[extra description/.code={\node[anchor=south east, draw=black, fill=white, font=\scriptsize, inner sep=1.5pt] at ([xshift=-2pt,yshift=2pt]rel axis cs:1,0) {DA};}, title={USA-road-t (sparse), Tb.~\ref{tab:dijkstra-road}}, xtick={524288,1048576,2097152,4194304,8388608,16777216}]
\addplot[black, mark=*, mark options={solid}] coordinates {(264346,0.131068) (321270,0.114199) (435666,0.11472) (1070376,0.300082) (1207945,0.362524) (1524453,0.723976) (1890815,0.787959) (2758119,1.13889) (3598623,1.53754) (6262104,2.35088) (14081816,7.1309) (23947347,12.4671)};
\addplot[red, mark=square*, mark options={solid}] coordinates {(264346,1.14829) (321270,1.44607) (435666,1.6874) (1070376,3.92764) (1207945,4.92721) (1524453,6.63158) (1890815,7.70639) (2758119,11.192) (3598623,15.2346) (6262104,25.439) (14081816,71.124)};
\addplot[blue, mark=triangle*, mark options={solid}] coordinates {(264346,14.0559) (321270,19.1877) (435666,21.8221) (1070376,51.0619) (1207945,61.3685) (1524453,80.1015)};
\addplot[olive, mark=star, mark options={solid}] coordinates {(264346,0.060595) (321270,0.039254) (435666,0.047951) (1070376,0.133844) (1207945,0.16801) (1524453,0.232975) (1890815,0.288692) (2758119,0.415398) (3598623,0.527822) (6262104,0.870441) (14081816,2.66508) (23947347,4.1279)};
\addplot[olive, densely dotted, mark=star, mark options={solid}] coordinates {(264346,0.019757) (321270,0.028782) (435666,0.031867) (1070376,0.093615) (1207945,0.122843) (1524453,0.171731) (1890815,0.207691) (2758119,0.300454) (3598623,0.393772) (6262104,0.598339) (14081816,2.02066) (23947347,3.90349)};
\addplot[orange, mark=pentagon*, mark options={solid}] coordinates {(264346,0.174618) (321270,0.21888) (435666,0.267412) (1070376,0.99459) (1207945,1.21821) (1524453,1.89006) (1890815,2.2849) (2758119,4.21578) (3598623,5.70425) (6262104,10.4146) (14081816,29.0306) (23947347,45.6287)};
\addplot[orange, densely dotted, mark=pentagon*, mark options={solid}] coordinates {(264346,0.180434) (321270,0.262996) (435666,0.324397) (1070376,0.694673) (1207945,1.00642) (1524453,1.21842) (1890815,1.2814) (2758119,2.04829) (3598623,3.01641) (6262104,4.15492) (14081816,13.197) (23947347,24.3674)};
\addplot[teal, mark=diamond*, mark options={solid}] coordinates {(264346,0.08189) (321270,0.093768) (435666,0.13974) (1070376,0.384251) (1207945,0.452418) (1524453,0.586231) (1890815,0.818575) (2758119,1.21177) (3598623,1.65251) (6262104,2.96931) (14081816,8.30146) (23947347,13.2389)};
\addplot[teal, densely dotted, mark=diamond*, mark options={solid}] coordinates {(264346,0.07962) (321270,0.088301) (435666,0.127884) (1070376,0.34699) (1207945,0.440318) (1524453,0.572907) (1890815,0.783572) (2758119,1.14142) (3598623,1.5956) (6262104,2.92486) (14081816,8.17117) (23947347,12.8883)};
\addplot[darkgray, mark=asterisk, mark options={solid}] coordinates {(264346,1.08009) (321270,2.66013) (435666,2.95658) (1070376,8.13669) (1207945,16.1652) (1524453,19.2807) (1890815,29.2873) (2758119,42.9133)};
\end{groupplot}
\pgfresetboundingbox\path ([yshift=0.55cm]group c1r1.outer west |- group c1r1.north) rectangle ([yshift=-0.55cm]group c3r1.outer east |- group c3r1.south);
\end{tikzpicture}

%% file: measurements/tex_figures/kruskal.tex
\begin{tikzpicture}
\begin{groupplot}[group style={group size=3 by 1, horizontal sep=0.7cm, vertical sep=1.1cm}, xmode=log, ymode=log, log basis x=2, ytickten={-6,...,6}, log basis y=10, yticklabel={$\pgfmathparse{int(round(\tick))}\pgfmathresult$}, after end axis/.code={\node[anchor=north east, font=\footnotesize, inner sep=1pt] at (rel axis cs:0,0) {$10$};}, grid=major, grid style={gray!25}, tick label style={font=\tiny}, label style={font=\scriptsize}, every axis plot/.append style={thick, mark size=1.5pt, legend image code/.code={\draw[mark repeat=2, mark phase=2] plot coordinates {(0cm,0cm) (0.15cm,0cm) (0.3cm,0cm)};}}, width=0.405\textwidth, height=4cm, title style={font=\scriptsize, align=center}]
\nextgroupplot[extra description/.code={\node[anchor=south east, draw=black, fill=white, font=\scriptsize, inner sep=1.5pt] at ([xshift=-2pt,yshift=2pt]rel axis cs:1,0) {KA};}, title={Square-n (sparse), Tb.~\ref{tab:kruskal-square}}, xtick={1024,2025,4096,8190,16384,32761,65536,131044,262144,524176,1048576,2096704}, xticklabels={$2^{10}$,,$2^{12}$,,$2^{14}$,,$2^{16}$,,$2^{18}$,,$2^{20}$,}]
\addplot[black, mark=*, mark options={solid}] coordinates {(1024,0.000355) (2025,0.000701) (4096,0.00147) (8190,0.002939) (16384,0.006277) (32761,0.014286) (65536,0.0372) (131044,0.064456) (262144,0.14605) (524176,0.353678) (1048576,0.765369) (2096704,1.67108)};
\label{leg-kruskal:sml}
\addplot[red, mark=square*, mark options={solid}] coordinates {(1024,0.013837) (2025,0.047851) (4096,0.249405) (8190,0.746027) (16384,3.85803) (32761,16.076) (65536,72.0001) (131044,303.465)};
\label{leg-kruskal:scala}
\addplot[blue, mark=triangle*, mark options={solid}] coordinates {(1024,0.024959) (2025,0.060038) (4096,0.15825) (8190,0.463894) (16384,1.56688) (32761,5.97335) (65536,22.5018) (131044,85.1163)};
\label{leg-kruskal:go}
\addplot[olive, mark=star, mark options={solid}] coordinates {(1024,0.000111) (2025,0.00023) (4096,0.00049) (8190,0.001119) (16384,0.002434) (32761,0.005132) (65536,0.011063) (131044,0.023581) (262144,0.056044) (524176,0.129404) (1048576,0.301369) (2096704,0.668916)};
\label{leg-kruskal:lemon_csr}
\addplot[orange, densely dotted, mark=pentagon*, mark options={solid}] coordinates {(1024,0.000757) (2025,0.001755) (4096,0.004385) (8190,0.00852) (16384,0.014516) (32761,0.044025) (65536,0.109338) (131044,0.284757) (262144,0.742527) (524176,1.76198) (1048576,4.05755) (2096704,9.80114)};
\label{leg-kruskal:jgrapht_hashgraph}
\addplot[teal, mark=diamond*, mark options={solid}] coordinates {(1024,0.000323) (2025,0.000493) (4096,0.000892) (8190,0.001733) (16384,0.003548) (32761,0.007492) (65536,0.016833) (131044,0.035155) (262144,0.078428) (524176,0.1598) (1048576,0.365424) (2096704,0.882395)};
\label{leg-kruskal:scipy_csr}
\addplot[darkgray, mark=asterisk, mark options={solid}] coordinates {(1024,0.004909) (2025,0.00884) (4096,0.018078) (8190,0.036172) (16384,0.094741) (32761,0.203566) (65536,0.561531) (131044,1.39956) (262144,3.25614) (524176,7.5555) (1048576,15.9385) (2096704,34.2393)};
\label{leg-kruskal:networkx}
\nextgroupplot[extra description/.code={\node[anchor=south east, draw=black, fill=white, font=\scriptsize, inner sep=1.5pt] at ([xshift=-2pt,yshift=2pt]rel axis cs:1,0) {KA};}, title={Long-n (sparse), Tb.~\ref{tab:kruskal-long}}, xtick={1024,2048,4096,8192,16384,32768,65536,131072,262144,524288,1048576,2097152}, xticklabels={$2^{10}$,,$2^{12}$,,$2^{14}$,,$2^{16}$,,$2^{18}$,,$2^{20}$,}]
\addplot[black, mark=*, mark options={solid}] coordinates {(1024,0.000351) (2048,0.000699) (4096,0.001404) (8192,0.002943) (16384,0.006156) (32768,0.014781) (65536,0.036819) (131072,0.064402) (262144,0.136128) (524288,0.368108) (1048576,0.738736) (2097152,1.67449)};
\addplot[red, mark=square*, mark options={solid}] coordinates {(1024,0.01398) (2048,0.063363) (4096,0.16156) (8192,0.918605) (16384,4.15101) (32768,16.5748) (65536,66.6569) (131072,281.6)};
\addplot[blue, mark=triangle*, mark options={solid}] coordinates {(1024,0.024605) (2048,0.059647) (4096,0.154387) (8192,0.457212) (16384,1.51368) (32768,5.35774) (65536,21.1935) (131072,81.3861) (262144,671.742)};
\addplot[olive, mark=star, mark options={solid}] coordinates {(1024,0.000111) (2048,0.000231) (4096,0.000487) (8192,0.001105) (16384,0.002395) (32768,0.005092) (65536,0.010916) (131072,0.023261) (262144,0.05504) (524288,0.12608) (1048576,0.291647) (2097152,0.680645)};
\addplot[orange, densely dotted, mark=pentagon*, mark options={solid}] coordinates {(1024,0.000781) (2048,0.001839) (4096,0.004383) (8192,0.00787) (16384,0.015437) (32768,0.041154) (65536,0.105256) (131072,0.291308) (262144,0.703332) (524288,1.7262) (1048576,3.99248) (2097152,9.60198)};
\addplot[teal, mark=diamond*, mark options={solid}] coordinates {(1024,0.000312) (2048,0.000496) (4096,0.000888) (8192,0.00171) (16384,0.003515) (32768,0.007357) (65536,0.015849) (131072,0.034774) (262144,0.073215) (524288,0.158343) (1048576,0.347761) (2097152,0.851764)};
\addplot[darkgray, mark=asterisk, mark options={solid}] coordinates {(1024,0.004821) (2048,0.008899) (4096,0.017849) (8192,0.035711) (16384,0.095513) (32768,0.215536) (65536,0.528833) (131072,1.39708) (262144,3.29264) (524288,6.78571) (1048576,15.5077) (2097152,34.1833)};
\nextgroupplot[extra description/.code={\node[anchor=south east, draw=black, fill=white, font=\scriptsize, inner sep=1.5pt] at ([xshift=-2pt,yshift=2pt]rel axis cs:1,0) {KA};}, title={dense connected rand., Tb.~\ref{tab:kruskal-dense}}, xtick={1024,2048,4096,8192,16384,32768}]
\addplot[black, mark=*, mark options={solid}] coordinates {(1024,0.011839) (2048,0.060555) (4096,0.291423) (8192,1.26422) (16384,7.22713) (32768,34.1337)};
\addplot[red, mark=square*, mark options={solid}] coordinates {(1024,13.3387) (2048,248.631)};
\addplot[blue, mark=triangle*, mark options={solid}] coordinates {(1024,5.07699) (2048,81.0544)};
\addplot[olive, mark=star, mark options={solid}] coordinates {(1024,0.003366) (2048,0.015356) (4096,0.087388) (8192,0.45597) (16384,2.02212) (32768,8.70956)};
\addplot[orange, densely dotted, mark=pentagon*, mark options={solid}] coordinates {(1024,0.017496) (2048,0.098634) (4096,0.697787) (8192,2.88174) (16384,12.8924) (32768,62.936)};
\addplot[teal, mark=diamond*, mark options={solid}] coordinates {(1024,0.004588) (2048,0.020677) (4096,0.090594) (8192,0.447644) (16384,2.23972) (32768,10.8722)};
\addplot[darkgray, mark=asterisk, mark options={solid}] coordinates {(1024,0.107573) (2048,0.588907) (4096,2.92348) (8192,13.9817) (16384,67.8538) (32768,324.28)};
\end{groupplot}
\pgfresetboundingbox\path ([yshift=0.55cm]group c1r1.outer west |- group c1r1.north) rectangle ([yshift=-0.55cm]group c3r1.east |- group c3r1.south);
\end{tikzpicture}

%% file: measurements/tex_figures/push_relabel.tex
\begin{tikzpicture}
\begin{groupplot}[group style={group size=3 by 1, horizontal sep=0.7cm, vertical sep=1.1cm}, xmode=log, ymode=log, log basis x=2, ytickten={-6,...,6}, log basis y=10, yticklabel={$\pgfmathparse{int(round(\tick))}\pgfmathresult$}, after end axis/.code={\node[anchor=north east, font=\footnotesize, inner sep=1pt] at (rel axis cs:0,0) {$10$};}, grid=major, grid style={gray!25}, tick label style={font=\tiny}, label style={font=\scriptsize}, every axis plot/.append style={thick, mark size=1.5pt, legend image code/.code={\draw[mark repeat=2, mark phase=2] plot coordinates {(0cm,0cm) (0.15cm,0cm) (0.3cm,0cm)};}}, width=0.41\textwidth, height=4cm, title style={font=\scriptsize, align=center}]
\nextgroupplot[extra description/.code={\node[anchor=south east, draw=black, fill=white, font=\scriptsize, inner sep=1.5pt] at ([xshift=-2pt,yshift=2pt]rel axis cs:1,0) {PR};}, title={RMF-long, Tb.~\ref{tab:push-relabel-rmf-long}}, xtick={4096,8192,16384,32768,65536,131072,262144,524288,1048576,2097152,4194304,8388608}, xticklabels={$2^{12}$,,$2^{14}$,,$2^{16}$,,$2^{18}$,,$2^{20}$,,$2^{22}$,}]
\addplot[orange, mark=pentagon*, mark options={solid}] coordinates {(4096,0.15) (9000,0.32) (15488,0.61) (30589,1.13333) (65536,2.4) (130321,6.80667) (270319,14.8933) (527067,31.8867) (1046528,80.4033) (2090912,200.44) (4147200,469.133)};
\label{leg-push_relabel:jgrapht}
\addplot[teal, solid, mark=diamond*, mark options={solid}] coordinates {(4096,0.080073) (9000,0.289944) (15488,0.816756) (30589,1.80023) (65536,4.69118) (130321,12.2453) (270319,44.2319) (527067,149.562) (1046528,791.564)};
\label{leg-push_relabel:scipy_prebuilt}
\addplot[black, mark=*, mark options={solid}] coordinates {(4096,0.047346) (9000,0.150828) (15488,0.376611) (30589,0.913864) (65536,2.12165) (130321,5.56504) (270319,29.3524) (527067,147.06) (1046528,508.568)};
\label{leg-push_relabel:sml_hl}
\addplot[black, mark=halfcircle*, mark options={solid, rotate=90}] coordinates {(4096,0.055506) (9000,0.201044) (15488,0.713521) (30589,2.01092) (65536,5.628) (130321,14.3853) (270319,100.129) (527067,466.902)};
\label{leg-push_relabel:sml_fifo}
\addplot[black, mark=*, mark options={solid, fill=white, scale=0.7}] coordinates {};
\addplot[olive, mark=star, mark options={solid}] coordinates {(4096,0.007055) (9000,0.021436) (15488,0.050192) (30589,0.104899) (65536,0.444692) (130321,1.08655) (270319,4.77632) (527067,22.7071) (1046528,51.5133) (2090912,266.608) (4147200,826.515)};
\label{leg-push_relabel:lemon_csr}
\addplot[olive, densely dotted, mark=star, mark options={solid}] coordinates {(4096,0.009205) (9000,0.033963) (15488,0.14133) (30589,0.240069) (65536,1.72281) (130321,5.1479) (270319,48.6485) (527067,300.007)};
\label{leg-push_relabel:lemon_linked}
\addplot[darkgray, mark=asterisk, mark options={solid}] coordinates {(4096,0.940675) (9000,1.98161) (15488,3.36189) (30589,6.77345) (65536,20.1153) (130321,48.5535) (270319,141.875) (527067,357.975) (1046528,923.261)};
\label{leg-push_relabel:networkx}
\addplot[brown, dashed, mark=x, mark options={solid}] coordinates {(4096,0.003731) (9000,0.009141) (15488,0.017814) (30589,0.038153) (65536,0.083529) (130321,0.232018) (270319,1.05517) (527067,1.47385) (1046528,4.52759) (2090912,6.56206) (4147200,19.4398) (8444736,90.1748)};
\label{leg-push_relabel:hipr}
\nextgroupplot[extra description/.code={\node[anchor=south east, draw=black, fill=white, font=\scriptsize, inner sep=1.5pt] at ([xshift=-2pt,yshift=2pt]rel axis cs:1,0) {PR};}, title={WASH-RLG-long, Tb.~\ref{tab:push-relabel-wash-rlg-long}}, xtick={4098,8194,16386,32770,65538,131074,262146,524290,1048578,2097154,4194306,8388610}, xticklabels={$2^{12}$,,$2^{14}$,,$2^{16}$,,$2^{18}$,,$2^{20}$,,$2^{22}$,}]
\addplot[orange, mark=pentagon*, mark options={solid}] coordinates {(4098,0.116667) (8194,0.28) (16386,0.6) (32770,1.22667) (65538,2.38) (131074,4.47333) (262146,11.66) (524290,29.44) (1048578,72.4433) (2097154,179.793) (4194306,458.757)};
\addplot[teal, solid, mark=diamond*, mark options={solid}] coordinates {(4098,0.027542) (8194,0.095721) (16386,0.33855) (32770,1.11642) (65538,2.59518) (131074,8.98165) (262146,33.9651) (524290,129.207) (1048578,512.182)};
\addplot[black, mark=*, mark options={solid}] coordinates {(4098,0.028411) (8194,0.086407) (16386,0.264429) (32770,0.889243) (65538,2.54294) (131074,6.88764) (262146,36.567) (524290,273.285)};
\addplot[black, mark=halfcircle*, mark options={solid, rotate=90}] coordinates {(4098,0.02682) (8194,0.120891) (16386,0.430349) (32770,1.60149) (65538,3.6679) (131074,8.75821) (262146,35.8178) (524290,298.912)};
\addplot[black, mark=*, mark options={solid, fill=white, scale=0.7}] coordinates {(4098,0.309812) (8194,1.23053) (16386,3.38657) (32770,10.204)};
\label{leg-push_relabel:sml_afp}
\addplot[olive, mark=star, mark options={solid}] coordinates {(4098,0.009486) (8194,0.041784) (16386,0.115929) (32770,0.640366) (65538,1.47314) (131074,3.30433) (262146,3.6309) (524290,90.2254) (1048578,139.13) (2097154,136.869)};
\addplot[olive, densely dotted, mark=star, mark options={solid}] coordinates {(4098,0.009228) (8194,0.041829) (16386,0.113191) (32770,0.598765) (65538,1.84755) (131074,3.05189) (262146,3.39484) (524290,92.8671) (1048578,137.896) (2097154,134.324)};
\addplot[darkgray, mark=asterisk, mark options={solid}] coordinates {(4098,0.823765) (8194,1.75696) (16386,2.95625) (32770,7.08882) (65538,14.1814) (131074,28.7601) (262146,64.385) (524290,153.855) (1048578,304.332) (2097154,567.603)};
\addplot[brown, dashed, mark=x, mark options={solid}] coordinates {(4098,0.004655) (8194,0.009732) (16386,0.018916) (32770,0.042037) (65538,0.079826) (131074,0.169566) (262146,0.34681) (524290,0.747906) (1048578,1.00703) (2097154,2.03652) (4194306,4.68679) (8388610,9.32696)};
\nextgroupplot[extra description/.code={\node[anchor=south east, draw=black, fill=white, font=\scriptsize, inner sep=1.5pt] at ([xshift=-2pt,yshift=2pt]rel axis cs:1,0) {PR};}, title={AK, Tb.~\ref{tab:push-relabel-ak}}, xtick={4102,8198,16390,32774,65542,131078,262150,524294,1048582}, xticklabels={$2^{12}$,,$2^{14}$,,$2^{16}$,,$2^{18}$,,$2^{20}$}]
\addplot[orange, mark=pentagon*, mark options={solid}] coordinates {(4102,0.46) (8198,0.93) (16390,2.63) (32774,7.39) (65542,32.53) (131078,391.75)};
\addplot[teal, solid, mark=diamond*, mark options={solid}] coordinates {(4102,0.294446) (8198,1.17212) (16390,3.1317) (32774,9.62681) (65542,35.4508) (131078,138.499) (262150,552.882)};
\addplot[black, mark=*, mark options={solid}] coordinates {(4102,0.119669) (8198,0.280498) (16390,1.60741) (32774,4.40805) (65542,14.9867) (131078,55.2833) (262150,222.297) (524294,882.608)};
\addplot[black, mark=halfcircle*, mark options={solid, rotate=90}] coordinates {(4102,0.072126) (8198,0.272989) (16390,1.08688) (32774,3.16055) (65542,9.13543) (131078,32.9751) (262150,128.64) (524294,510.092)};
\addplot[black, mark=*, mark options={solid, fill=white, scale=0.7}] coordinates {(4102,0.391722) (8198,1.4322) (16390,3.69084) (32774,11.8954)};
\addplot[olive, mark=star, mark options={solid}] coordinates {(4102,0.002629) (8198,0.00652) (16390,0.018862) (32774,0.057679) (65542,0.197822) (131078,0.729799) (262150,3.41261) (524294,11.8758) (1048582,43.7666)};
\addplot[olive, densely dotted, mark=star, mark options={solid}] coordinates {(4102,0.006238) (8198,0.013769) (16390,0.047474) (32774,0.089816) (65542,0.241997) (131078,0.792879) (262150,3.68434) (524294,11.8762) (1048582,43.0215)};
\addplot[darkgray, mark=asterisk, mark options={solid}] coordinates {(4102,8.81189) (8198,32.0347) (16390,127.483) (32774,523.362)};
\addplot[brown, dashed, mark=x, mark options={solid}] coordinates {(4102,0.013749) (8198,0.05695) (16390,0.232919) (32774,0.888729) (65542,2.18976) (131078,6.96297) (262150,28.3034) (524294,110.619) (1048582,440.559)};
\end{groupplot}
\pgfresetboundingbox\path ([yshift=0.55cm]group c1r1.outer west |- group c1r1.north) rectangle ([yshift=-0.55cm]group c3r1.outer east |- group c3r1.south);
\end{tikzpicture}

%% file: measurements/tex_figures/hungarian.tex
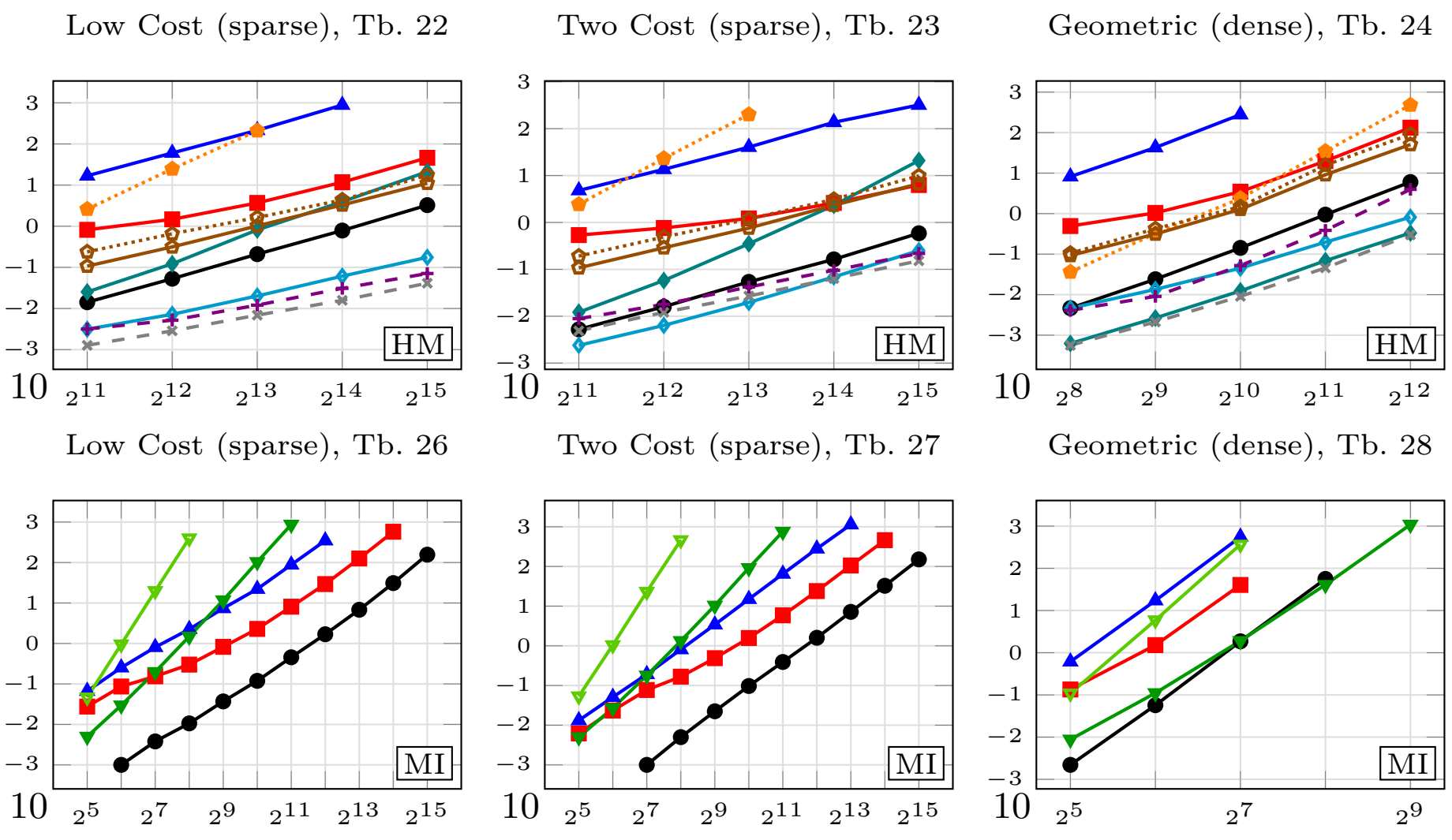
\begin{figure}[!t]
\centering
\begin{tikzpicture}
\begin{groupplot}[group style={group size=3 by 2, horizontal sep=0.7cm, vertical sep=1.1cm}, xmode=log, ymode=log, log basis x=2, ytickten={-6,...,6}, log basis y=10, yticklabel={$\pgfmathparse{int(round(\tick))}\pgfmathresult$}, after end axis/.code={\node[anchor=north east, font=\footnotesize, inner sep=1pt] at (rel axis cs:0,0) {$10$};}, grid=major, grid style={gray!25}, tick label style={font=\tiny}, label style={font=\scriptsize}, every axis plot/.append style={thick, mark size=1.5pt, legend image code/.code={\draw[mark repeat=2, mark phase=2] plot coordinates {(0cm,0cm) (0.15cm,0cm) (0.3cm,0cm)};}}, width=0.41\textwidth, height=4cm, title style={font=\scriptsize, align=center}]
\nextgroupplot[extra description/.code={\node[anchor=south east, draw=black, fill=white, font=\scriptsize, inner sep=1.5pt] at ([xshift=-2pt,yshift=2pt]rel axis cs:1,0) {HM};}, title={Low Cost (sparse), Tb.~\ref{tab:hungarian-low-cost}}, xtick={2048,4096,8192,16384,32768}]
\addplot[black, mark=*, mark options={solid}] coordinates {(2048,0.014186) (4096,0.05298) (8192,0.209357) (16384,0.785678) (32768,3.25274)};
\label{leg-hungarian:sml}
\addplot[red, mark=square*, mark options={solid}] coordinates {(2048,0.817114) (4096,1.47188) (8192,3.68717) (16384,11.7344) (32768,46.042)};
\label{leg-hungarian:scala}
\addplot[blue, mark=triangle*, mark options={solid}] coordinates {(2048,16.8341) (4096,60.3652) (8192,214.712) (16384,884.901)};
\label{leg-hungarian:go}
\addplot[teal, mark=diamond*, mark options={solid}] coordinates {(2048,0.025265) (4096,0.122225) (8192,0.812698) (16384,3.93239) (32768,21.2498)};
\label{leg-hungarian:scipy_lsa}
\addplot[cyan!80!black, mark=diamond, mark options={solid}] coordinates {(2048,0.003133) (4096,0.007229) (8192,0.020151) (16384,0.061128) (32768,0.17507)};
\label{leg-hungarian:scipy_mwfbm_csr}
\addplot[orange, densely dotted, mark=pentagon*, mark options={solid}] coordinates {(2048,2.6033) (4096,24.7188) (8192,209.532)};
\label{leg-hungarian:jgrapht_kuhnmunkres_threadgc}
\addplot[orange!60!black, mark=pentagon, mark options={solid}] coordinates {(2048,0.107053) (4096,0.314313) (8192,1.01779) (16384,3.28262) (32768,11.0783)};
\label{leg-hungarian:jgrapht_mwbm_csr}
\addplot[orange!60!black, densely dotted, mark=pentagon, mark options={solid}] coordinates {(2048,0.238096) (4096,0.655124) (8192,1.62808) (16384,4.36142) (32768,17.2903)};
\label{leg-hungarian:jgrapht_mwbm_hashgraph}
\addplot[gray, dashed, mark=x, mark options={solid}] coordinates {(2048,0.001272) (4096,0.002859) (8192,0.006885) (16384,0.015795) (32768,0.041176)};
\label{leg-hungarian:csa_q_csr}
\addplot[violet, dashed, mark=+, mark options={solid}] coordinates {(2048,0.003167) (4096,0.0052) (8192,0.012068) (16384,0.030874) (32768,0.070918)};
\label{leg-hungarian:ortools}
\nextgroupplot[extra description/.code={\node[anchor=south east, draw=black, fill=white, font=\scriptsize, inner sep=1.5pt] at ([xshift=-2pt,yshift=2pt]rel axis cs:1,0) {HM};}, title={Two Cost (sparse), Tb.~\ref{tab:hungarian-two-cost}}, xtick={2048,4096,8192,16384,32768}]
\addplot[black, mark=*, mark options={solid}] coordinates {(2048,0.005246) (4096,0.015934) (8192,0.053997) (16384,0.163228) (32768,0.587905)};
\addplot[red, mark=square*, mark options={solid}] coordinates {(2048,0.53725) (4096,0.762624) (8192,1.21729) (16384,2.605) (32768,6.28129)};
\addplot[blue, mark=triangle*, mark options={solid}] coordinates {(2048,4.80614) (4096,13.5903) (8192,40.3706) (16384,136.785) (32768,319.761)};
\addplot[teal, mark=diamond*, mark options={solid}] coordinates {(2048,0.01216) (4096,0.057745) (8192,0.351081) (16384,2.31961) (32768,20.8969)};
\addplot[cyan!80!black, mark=diamond, mark options={solid}] coordinates {(2048,0.002395) (4096,0.006368) (8192,0.019671) (16384,0.068926) (32768,0.251225)};
\addplot[orange, densely dotted, mark=pentagon*, mark options={solid}] coordinates {(2048,2.43141) (4096,23.2663) (8192,200.613)};
\addplot[orange!60!black, mark=pentagon, mark options={solid}] coordinates {(2048,0.10759) (4096,0.288194) (8192,0.757764) (16384,2.33276) (32768,6.68351)};
\addplot[orange!60!black, densely dotted, mark=pentagon, mark options={solid}] coordinates {(2048,0.191611) (4096,0.488905) (8192,1.18314) (16384,3.11317) (32768,9.88865)};
\addplot[gray, dashed, mark=x, mark options={solid}] coordinates {(2048,0.004878) (4096,0.012093) (8192,0.027026) (16384,0.063723) (32768,0.152077)};
\addplot[violet, dashed, mark=+, mark options={solid}] coordinates {(2048,0.008857) (4096,0.017964) (8192,0.041974) (16384,0.095382) (32768,0.218493)};
\nextgroupplot[extra description/.code={\node[anchor=south east, draw=black, fill=white, font=\scriptsize, inner sep=1.5pt] at ([xshift=-2pt,yshift=2pt]rel axis cs:1,0) {HM};}, title={Geometric (dense), Tb.~\ref{tab:hungarian-geometric}}, xtick={256,512,1024,2048,4096}]
\addplot[black, mark=*, mark options={solid}] coordinates {(256,0.004642) (512,0.024008) (1024,0.142642) (2048,0.944814) (4096,6.02971)};
\addplot[red, mark=square*, mark options={solid}] coordinates {(256,0.494005) (512,1.04324) (1024,3.47947) (2048,19.8819) (4096,132.617)};
\addplot[blue, mark=triangle*, mark options={solid}] coordinates {(256,8.19934) (512,42.8226) (1024,278.566)};
\addplot[teal, mark=diamond*, mark options={solid}] coordinates {(256,0.000623) (512,0.002654) (1024,0.012354) (2048,0.068873) (4096,0.332708)};
\addplot[cyan!80!black, mark=diamond, mark options={solid}] coordinates {(256,0.004616) (512,0.013483) (1024,0.045101) (2048,0.197879) (4096,0.816661)};
\addplot[orange, densely dotted, mark=pentagon*, mark options={solid}] coordinates {(256,0.036336) (512,0.320591) (1024,2.38193) (2048,34.6229) (4096,480.288)};
\addplot[orange!60!black, mark=pentagon, mark options={solid}] coordinates {(256,0.092566) (512,0.313217) (1024,1.27957) (2048,9.12891) (4096,50.8008)};
\addplot[orange!60!black, densely dotted, mark=pentagon, mark options={solid}] coordinates {(256,0.105191) (512,0.424793) (1024,1.48789) (2048,15.7404) (4096,90.7295)};
\addplot[gray, dashed, mark=x, mark options={solid}] coordinates {(256,0.000561) (512,0.002122) (1024,0.009062) (2048,0.045822) (4096,0.299638)};
\addplot[violet, dashed, mark=+, mark options={solid}] coordinates {(256,0.00409) (512,0.009073) (1024,0.052178) (2048,0.385692) (4096,3.97262)};
\nextgroupplot[extra description/.code={\node[anchor=south east, draw=black, fill=white, font=\scriptsize, inner sep=1.5pt] at ([xshift=-2pt,yshift=2pt]rel axis cs:1,0) {MI};}, title={Low Cost (sparse), Tb.~\ref{tab:mi-low-cost}}, xtick={32,64,128,256,512,1024,2048,4096,8192,16384,32768}, xticklabels={$2^{5}$,,$2^{7}$,,$2^{9}$,,$2^{11}$,,$2^{13}$,,$2^{15}$}]
\addplot[black, mark=*, mark options={solid}] coordinates {(64,0.001) (128,0.0038) (256,0.0106) (512,0.037) (1024,0.1192) (2048,0.4556) (4096,1.6904) (8192,6.8064) (16384,31.0394) (32768,156.572)};
\addplot[red, mark=square*, mark options={solid}] coordinates {(32,0.0276) (64,0.0864) (128,0.1542) (256,0.3002) (512,0.8258) (1024,2.291) (2048,8.1178) (4096,29.336) (8192,125.85) (16384,576.958)};
\addplot[blue, mark=triangle*, mark options={solid}] coordinates {(32,0.0666) (64,0.253) (128,0.8068) (256,2.2446) (512,7.3776) (1024,22.1328) (2048,88.0816) (4096,348.521)};
\addplot[green!60!black, mark=triangle*, mark options={solid, rotate=180}] coordinates {(32,0.005) (64,0.0296) (128,0.1982) (256,1.4856) (512,11.7114) (1024,103.454) (2048,867.69)};
\label{leg-hungarian:sage_oracle}
\addplot[green!40!lime!80!black, mark=triangle, mark options={solid, rotate=180}] coordinates {(32,0.0468) (64,0.9538) (128,19.8766) (256,398.347)};
\label{leg-hungarian:sage_plain}
\nextgroupplot[extra description/.code={\node[anchor=south east, draw=black, fill=white, font=\scriptsize, inner sep=1.5pt] at ([xshift=-2pt,yshift=2pt]rel axis cs:1,0) {MI};}, title={Two Cost (sparse), Tb.~\ref{tab:mi-two-cost}}, xtick={32,64,128,256,512,1024,2048,4096,8192,16384,32768}, xticklabels={$2^{5}$,,$2^{7}$,,$2^{9}$,,$2^{11}$,,$2^{13}$,,$2^{15}$}]
\addplot[black, mark=*, mark options={solid}] coordinates {(128,0.001) (256,0.005) (512,0.0224) (1024,0.0972) (2048,0.3912) (4096,1.5948) (8192,7.1722) (16384,32.5852) (32768,150.648)};
\addplot[red, mark=square*, mark options={solid}] coordinates {(32,0.0062) (64,0.0234) (128,0.077) (256,0.1672) (512,0.4882) (1024,1.5732) (2048,5.8598) (4096,23.9586) (8192,105.656) (16384,459.337)};
\addplot[blue, mark=triangle*, mark options={solid}] coordinates {(32,0.0132) (64,0.051) (128,0.1914) (256,0.8102) (512,3.4196) (1024,14.9722) (2048,64.9566) (4096,279.586) (8192,1136.59)};
\addplot[green!60!black, mark=triangle*, mark options={solid, rotate=180}] coordinates {(32,0.005) (64,0.027) (128,0.1792) (256,1.3284) (512,10.3626) (1024,91.5302) (2048,754.667)};
\addplot[green!40!lime!80!black, mark=triangle, mark options={solid, rotate=180}] coordinates {(32,0.0536) (64,1.0428) (128,23.1074) (256,461.852)};
\nextgroupplot[extra description/.code={\node[anchor=south east, draw=black, fill=white, font=\scriptsize, inner sep=1.5pt] at ([xshift=-2pt,yshift=2pt]rel axis cs:1,0) {MI};}, title={Geometric (dense), Tb.~\ref{tab:mi-geometric}}, xtick={32,64,128,256,512,1024,2048,4096}, xticklabels={$2^{5}$,,$2^{7}$,,$2^{9}$,,$2^{11}$,}]
\addplot[black, mark=*, mark options={solid}] coordinates {(32,0.0022) (64,0.057) (128,1.8638) (256,56.0898)};
\addplot[red, mark=square*, mark options={solid}] coordinates {(32,0.1346) (64,1.514) (128,40.358)};
\addplot[blue, mark=triangle*, mark options={solid}] coordinates {(32,0.6134) (64,17.21) (128,555.577)};
\addplot[green!60!black, mark=triangle*, mark options={solid, rotate=180}] coordinates {(32,0.0088) (64,0.1108) (128,1.9026) (256,40.7746) (512,1088.74)};
\addplot[green!40!lime!80!black, mark=triangle, mark options={solid, rotate=180}] coordinates {(32,0.1088) (64,5.8934) (128,366.834)};
\end{groupplot}
\end{tikzpicture}
\caption{\ref{leg-hungarian:sml}~SML, \ref{leg-hungarian:scala}~Scala, \ref{leg-hungarian:go}~Go, \ref{leg-hungarian:scipy_lsa}~SciPy LSA, \ref{leg-hungarian:scipy_mwfbm_csr}~SciPy MWFBM CSR, \ref{leg-hungarian:jgrapht_kuhnmunkres_threadgc}~JGraphT KM, \ref{leg-hungarian:jgrapht_mwbm_csr}~JGraphT MWBM CSR, \ref{leg-hungarian:jgrapht_mwbm_hashgraph}~JGraphT MWBM hash-backed, \ref{leg-hungarian:csa_q_csr}~CSA CSR, \ref{leg-hungarian:ortools}~OR-Tools, NetworkX wrapps Scipy, skipped.\ref{leg-hungarian:sage_oracle}~Sage (our oracle), \ref{leg-hungarian:sage_plain}~Sage (plain).}
\label{fig:hungarian}
\label{fig:mi}
\end{figure}

%% file: measurements/tex_tables/mcf_summary.tex
\begin{table}[!t]
\centering
\scriptsize
\setlength{\tabcolsep}{2pt}
\caption{CV derived from Tables~\ref{tab:mcf-netgen-deg}, \ref{tab:mcf-netgen-lo-sr}, \ref{tab:mcf-road-paths}, \ref{tab:hybrid-low-cost}, \ref{tab:hybrid-two-cost}, \ref{tab:hybrid-geometric} in the appendix.}
\label{tab:mcf-summary}
\begin{tabular}{@{}llrrrrrrrrrrr@{}}
\toprule
Family & Size & \llap{$\backslash\,j$\quad \quad}1 & 2 & 3 & 4 & 5 & 6 & 7 & 8 & 9 & \dots & 12 \\
\midrule
\texttt{netgen\_deg} & $n = 4096$, $m = 2^j n$ & 1.18 & 1.24 & 1.28 & 1.31 & 1.50 & 1.68 & 1.88 & 2.23 & 2.61 & \dots & 3.69 \\
\texttt{netgen\_lo\_sr} & $n = 2^{j+7}$, $m = n^{1.5}$ & 3.17 & 3.34 & 3.40 & 2.94 & 2.96 & 2.38 & 2.22 & 1.89 & 1.80 &  &  \\
\texttt{road\_paths} & $n \approx 1.5\cdot 10^4 \cdot 2^j$, $m \approx 2.4 n$ &  & 1.91 & 1.63 & 1.40 & 1.22 & 1.29 & 1.16 &  &  &  &  \\
\midrule
LC & $n = 2^{j+8}$, $m = n \log_2 n$ &  &  & 1.72 & 1.68 & 1.61 & 1.62 & 1.65 & 1.64 & 1.51 & \dots & 1.26 \\
TC & $n = 2^{j+8}$, $m = n \log_2 n$ &  &  & 1.18 & 1.16 & 1.16 & 1.16 & 1.18 & 1.18 & 1.10 & \dots & 1.05 \\
GEO & $n = 2^{j+7}$, $m = n^2/4$ & 2.21 & 2.35 & 2.43 & 2.36 & 2.13 & 1.97 & 1.69 &  &  &  &  \\
\bottomrule
\end{tabular}
\end{table}

%% file: paper_ai_development.tex
\section{AI-Assisted Development}
\label{sec:ai}

Our verification methodology developed on a small number of algorithms: the graph representation and modelling loops as tail-recursive functions for the blossom algorithm~\cite{BlossomIsabelleJAR}, and the graph ADTs, the invariant reasoning, and the imperative refinement for depth- and breadth-first search~\cite{fdsBook}; it was then refined for minimum-cost flows~\cite{ScalingIsabelle}, algorithms for matroids and greedoids~\cite{MatroidIsabelle}, and primal-dual algorithms~\cite{abdulazizFormalPrimalDualAlgorithm2026b}.
Our initial intention behind this methodology was pedagogical, e.g.\ using basic Isabelle/HOL formalisms like locales and recursive functions, instead of complex formalisms like non-determinism monads. 
It turns out that, according to our experience, LLM-generated algorithms and specifications, when done according to this methodology, are easier to read and to review than what an LLM would output by default.
This mitigates the well-documented difficulty of ensuring that LLM-generated definitions, code, and theorem statements are aligned with the user's request~\cite{zhongImpossibleBench2026,aggarwalAlphaVerus2025,luFormalAlign2025}.
We used two methods to improve this alignment, one for definitions and specifications, and one for proofs.

The first method is standard: the LLM works in phases, and we review its design and implementation decisions, in the form of an LLM-generated plan, before it implements them~\cite{jiangSelfPlanningCodeGeneration2024,bairiCodePlan2024,jiangDraftSketchProve2023,songLeanCopilot2025}.
At that stage, we remove the most obvious flaws, e.g.\ wrong definitions or specifications, reinvented definitions, lemmas that are not reused, and deviations from our formalisation idioms.

For proofs, alignment is less a matter of correctness than of readability.
This matters e.g.\ to understand the correctness argument when one wants to further improve an algorithm.
Since in our library proofs are more than eight times as long as the definitions and specifications, reviewing all LLM-generated proofs is infeasible.
Instead, we compute a statistical signature of our own proofs, e.g.\ the share of calls to automated Isabelle proof methods, such as \texttt{auto}, among all proof steps, and prompt the LLM to restyle its already working proofs towards that signature, since LLMs write verbose proofs with many low-level steps~\cite{guProofOptimizer2026,ahujaImProver2025}.
For instance, in the LLM-generated proofs of the Hopcroft-Karp algorithm, \texttt{auto} accounts for 10.7\% of the proof method calls, against 61.2\% and 50.7\% in our proofs of depth- and breadth-first search, and 38.0\% of the lemmas have a single-step automated proof, against 76.5\% and 63.1\%; after restyling, 69.2\% of Hopcroft-Karp's proofs have a single-step proof (\cref{tab:hk-style} in \cref{app:style-statistics}).
To our knowledge, this is the first use of a statistical signature of one's own verified proofs to guide the restyling of LLM-generated proofs.

In addition to alignment, another concern we had related to using LLMs to generate proofs was attribution.
We label parts of theory files at the lowest granularity we believe is practical: we label sections of Isabelle/HOL files as developed by humans, `collaboratively' with an LLM, or by an LLM.
We label a part as collaborative, e.g.\ if a human prompted the LLM in detail on how to formalise a concept, revised it, or manually formalised parts of it. Parts labelled as produced by an LLM alone are mostly auxiliary lemmas and low-level proofs.
Counting the lines labelled that way, more than 55\% of the lines of the library were developed purely by humans~(\cref{tab:ai-algorithms} in \cref{app:ai}).
This granularity gives data for further studies of the formalisation styles of LLMs and humans, something which we plan to do in the future.

\todo{Examples:
\begin{itemize}
\item Full development (network simplex).
  \todo{Thomas}
\item Executability (Blossom).
\item Refinement to imperative (BFS).
  \todo{Thomas, Thomas says: I think the bfs was still manually, HM and unweighted MI were refined by AI}
\item Full autoformalisation.
\item Metrics to evaluate the similarity of formalisations (\cref{tab:hk-style}).
\end{itemize}}

%% file: paper_conclusion.tex
\section{Discussion}
\label{sec:Discussion}

\paragraph{Graph libraries.}
In unverified graph libraries, such as LEDA~\cite{LEDAbook}, LEMON~\cite{dezsoLEMON2011}, Boost~\cite{BoostGraphLibrary}, and JGraphT~\cite{michailJGraphTJavaLibrary2020}, algorithms are written against an interface of graph operations and work with any implementation of that interface.
Our ADT locales have the same role, and every interface has a specification, against which the correctness of every algorithm that uses it is proved.
Algorithms in LEDA return the solution alongside a certificate that a simple checker program tests, which moves the trust to the checker, which can itself be wrong~\cite{matchingCertIsabelle}.
The graph library for Isabelle/HOL by Noschinski~\cite{noschinskiGraphLib} and the graph theory in Lean's mathlib~\cite{mathlib} formalise the mathematical theory of graphs without executable algorithms.
Verification, on the other hand, has mostly been applied to one or a small family of algorithms at a time, each development with its own graph theory and representation, e.g.\ the Ford--Fulkerson method in Mizar~\cite{FordFulkersonMizar}, 
various graph algorithms in Isabelle/HOL~\cite{LammichFlows,Kruskal-AFP,primsLammichNipkow}, and C implementations in Coq~\cite{graphAlgosinCoq}.
In contrast, all algorithms in our library are stated in terms of the same few representations, which are connected by adaptors (\cref{sec:graphs}).

\paragraph{Refinement.}
Obtaining efficient verified implementations by stepwise refinement~\cite{refinementWirth} and data refinement~\cite{hoareRefinement} is an established approach in Isabelle/HOL.
Its most developed instance is Lammich's Isabelle Refinement Framework~\cite{lammichMonads,lammichMonadsII}, 
which was used for graph algorithms, too~\cite{LammichFlows,Kruskal-AFP}.
Our approach differs: 1.\ instead of a nondeterministic monad, we write algorithms purely functionally, 2.\ instead of automated data refinement to the Isabelle Collections Framework~\cite{collectionsFramework,lammichContainers}, we apply data refinement by locale interpretations~\cite{LocalesBallarin}, and 3.\ the resulting functional implementation is the intermediate step from which the imperative implementation is refined (\cref{sec:imperative}).
Our functional implementations have to be deterministic, but we can state them as ordinary recursive functions and conduct proofs in the style of textbooks on verified functional algorithms~\cite{fdsBook}.
Alternatively, others verify the low-level code that is actually run, either directly, e.g.\ C implementations of graph algorithms~\cite{graphAlgosinCoq}, 
or by abstracting C programs to higher-level specifications~\cite{simplIsabelle,AutoCorres}.

\paragraph{LLM-based development of formal libraries.}
Theorem proving with LLMs has mostly been studied as proving individual theorems against a fixed library, using a variety of techniques~\cite{RingerProofRepair,jiangDraftSketchProve2023,deepseekProver,provingRecursivelyLLMs,songLeanCopilot2025,zhouDontTrustVerify2023,luFormalAlign2025}.
Developing a library aided by AI is a different task, since much of the work is in choosing definitions, representations, and the statements of lemmas so that later developments can reuse them, and none of these choices is checked by the proof checker.
In our development (\cref{sec:ai}), this is where the decisions of the LLM had to be reviewed, and the methodology of \cref{sec:methodology} gave that review a fixed reference: a definition, a specification, or a proof that does not follow its idioms is easy to identify.

The restyling of LLM-generated proofs in \cref{sec:ai} is related to recent work on automated proof optimisation.
ImProver~\cite{ahujaImProver2025} and ProofOptimizer~\cite{guProofOptimizer2026} optimise Lean proofs that already check w.r.t.\ a single criterion, e.g.\ length.
Our guideline is derived from a statistical signature of our own verified proofs, and the aim of the restyling is that the LLM-generated proofs match the style of those proofs, rather than that they minimise one measure.
Measuring style is not new~\cite{blanchetteMiningAFP2015}, but to our knowledge, it has not previously been used to guide or to evaluate LLM-generated proofs.



\paragraph{Conclusion.}
We presented a library of formally verified graph algorithms in Isabelle/HOL.
For simple algorithms, we are within an order of magnitude of standard libraries such as SciPy~\cite{scipyManual} or LEMON~\cite{dezsoLEMON2011}, or even on par, whereas for more complex algorithms, which come with much more room for heuristics and tuning, it is harder as our code is sometimes two orders of magnitude slower. We plan to add more optimisations, which come with more low-level details, but the effort should be greatly reduced by AI. Keeping the library's style, especially maintaining the abstraction barriers, is a challenge.
Also, some of the complex implementations in the libraries we benchmarked against have been tuned for decades, whereas we started including efficient code less than a year ago.

The algorithms share one methodology, so that a lemma or a data structure developed for one algorithm is available to the others.
The methodology also allowed us to develop about a third of the library collaboratively with LLMs while still being able to review what they produced.
Directions for future work include: 1.\ proving running-time bounds for all algorithms of the library in the model of \cref{sec:other-methodology}, as we did for Orlin's algorithm, and relating them to a concrete model of computation using our framework~\cite{cookLevinFramework}, and 2.\ refining the imperative implementations further to LLVM~\cite{isabelleLLVM,LammichLLVM}.

%% file: paper_appendix.tex
\clearpage
\appendix

\section{Verified Algorithms and Graph Data Types}
\label{app:library}

\Cref{tab:algorithms} gives the details of the verification of the algorithms of
\cref{tab:library}, and \cref{tab:graph-adts} describes the graph ADTs of
\cref{sec:graph-impl}.

\begin{table}[!tbp]
  \caption{The verified algorithms. \emph{Pat.}: loops are modelled as in
    \cref{sec:proofs} (\checkmark), in part ((\checkmark)), or not ($-$), or
    the algorithm is an instance of another one (inst.). \emph{Inv.}: named
    invariants of the loops, without invariants of data structures; numbers in
    parentheses are inherited. \emph{Proved}: termination (T), correctness (C),
    optimality (O), approximation or competitive ratio (A); running-time
    analyses are not included. \emph{Lines}: theories specific to the
    algorithm, without running-time analyses, executable instances,
    refinements, as well as the abstract graph-theoretic reasoning. Daggers mark algorithms only
    in development branches. The Isabelle logo links to the main theorem in the
    formal development.}
  \label{tab:algorithms}
  \centering
  \scriptsize
  \renewcommand{\arraystretch}{1.0}
  \setlength{\tabcolsep}{3pt}
  \begin{tabular}{@{}p{30mm}cccp{49mm}r@{}}
    \toprule
    Algorithm & Pat. & Inv. & Proved & \raggedright Main theorem & Lines\tabularnewline
    \midrule
    \multicolumn{6}{@{}l}{\emph{Graph search and shortest paths}}\tabularnewline
    \raggedright Depth-first search~\isalink{Set_Graphs/Graph_Algorithms/DFS.thy}{739}{762}
      & \checkmark & 5 & T, C
      & \raggedright Finds a path from the source to the target iff one exists
      & 877\tabularnewline
    \raggedright Cycle detection~\isalink{Set_Graphs/Graph_Algorithms/DFS_Cycles.thy}{1185}{1200}
      & \checkmark & 16 & T, C
      & \raggedright Reports a cycle iff the graph has one
      & 3,305\tabularnewline
    \raggedright Directed cycle detection$^\dagger$~\isalink{Set_Graphs/Graph_Algorithms/DFS_DirCycle.thy}{404}{406}\,\isalink{Set_Graphs/Graph_Algorithms/DFS_DirCycle.thy}{676}{678}
      & \checkmark & 5 (1) & T, C
      & \raggedright Reports a cycle iff the searched part has one
      & 697\tabularnewline
    \raggedright Breadth-first search~\isalink{Set_Graphs/Graph_Algorithms/BFS_2.thy}{1504}{1524}
      & \checkmark & 13 & T, C
      & \raggedright Visits exactly the reachable vertices, along shortest paths
      & 1,805\tabularnewline
    \raggedright BFS with distances$^\dagger$~\isalink{Set_Graphs/Graph_Algorithms/BFS_3.thy}{2901}{2917}
      & \checkmark & 15 & T, C
      & \raggedright Also computes distances and shortest-path parents
      & 2,901\tabularnewline
    \raggedright Bellman--Ford~\isalink{Set_Graphs/Graph_Algorithms/Bellman_Ford.thy}{1999}{2003}
      & $-$ & 8 & C, O
      & \raggedright Without negative cycles, computes shortest paths
      & 2,837\tabularnewline
    \raggedright Dijkstra$^\dagger$~\isalink{Set_Graphs/Graph_Algorithms/Dijkstra.thy}{1645}{1647}
      & \checkmark & 14 & T, C, O
      & \raggedright Computes shortest paths to the settled vertices
      & 2,865\tabularnewline
    \midrule
    \multicolumn{6}{@{}l}{\emph{Greedy algorithms and spanning trees}}\tabularnewline
    \raggedright Best-in-greedy~\isalink{Matroids_Greedy/BestInGreedy_stablesort.thy}{629}{632}
      & \checkmark & 4 & T, C, A, O
      & \raggedright Rank-quotient approximation; optimal for all weights iff matroid
      & 5,750\tabularnewline
    \raggedright Kruskal~\isalink{Spanning_Trees/Kruskal_Functional.thy}{1097}{1098}
      & inst. & (4) & T, C, O
      & \raggedright Computes a maximum-weight spanning forest
      & 1,559\tabularnewline
    \raggedright Greedoid greedy~\isalink{Greedoids/Greedoids_Optimisation.thy}{1595}{1598}
      & $-$ & 2 & T, C, O
      & \raggedright Optimal for all modular weights iff strong exchange
      & 2,217\tabularnewline
    \raggedright Jarn\'{\i}k--Prim~\isalink{Spanning_Trees/Jarnik_Prim.thy}{497}{499}
      & inst. & (2) & T, C, O
      & \raggedright Computes a maximum-weight spanning tree
      & 1,277\tabularnewline
    \midrule
    \multicolumn{6}{@{}l}{\emph{Flows}}\tabularnewline
    \raggedright Successive shortest paths~\isalink{Mincost_Flow_Algorithms/Successive_Shortest_Path.thy}{1056}{1060}
      & \checkmark & 3 & T, C, O
      & \raggedright Computes a minimum-cost flow or reports none
      & 1,158\tabularnewline
    \raggedright Capacity scaling~\isalink{Mincost_Flow_Algorithms/Scaling.thy}{1624}{1629}
      & \checkmark & 3 & T, C, O
      & \raggedright As for successive shortest paths
      & 1,785\tabularnewline
    \raggedright Orlin~\isalink{Mincost_Flow_Algorithms/Orlins.thy}{2837}{2844}
      & \checkmark & 30 & T, C, O
      & \raggedright As for successive shortest paths
      & 12,080\tabularnewline
    \raggedright Dinic~\isalink{Dinic/Dinic.thy}{426}{428}
      & \checkmark & 1 & T, C, O
      & \raggedright Computes a maximum flow
      & 1,455\tabularnewline
    \raggedright Blocking flow~\isalink{Dinic/Compute_Blocking_Simple.thy}{587}{597}
      & (\checkmark) & 3 & T, C
      & \raggedright Computes a blocking flow
      & 713\tabularnewline
    \raggedright DFS with backtracking~\isalink{Dinic/DFS_Collect_Backtrack.thy}{577}{583}
      & \checkmark & 5 & T, C
      & \raggedright Removed edges lie on no source--target path
      & 746\tabularnewline
    \raggedright Push-relabel$^\dagger$~\isalink{Push_Relabel/Push_Relabel_Generic.thy}{1971}{1971}
      & \checkmark & 7 & T, C, O
      & \raggedright Computes a maximum flow
      & 4,897\tabularnewline
    \raggedright Network simplex$^\dagger$~\isalink{Mincost_Flow_Algorithms/Network_Simplex_Final.thy}{479}{482}
      & \checkmark & 7 & T, C, O
      & \raggedright Minimum-cost flow or infinite negative cycle
      & 7,582\tabularnewline
    \raggedright Acyclic flows$^\dagger$~\isalink{Mincost_Flow_Algorithms/Acyclic_Flow.thy}{5449}{5453}
      & \checkmark & 9 & T, C
      & \raggedright Makes a flow acyclic, not raising its cost
      & 10,691\tabularnewline
    \midrule
    \multicolumn{6}{@{}l}{\emph{Matchings}}\tabularnewline
    \raggedright Edmonds' blossom~\isalink{Edmonds_Blossom_Executable/Find_Max_Matching.thy}{321}{323}
      & (\checkmark) & 15 & T, C, O
      & \raggedright Computes a maximum-cardinality matching
      & 10,460\tabularnewline
    \raggedright Weighted blossom$^\dagger$~\isalink{Edmonds_Weighted_Blossom/Naive_Weighted_Blossom.thy}{3704}{3709}
      & (\checkmark) & 11 & C, O
      & \raggedright Min-weight perfect matching or reports none, if it terminates
      & 5,801\tabularnewline
    \raggedright Edmonds--Gallai decomposition$^\dagger$~\isalink{Edmonds_Gallai/Edmonds_Gallai_Computation.thy}{574}{580}
      & $-$ & 0 & T, C
      & \raggedright Computes the decomposition from a maximum matching
      & 5,007\tabularnewline
    \raggedright Hopcroft--Karp$^\dagger$~\isalink{Hopcroft_Karp/Hopcroft_Karp_Top_Loop.thy}{395}{396}
      & \checkmark & 7 & T, C, O
      & \raggedright Computes a maximum-cardinality bipartite matching
      & 2,381\tabularnewline
    \raggedright Hungarian method~\isalink{Hungarian_Method/Hungarian_Method_Top_Loop.thy}{571}{573}
      & \checkmark & 8 & T, C, O
      & \raggedright Min-weight perfect matching or reports none
      & 4,123\tabularnewline
    \raggedright Primal-dual (Egerv\'ary)~\isalink{Hungarian_Method/Naive_Primal_Dual.thy}{1239}{1242}
      & (\checkmark) & 0 & T, C, O
      & \raggedright Max-weight matching and an optimal dual
      & 1,863\tabularnewline
    \raggedright RANKING~\isalink{RANKING/Ranking_Probabilistic.thy}{1967}{1967}
      & $-$ & 0 & A
      & \raggedright Competitive ratio tending to $\rankingratio$
      & 6,973\tabularnewline
    \raggedright RANKING, primal-dual~\isalink{Primal_Dual_Bipartite_Matching/Ranking_Order_Probabilistic.thy}{1298}{1312}
      & $-$ & 0 & A
      & \raggedright Ratio $\rankingratio$, also vertex-weighted
      & 4,861\tabularnewline
    \raggedright AdWords~\isalink{Primal_Dual_Bipartite_Matching/AdWords.thy}{1352}{1354}
      & $-$ & 0 & C, A
      & \raggedright Ratio in terms of bids and budgets
      & 1,347\tabularnewline
    \midrule
    \multicolumn{6}{@{}l}{\emph{Matroid intersection}}\tabularnewline
    \raggedright Matroid intersection~\isalink{Matroid_Intersection/Matroid_Intersection_Path_Loop.thy}{336}{338}
      & \checkmark & 1 & T, C, O
      & \raggedright Max-cardinality common independent set
      & 2,646\tabularnewline
    \raggedright Weighted matroid intersection$^\dagger$~\isalink{Matroid_Intersection/Matroid_Weighted_Intersection_Path_Loop.thy}{1144}{1147}
      & \checkmark & 1 & T, C, O
      & \raggedright Max-weight common independent set
      & 5,274\tabularnewline
    \raggedright Matroid partitioning$^\dagger$~\isalink{Matroid_Intersection/Matroid_Partitioning.thy}{899}{900}
      & $-$ & 0 & C, O
      & \raggedright Optimal partition from a max-cardinality common independent set
      & 1,215\tabularnewline
    \bottomrule
  \end{tabular}
\end{table}

\begin{table}[!tbp]
  \caption{The graph ADTs of the library and their implementations. The costs
    are the standard bounds of the data structures, where $\numverts$ is the
    number of vertices, $\numedges$ the number of edges, and $\vdeg$ the degree
    of the vertex concerned; the library does not verify them. Daggers mark ADTs
    that are only in development branches of the library.}
  \label{tab:graph-adts}
  \centering
  \scriptsize
  \renewcommand{\arraystretch}{1.15}
  \setlength{\tabcolsep}{3pt}
  \renewcommand{\_}{\textunderscore\allowbreak}
  \begin{tabular}{@{}p{22mm}p{21mm}p{20mm}p{25mm}p{25mm}@{}}
    \toprule
    \raggedright ADT & \raggedright Graph; abstraction & \raggedright Implementation & \raggedright Costs & \raggedright Reasons and users\tabularnewline
    \midrule
    \raggedright \isainl{Set_Graphs/Directed_Graphs/Pair_Graph_Specs.thy}{243}{251}{Pair\_Graph\_Specs}
      & \raggedright Directed; vertex to set of out-neighbours;
        \texttt{digraph\_abs}
      & \raggedright Red-black trees~\isalink{Set_Graphs/Directed_Graphs/Pair_Graph_RBT.thy}{54}{55}
      & \raggedright Looking up a neighbourhood, adding or deleting an edge:
        $\bigO{\log \numverts + \log \vdeg}$; choosing a neighbour:
        $\bigO{1}$
      & \raggedright Purely functional; vertices of any linearly ordered type;
        e.g.\ depth-first and breadth-first search, Dinic's algorithm,
        Prim's algorithm\tabularnewline
    & & \raggedright AVL trees~\isalink{Set_Graphs/Directed_Graphs/Pair_Graph_AVL.thy}{68}{72}
      & \raggedright As for red-black trees
      & \raggedright As for red-black trees; no algorithm in the main branch
        uses it\tabularnewline
    \midrule
    \raggedright \isainl{Set_Graphs/Undirected_Graphs/Pair_Graph_U_Specs.thy}{498}{502}{Pair\_Graph\_U\_Specs}
      & \raggedright Undirected; both orientations of every edge;
        \texttt{ugraph\_abs}
      & \raggedright Red-black trees
      & \raggedright As for \texttt{Pair\_Graph\_Specs}, with every edge added
        or deleted twice
      & \raggedright Reuses the directed ADT and its lemmas; e.g.\ Kruskal's
        algorithm\tabularnewline
    \midrule
    \raggedright \isainl{Mincost_Flow_Algorithms/Orlins_Preparation.thy}{175}{185}{Adj\_Map\_Specs2}
      & \raggedright Directed, over a map without deletion;
        \texttt{digraph\_abs}
      & \raggedright Red-black trees
      & \raggedright Looking up a neighbourhood, adding an edge:
        $\bigO{\log \numverts + \log \vdeg}$
      & \raggedright Requires no deletion from the map; the forest that
        Orlin's algorithm maintains\tabularnewline
    \midrule
    \raggedright \isainl{Set_Graphs/Directed_Graphs/Refinements/Pair_Graph_Imperative.thy}{13}{19}{Pair\_Graph\_Imperative}
      & \raggedright Directed, on the heap; vertex to list of out-neighbours;
        \texttt{graph\_assn}
      & \raggedright Array maps and lists~\isalink{Set_Graphs/Graph_Algorithms/Refinements/DFS_Imperative.thy}{239}{240}
      & \raggedright Looking up a neighbourhood: $\bigO{1}$; adding an edge:
        $\bigO{\vdeg}$, since a neighbourhood is a list without duplicates
      & \raggedright In-place updates; vertices must be numbers below a fixed
        bound; e.g.\ imperative depth-first search\tabularnewline
    & & \raggedright Hash maps and lists~\isalink{Set_Graphs/Graph_Algorithms/Refinements/DFS_Imperative.thy}{254}{255}
      & \raggedright Looking up a neighbourhood: expected $\bigO{1}$; adding an
        edge: $\bigO{\vdeg}$
      & \raggedright In-place updates; vertices of any hashable type; e.g.\
        imperative depth-first search\tabularnewline
    \midrule
    \raggedright \texttt{Multi\_Graph\_Specs}$^\dagger$
      & \raggedright Multigraph; vertex to set of its edges;
        \texttt{multigraph\_abs}
      & \raggedright Arrays (CSR)
      & \raggedright Accessing the edges of a vertex: $\bigO{1}$
      & \raggedright Parallel edges; e.g.\ breadth-first search on
        multigraphs\tabularnewline
    \midrule
    \raggedright \isainl{Data_Structures/Iterable_Set_Specs.thy}{145}{160}{indexed\_iterable\_set}$^\dagger$
      & \raggedright One set per index, e.g.\ per vertex, each with a cursor;
        \texttt{idx\_abstract}
      & \raggedright Arrays (CSR), \isainl{Set_Graphs/Directed_Graphs/CSR_Graph_Spec.thy}{347}{348}{csr\_graph} and, with weights,
        \isainl{Set_Graphs/Directed_Graphs/Refinements/CSR_Graph.thy}{396}{404}{weighted\_csr}
      & \raggedright Element and weight at a cursor, advancing or resetting a
        cursor: $\bigO{1}$; building the arrays:
        $\bigO{\numverts+\numedges}$
      & \raggedright A scan resumes where it stopped; the edges of a vertex
        are contiguous in memory, which is cache-efficient, but the graph
        cannot change; e.g.\ the Hungarian method\tabularnewline
    \midrule
    \raggedright \isainl{Set_Graphs/Directed_Graphs/Edge_Iterators.thy}{17}{37}{outgoing\_edge\_iterator}, \isainl{Set_Graphs/Directed_Graphs/Edge_Iterators.thy}{48}{68}{ingoing\_edge\_iterator}$^\dagger$
      & \raggedright Multigraph whose out- or in-edges at every vertex are an
        \texttt{indexed\_iterable\_set}; \texttt{out\_abstract},
        \texttt{in\_abstract}
      & \raggedright Arrays (CSR)
      & \raggedright As for \texttt{indexed\_iterable\_set}
      & \raggedright As for \texttt{indexed\_iterable\_set}; e.g.\ Dijkstra's
        algorithm, the network simplex method\tabularnewline
    \bottomrule
  \end{tabular}
\end{table}

\clearpage
\section{Running Times}
\label{app:running-times}
\raggedbottom

\Cref{tab:dfs-square,tab:dfs-long,tab:dfs-dense,tab:dfs-road} give the running
times for depth-first search,
\cref{tab:bfs-square,tab:bfs-long,tab:bfs-dense,tab:bfs-road} those for breadth-first search,
\cref{tab:dijkstra-square,tab:dijkstra-long,tab:dijkstra-road} those for Dijkstra's algorithm,
\cref{tab:kruskal-square,tab:kruskal-long,tab:kruskal-dense} those for Kruskal's algorithm,
\cref{tab:push-relabel-rmf-long,tab:push-relabel-wash-rlg-long,tab:push-relabel-ak} those for push-relabel,
\cref{tab:hungarian-low-cost,tab:hungarian-two-cost,tab:hungarian-geometric}
those for the Hungarian method and
\cref{tab:mi-low-two-cost,tab:mi-low-cost,tab:mi-two-cost,tab:mi-geometric}
those for matroid intersection. Each row is an instance size~$n$ (a network in the road tables) and each column
a solver. 

An entry is the mean CPU time in seconds of the solver call over the
instances of that size, where $<$0.001 is a mean below a millisecond. Next to
it (below it in all but the matroid intersection tables), in parentheses, is that mean normalised
by SML (by SML CSR in the DFS tables, by the imperative SML export in the BFS tables, by SML HL in the push-relabel tables), i.e.\ divided by the SML mean for the same size, so SML's own ratio is
$(1.00)$ and a value below~$1$ means the solver was faster than SML.
A superscript $k/m$ marks a mean over only the $k$ of $m$ instances that
finished. TO means that no instance finished within 1200\,s (larger sizes were then not run), OOM that the
solver ran out of memory.
OVF means that the AFP export's integer index overflowed, n/a that the instances violate its precondition of no antiparallel arcs, and $(\text{--})$ that the SML mean is $0$ or missing,
so no ratio can be given.

Measurements in Java and Scala are done after a dummy call to the solver enforcing a JVM warm-up.

For each instance family, the table captions cite its generator or its
instance files online.
We use the settings prescribed by the DIMACS challenges.
The only exception is the dense-n family~\cref{tab:dfs-dense,tab:bfs-dense,tab:kruskal-dense}, whose random graphs
come from our own instantiation of a DIMACS generator~\citecode{ch9CodeSuite}: connect all vertices in the order of a random
permutation to ensure strong connectivity. Then, for $m-n$ times, we randomly
draw an edge, retry if a self-loop is drawn, and insert the resulting random
edge into the graph. For a number of vertices $n$, we set $m=n^2/16$ arcs, so
the average degree grows from 64 at $n=1024$ to 2048 at $n=32768$.
Weights are between $0$ and $n$.

For BFS and DFS, whose graphs come with weights, the weights are ignored.
We select a source randomly and explore all reachable vertices.
For Dijkstra's algorithm, we select a source randomly and search for a randomly selected target.
The only exception is SciPy, which by its functionality explores the whole graph.
For the reachability problems, a graph with a source and, if applicable, a target constitutes a problem instance.
All solvers were given the same instance, i.e.\ the selected sources (and targets) are consistent among columns.

For Kruskal's algorithm, which reuses the graphs for path search, we treat edges as undirected and the parser always picks the minimum-weight edge between two vertices.

The tables for the hybrid solvers,
\cref{tab:mcf-gridgen-deg,tab:mcf-netgen-deg,tab:mcf-netgen-lo-sr,tab:mcf-road-paths,tab:mcf-goto-8}
for minimum-cost flows and
\cref{tab:hybrid-low-cost,tab:hybrid-two-cost,tab:hybrid-geometric} for
weighted bipartite matching, use a different measurement methodology, which
is described with them below.

\input{measurements/tex_tables/dfs_square}
\input{measurements/tex_tables/dfs_long}
\input{measurements/tex_figures/dfs_long}
\input{measurements/tex_tables/dfs_dense}
\input{measurements/tex_tables/dfs_road}

\input{measurements/tex_tables/bfs_square}
\input{measurements/tex_tables/bfs_long}
\input{measurements/tex_figures/bfs_long}
\input{measurements/tex_tables/bfs_dense}
\input{measurements/tex_tables/bfs_road}

\input{measurements/tex_tables/dijkstra_square}
\input{measurements/tex_tables/dijkstra_long}
\input{measurements/tex_tables/dijkstra_road}

\input{measurements/tex_tables/kruskal_square}
\input{measurements/tex_tables/kruskal_long}
\input{measurements/tex_tables/kruskal_dense}

\input{measurements/tex_tables/push_relabel_rmf-long}
\input{measurements/tex_tables/push_relabel_wash-rlg-long}
\input{measurements/tex_tables/push_relabel_ak}

\input{measurements/tex_tables/hungarian_low-cost}
\input{measurements/tex_tables/hungarian_two-cost}
\input{measurements/tex_tables/hungarian_geometric}
\clearpage
\paragraph{$b$-Matching.}
For $b$-matching, we set the capacities $b(x) = \min \lbrace \beta, \deg(x)\rbrace$, where $\beta$ is selected uniformly at random between $1$ and $20$ and $\deg(x)$ is the vertex degree.

\input{measurements/tex_tables/mi_low-two-cost}
\input{measurements/tex_tables/mi_low-cost}
\input{measurements/tex_tables/mi_two-cost}
\input{measurements/tex_tables/mi_geometric}
\input{measurements/tex_figures/mi_appendix}

\clearpage
\paragraph{Hybrid Solver for Minimum-Cost Flows.}
We evaluated on 18 families~\cite{kovacsMinimumcostFlowAlgorithms2015} in total, generated by NETGEN~\cite{klingmanNETGENProgramGenerating1974}, GRIDGEN, GOTO~\cite{goldbergImplementingScalingPushRelabel1993} and GRIDGRAPH or converted from road networks~\cite{demetrescuShortestPathProblem2009} and vision instances.
{\renewcommand\nobreakspace{\ }
These are \texttt{goto\_8} \citecode{mcfDataGoto8}, \texttt{goto\_sr} \citecode{mcfDataGotoSr}, \texttt{grid\_long} \citecode{mcfDataGridLong}, \texttt{grid\_square} \citecode{mcfDataGridSquare}, \texttt{grid\_wide} \citecode{mcfDataGridWide}, \texttt{gridgen\_8} \citecode{mcfDataGridgen8}, \texttt{gridgen\_deg} \citecode{mcfDataGridgenDeg}, \texttt{gridgen\_sr} \citecode{mcfDataGridgenSr}, \texttt{netgen\_8} \citecode{mcfDataNetgen8}, \texttt{netgen\_deg} \citecode{mcfDataNetgenDeg}, \texttt{netgen\_lo\_8} \citecode{mcfDataNetgenLo8}, \texttt{netgen\_lo\_sr} \citecode{mcfDataNetgenLoSr}, \texttt{netgen\_sr} \citecode{mcfDataNetgenSr}, \texttt{road\_flow} \citecode{mcfDataRoad}, \texttt{road\_paths} \citecode{mcfDataRoad}, \texttt{vision\_inv} \citecode{mcfDataVision}, \texttt{vision\_prop} \citecode{mcfDataVision} and \texttt{vision\_rnd} \citecode{mcfDataVision}.}
For reasons of brevity, we only display 5 of them, \texttt{gridgen\_deg}, \texttt{netgen\_deg}, \texttt{netgen\_lo\_sr} (which performed worst), \texttt{road\_paths} (within the range of $\leq10\,\%$ to $\leq30\,\%$ for all others) and \texttt{goto\_8} (with exceptionally good performance), in \cref{tab:mcf-gridgen-deg,tab:mcf-netgen-deg,tab:mcf-netgen-lo-sr,tab:mcf-road-paths,tab:mcf-goto-8}.
The first two proceed by density at fixed node count, the third by node count with increasing density, and the last two by the number of nodes at fixed density.
As oracles we use network simplex~\citecode{lemonNetworkSimplex}, cost scaling~\citecode{lemonCostScaling}, capacity scaling and successive shortest paths~\citecode{lemonCapacityScaling}, whose code was taken from LEMON.
For every family, size and algorithm, five different instances were run once each, and every time is the mean of these five runs.
All times are compute times in milliseconds, measured after parsing.
\textit{Oracle} is the C++ solver's initialisation and main loop, \textit{Cert.}\ the emission of the certificate and \textit{Check} the verified check.
\textit{Glue} is what is left of the solver's entire compute time \textit{Total}.
The last column is the relative cost of verification, i.e.\ Total divided by Oracle.
For each size, the row of the algorithm whose oracle is fastest is shaded grey.
Note that we here use a higher timeout of 2 hours.

\input{measurements/tex_tables/mcf_hybrid}

\paragraph{Hybrid Solver for Weighted Bipartite Matching.}
We ran an analogous hybrid solver with CSA~\citecode{goldbergCSACode} as oracle on the families of \cref{tab:hungarian-low-cost,tab:hungarian-two-cost,tab:hungarian-geometric}, with columns as above and each time the mean over ten instances, each the median of three runs (\cref{tab:hybrid-low-cost,tab:hybrid-two-cost,tab:hybrid-geometric}).

\input{measurements/tex_tables/hybrid_low-cost}
\input{measurements/tex_tables/hybrid_two-cost}
\input{measurements/tex_tables/hybrid_geometric}

\bibliographystylecode{splncs04}
\bibliographycode{thomas_code_refs}

\flushbottom
\section{AI-Assisted Development}
\label{app:ai}

\begin{table}[htbp]
\centering
\scriptsize
\begin{tabular}{@{}lccc@{}}
\toprule
Algorithm & Abstract & Functional & Imperative\\
\midrule
\multicolumn{4}{@{}l}{\emph{Graph search and shortest paths}}\\
Depth-first search & $-$ & \xmark & (\checkmark CSR version only)\\
Cycle detection & $-$ & \xmark & $-$\\
Directed cycle detection & \checkmark & \checkmark & \checkmark\\
Breadth-first search & $-$ & \xmark & \xmark\\
BFS with distances & $-$ & \xmark & $-$\\
Bellman--Ford & $-$ & \xmark & $-$\\
Dijkstra & $-$ & \checkmark & \checkmark\\
\midrule
\multicolumn{4}{@{}l}{\emph{Greedy algorithms and spanning trees}}\\
Best-in-greedy & $-$ & \xmark & \checkmark\\
Kruskal & $-$ & \xmark & \checkmark\\
Greedoid greedy & $-$ & \xmark & $-$\\
Jarn\'{\i}k--Prim & $-$ & \xmark & $-$\\
\midrule
\multicolumn{4}{@{}l}{\emph{Flows}}\\
DFS with backtracking & $-$ & \xmark & $-$\\
Blocking flow & $-$ & \xmark & $-$\\
Dinic & $-$ & \xmark & $-$\\
Push-relabel & $-$ & \checkmark & \checkmark\\
Successive shortest paths & \xmark & $-$ & $-$\\
Capacity scaling & \xmark & $-$ & $-$\\
Orlin & $-$ & \xmark & $-$\\
Acyclic flows & $-$ & \checkmark & \checkmark\\
Network simplex & $-$ & \checkmark & \checkmark\\
\midrule
\multicolumn{4}{@{}l}{\emph{Matchings}}\\
Edmonds' blossom & \xmark & \checkmark & \checkmark\\
Hopcroft--Karp & $-$ & \checkmark & \checkmark\\
Hungarian method & $-$ & \xmark (shortcut \checkmark)& \checkmark\\
Primal-dual (Egerv\'ary) & \xmark & $-$ & $-$\\
RANKING & \xmark & $-$ & $-$\\
RANKING, primal-dual & \xmark & $-$ & $-$\\
AdWords & \xmark & $-$ & $-$\\
\midrule
\multicolumn{4}{@{}l}{\emph{Matroids}}\\
Matroid intersection & $-$ & \xmark & \checkmark\\
Weighted matroid intersection & $-$ & \checkmark & \checkmark\\
Matroid partitioning & \checkmark & $-$ & $-$\\
\bottomrule
\end{tabular}
\caption{\todo{Fill the cells marked ?.} The parts of each algorithm that were developed or verified using AI.
\checkmark means AI contribution, \xmark only human contributors, and $-$ signifies that this variant of the algorithm is not present.
Readers can obtain a more fine-grained understanding of the extent of AI contribution by reading through the theory scripts, which have labels indicating human/human-AI/AI contributions.}
\label{tab:ai-algorithms}
\end{table}

\subsection{Style Statistics}
\label{app:style-statistics}

\begin{table}[htbp]
\centering
\scriptsize
\setlength{\tabcolsep}{4pt}
\begin{tabular}{@{}lrrrr@{}}
\toprule
Statistic & DFS & BFS & Before & After \\
\midrule
\multicolumn{5}{@{}l}{\emph{Size}} \\
Proof lines & 323 & 1041 & 1342 & 1055 \\
Definition and specification lines & 355 & 552 & 472 & 917 \\
Lemmas & 81 & 103 & 108 & 182 \\
\quad of which loop-template lemmas & 65 & 83 & 3 & 92 \\
\midrule
\multicolumn{5}{@{}l}{\emph{Proof form (share of lemmas)}} \\
Single-step proofs & 76.5\% & 63.1\% & 38.0\% & 69.2\% \\
Structured (\texttt{proof}\,\dots\,\texttt{qed}) & 21.0\% & 27.2\% & 61.1\% & 27.5\% \\
With \texttt{apply} & 2.5\% & 9.7\% & 0.0\% & 3.3\% \\
\midrule
\multicolumn{5}{@{}l}{\emph{Proof length (non-blank lines)}} \\
Median & 1 & 2 & 8 & 2 \\
Mean & 3.7 & 9.9 & 12.0 & 5.6 \\
90th percentile & 8 & 31 & 30 & 15 \\
Proof lines per statement line & 1.50 & 3.15 & 4.43 & 2.02 \\
Steps per structured proof & 2.4 & 8.4 & 8.7 & 4.1 \\
\midrule
\multicolumn{5}{@{}l}{\emph{Handling of facts (per lemma)}} \\
Labelled \texttt{have}/\texttt{show}/\texttt{obtain} & 0.01 & 0.21 & 2.29 & 0.12 \\
\texttt{[OF \dots]} & 0.26 & 0.61 & 2.10 & 0.74 \\
\texttt{using} & 0.44 & 1.66 & 3.10 & 0.79 \\
Facts quoted as cartouches & 0.07 & 1.64 & 0.02 & 0.19 \\
Named assumptions & 0.00 & 0.00 & 0.51 & 0.00 \\
\midrule
\multicolumn{5}{@{}l}{\emph{Proof methods (share of all method calls)}} \\
\texttt{auto} & 61.2\% & 50.7\% & 10.7\% & 40.2\% \\
\texttt{force} + \texttt{fastforce} & 9.3\% & 12.4\% & 0.4\% & 1.9\% \\
\texttt{simp} & 2.2\% & 4.3\% & 29.0\% & 11.6\% \\
\texttt{rule} & 10.1\% & 7.5\% & 23.8\% & 13.8\% \\
\texttt{metis} + \texttt{smt} & 0.0\% & 4.1\% & 0.0\% & 0.0\% \\
\midrule
\multicolumn{5}{@{}l}{\emph{Closing calls inside structured proofs}} \\
\texttt{by (rule \dots)} & 2.6\% & 1.0\% & 25.9\% & 14.2\% \\
Bare \texttt{by simp}/\texttt{auto}/\texttt{blast}/\dots & 20.5\% & 21.6\% & 46.4\% & 34.7\% \\
\midrule
\multicolumn{5}{@{}l}{\emph{Method modifiers (per lemma)}} \\
\texttt{elim!:} & 0.38 & 0.50 & 0.00 & 0.14 \\
\texttt{intro!:} & 0.46 & 0.49 & 0.00 & 0.16 \\
\midrule
\multicolumn{5}{@{}l}{\emph{Statements}} \\
Plain form $[\![A;\,B]\!] \Longrightarrow C$ & 79.0\% & 77.7\% & 57.4\% & 79.1\% \\
$[\![$ per statement & 0.69 & 0.79 & 0.00 & 0.34 \\
Lemmas with an attribute & 80.2\% & 67.0\% & 6.5\% & 45.6\% \\
\midrule
\multicolumn{5}{@{}l}{\emph{Layout}} \\
\texttt{by} opening its own line & 100\% & 95.0\% & 16.8\% & 100\% \\
\texttt{using} opening its own line & 100\% & 99.4\% & 34.9\% & 100\% \\
\texttt{proof-}/\texttt{proof(} (no space) & 4/15 & 15/43 & 0/0 & 30/28 \\
\texttt{proof -}/\texttt{proof (} (space) & 0/0 & 0/0 & 41/52 & 0/0 \\
Line length, 95th percentile & 119 & 117 & 99 & 104 \\
\midrule
\multicolumn{5}{@{}l}{\emph{Definitions and program state}} \\
State \texttt{record}s & 1 & 1 & 0 & 2 \\
Loops as \texttt{function (domintros)} over one record & 1 & 1 & 0 & 2 \\
Loops written as \texttt{fun} & 0 & 0 & 1 & 0 \\
Executable twins (\texttt{partial\_function}) & 1 & 1 & 1 & 2 \\
\texttt{named\_theorems} declared & 9 & 9 & 0 & 14 \\
Untyped \texttt{definition}s & 22/22 & 23/23 & 24/26 & 50/50 \\
\midrule
\multicolumn{5}{@{}l}{\emph{Invariants}} \\
\texttt{invar\_*} definitions & 5 & 13 & 0 & 8 \\
Complete families (elim., intro., per branch, loop) & 5 & 7 & 0 & 8 \\
Uses of \texttt{[invar\_holds\_intros]} & 26 & 33 & 0 & 33 \\
\midrule
\multicolumn{5}{@{}l}{\emph{Naming}} \\
Mean lemma-name length (characters) & 18.2 & 22.1 & 14.2 & 22.8 \\
Mean lemma-name parts & 4.0 & 4.5 & 3.0 & 4.0 \\
Constants with the prefix \texttt{hk\_} & -- & -- & 11 & 0 \\
Global abbreviations named with $\le 3$ characters & 0 & 0 & 9 & 0 \\
\midrule
\multicolumn{5}{@{}l}{\emph{Layering}} \\
Interpretations in algorithm theories (global/plain) & 0/0 & 0/0 & 2/3 & 0/0 \\
Interpretations in the sub-procedure theory (global/plain) & -- & 0/0 & -- & 4/3$^{a}$ \\
Interpretations in the example theory (global/plain) & 1/0 & 2/0 & 1/1 & 1/1 \\
\texttt{sublocale} commands & 0 & 0 & 0 & 0 \\
Algorithm theories importing concrete data structures & 0 & 0 & 1 & 0 \\
Locales with qualified data-structure parents & 1 & 1 & 0 & 2 \\
Sub-procedure contracts in the definitions locale & -- & yes & no & yes \\
\bottomrule
\end{tabular}

\smallskip
{\scriptsize $^{a}$\,All in \texttt{Hopcroft\_Karp\_Subprocedures}, a recorded deviation: the phase is implemented there at the concrete operations, since the alternating BFS and the incremental DFS of the Micali--Vazirani development are used through their concrete instances.\par}
\caption{Style statistics of the depth- and breadth-first search theories (the ground truth) and of the Hopcroft--Karp theories before and after the restyling.}
\label{tab:hk-style}
\end{table}

{\sloppy
To make the stylistic conventions of the library measurable, we use a set of statistics that are computed mechanically from the theory sources.
The depth-first and breadth-first search theories (\texttt{DFS.thy} and \texttt{BFS\_2.thy}) serve as the reference: their values define the target range.
The other columns of Table~\ref{tab:hk-style} give the values for the Hopcroft--Karp development before and after the restyling.
Both columns cover five theories.
The restyling renamed and reorganised them: the characterisation of augmenting paths, the top loop, a new generic search for disjoint paths, the implementation of a phase, and the executable example.
For the layering statistics, the reference columns cover the whole families, i.e.\ also \texttt{DFS\_Example}, \texttt{BFS\_Subprocedures} and \texttt{BFS\_Example}.

All statistics are computed by two scripts.
They do not use Isabelle; they work on the source text, as follows:
\begin{enumerate}
\item Comments are removed.
\item The contents of inner-syntax strings and cartouches are masked, so that keywords occurring inside terms are not counted.
\item Each theory is split into its top-level commands.
\item Each \emph{lemma} (a \texttt{lemma}, \texttt{theorem}, \texttt{corollary} or \texttt{proposition} command; \texttt{lemmas} is not counted) is split into its statement and its proof.
  The proof begins at the first proof keyword (\texttt{by}, \texttt{apply}, \texttt{proof}, \texttt{using}, \texttt{unfolding}, \texttt{including}, \texttt{supply}, \texttt{subgoal}).
\end{enumerate}
Unless stated otherwise, ``per lemma'' means a total over all proofs divided by the number of lemmas.
A share of method calls is taken over all method calls in the group.

\paragraph{Size.}
Proofs and specifications are counted separately, in non-blank lines.
Comments, top-level \texttt{text} blocks and section headings are not counted in either.
\begin{description}
\item[Proof lines] The lines of all proofs: the proofs of lemmas, the proofs of \texttt{interpretation}, \texttt{global\_interpretation} and \texttt{sublocale} commands, and the proofs of \texttt{function} definitions.
\item[Definition and specification lines] The lines of all lemma statements, all definitions (\texttt{definition}, \texttt{abbreviation}, \texttt{fun}, \texttt{function}, \texttt{partial\_function}, \texttt{primrec}, \texttt{record}, \texttt{datatype}), the headers of locales and contexts (parents, \texttt{fixes}, \texttt{assumes}), and the instantiation part of interpretations.
\item[Lemmas] The number of lemma commands as defined above.
  It is the denominator of all per-lemma figures.
\item[Loop-template lemmas] The lemmas that the loop template of the reference theories prescribes, recognised by their names.
  They are:
  \begin{itemize}
  \item branch conditions;
  \item the four standard lemmas \texttt{\_cases}, \texttt{\_simps}, \texttt{\_induct} and \texttt{\_domintros};
  \item the elimination, introduction and preservation lemmas of the invariants;
  \item termination;
  \item the initial state, the return branches, the final lemmas and the executable twin.
  \end{itemize}
  They are short and uniform.
  In the reference theories they make up about four fifths of all lemmas.
\end{description}

\paragraph{Proof form.}
These statistics classify each proof by the commands it uses.
They indicate how much of the reasoning is delegated to a single call of an automatic method.
\begin{description}
\item[Single-step proofs] The share of lemmas whose proof contains no \texttt{proof} and no \texttt{apply}.
  Such a proof consists of one \texttt{by} call, possibly preceded by \texttt{using} or \texttt{unfolding}, and possibly with two methods (\texttt{by}~$m_1$~$m_2$).
\item[Structured] The share of lemmas whose proof contains at least one \texttt{proof}, i.e.\ an Isar proof with intermediate steps.
\item[With \texttt{apply}] The share of lemmas whose proof is a sequence of \texttt{apply} steps, with no \texttt{proof}.
  A structured proof that uses \texttt{apply} inside it, as the loop lemmas of DFS do, is counted as structured.
\end{description}

\paragraph{Proof length.}
The length of a proof is its number of non-blank lines, from its first proof keyword to the end of the command.
\begin{description}
\item[Median, mean and 90th percentile] The distribution of proof lengths over all lemmas.
  A median of one or two lines means that most lemmas are proved by one call.
\item[Proof lines per statement line] The total number of proof lines divided by the total number of statement lines.
  It relates the effort spent on proofs to the size of what is proved.
\item[Steps per structured proof] The mean number of \texttt{have}, \texttt{show}, \texttt{hence}, \texttt{thus} and \texttt{obtain} commands in a structured proof.
  It measures how finely a structured proof is subdivided.
\end{description}

\paragraph{Handling of facts.}
These statistics describe how intermediate and assumed facts are named and passed on.
Explicit names and explicit forward application point to proofs that spell out book-keeping which the automatic methods could find themselves.
\begin{description}
\item[Labelled \texttt{have}/\texttt{show}/\texttt{obtain}] The number of intermediate facts given a name (\texttt{have h:~\dots}, also with \texttt{show}, \texttt{obtain}, \texttt{hence} and \texttt{thus}), per lemma.
\item[{\texttt{[OF \dots]}}] The number of explicit instantiations of a theorem with given facts, per lemma.
\item[\texttt{using}] The number of \texttt{using} commands, per lemma.
  Each one inserts named facts into a goal.
\item[Facts quoted as cartouches] The number of cartouches in proofs that do not open a \texttt{text} block or a comment, per lemma.
  These are mostly references to earlier facts by their statement.
  The count also includes cartouches used as method arguments, so it is approximate.
\item[Named assumptions] The number of assumptions given a name in the statement (\texttt{assumes h:~\dots} or \texttt{and h:~\dots}), per lemma.
  The reference theories refer to assumptions by position (\texttt{assms(2)}) or by statement instead.
\end{description}

\paragraph{Proof methods.}
For each \texttt{by}, \texttt{apply} and \texttt{proof}, the name of the first method that follows is recorded.
Each row gives the share of one method among all recorded calls.
\begin{description}
\item[\texttt{auto}, \texttt{force} and \texttt{fastforce}] The automatic methods that combine simplification with classical reasoning.
  They are the usual closing methods in the reference theories.
\item[\texttt{simp}] Simplification alone.
\item[\texttt{rule}] The application of one named rule.
  A high share indicates step-by-step forward reasoning.
\item[\texttt{metis} and \texttt{smt}] Calls to proof reconstruction from external provers.
  They are excluded by the rules of this project, although the BFS theory still contains some.
\end{description}

\paragraph{Closing calls inside structured proofs.}
These statistics are computed over the \texttt{by} calls inside structured proofs only.
\begin{description}
\item[\texttt{by (rule \dots)}] The share of calls that apply a single rule.
\item[Bare \texttt{by simp}/\texttt{auto}/\texttt{blast}/\dots] The share of calls to an automatic method without any arguments (\texttt{simp}, \texttt{auto}, \texttt{blast}, \texttt{force}, \texttt{fastforce}, \texttt{linarith}, \texttt{presburger}, \texttt{simp\_all}, \texttt{arith}, \texttt{fast}).
  A high share indicates many small steps, each of which is trivial.
\end{description}

\paragraph{Method modifiers.}
The number of occurrences of a modifier in proofs, per lemma.
\begin{description}
\item[\texttt{elim!:} and \texttt{intro!:}] Rules declared as safe elimination and introduction rules for a single call.
  In the reference theories these modifiers mostly carry the named theorem collections (\texttt{call\_cond\_elims}, \texttt{invar\_props\_elims}, \texttt{invar\_props\_intros}).
  Their frequency therefore shows how far proofs are driven by these collections rather than by individual facts.
\end{description}

\paragraph{Statements.}
\begin{description}
\item[Plain form] The share of lemmas stated as a single proposition, without \texttt{assumes}/\texttt{shows}.
  Premises are then written as $[\![A;\,B]\!] \Longrightarrow C$.
\item[$[\![$ per statement] The number of premise brackets in statements, per lemma.
  It shows how often several premises are stated in this compact form.
\item[Lemmas with an attribute] The share of lemmas declared with an attribute, such as \texttt{[simp]} or \texttt{[invar\_holds\_intros]}.
  Attributes make a lemma available to later automatic calls, so this share measures how far the theory prepares its own automation.
\end{description}

\paragraph{Layout.}
\begin{description}
\item[\texttt{by} (\texttt{using}) opening its own line] The share of occurrences of \texttt{by} (\texttt{using}) inside proofs that begin a new line.
  The keyword that opens a proof is counted as beginning a line, so the figure mainly reflects the steps inside structured proofs.
  The reference theories place \texttt{using} and \texttt{by} on their own indented lines; they do not append them to the stated fact.
\item[\texttt{proof-}/\texttt{proof(} versus \texttt{proof -}/\texttt{proof (}] The number of occurrences of each spelling.
  The reference theories write these keywords without a space.
\item[Line length, 95th percentile] The length, in characters, that 95\% of the non-blank lines do not exceed.
\end{description}

\paragraph{Definitions and program state.}
\begin{description}
\item[State \texttt{record}s] The number of \texttt{record} commands.
  In the reference theories, a record holds the complete state of an algorithm.
\item[Loops as \texttt{function (domintros)} over one record] The number of loops defined by the function package with domain introduction rules generated, whose single argument is a state record.
  The number of arguments is counted by the script; that the argument is a record was checked in the source.
\item[Loops written as \texttt{fun}] Recursive algorithm definitions by \texttt{fun}, which the reference theories do not use for loops.
\item[Executable twins] The number of \texttt{partial\_function (tailrec)} definitions.
  Each is an executable copy of a loop, related to it by a lemma.
\item[\texttt{named\_theorems} declared] The number of named theorem collections declared, such as \texttt{invar\_holds\_intros} and \texttt{call\_cond\_elims}.
  They are the means by which proofs refer to whole families of lemmas.
\item[Untyped \texttt{definition}s] The number of definitions given as an equation without a type signature, out of all definitions.
\end{description}

\paragraph{Invariants.}
\begin{description}
\item[\texttt{invar\_*} definitions] The number of definitions whose name begins with \texttt{invar\_}, i.e.\ the number of separately stated invariants of the algorithm states.
\item[Complete families] The number of these invariants that have the full family of lemmas of the reference theories:
  \begin{itemize}
  \item an elimination form \texttt{invar\_$x$\_props};
  \item an introduction form \texttt{invar\_$x$\_intro};
  \item at least one preservation lemma for a branch of the loop (\texttt{invar\_$x$\_holds\_$i$});
  \item the preservation lemma for the whole loop (\texttt{invar\_$x$\_holds}).
  \end{itemize}
\item[{Uses of \texttt{[invar\_holds\_intros]}}] The number of lemmas registered in the collection of preservation lemmas.
\end{description}

\paragraph{Naming.}
\begin{description}
\item[Mean lemma-name length and parts] The mean number of characters, and the mean number of parts separated by underscores, over all named lemmas.
  Short names with few parts point to names that do not say what the lemma is about.
\item[Constants with the prefix \texttt{hk\_}] The number of defined constants whose names begin with \texttt{hk\_}.
  The library writes the name of an algorithm out in full.
\item[Global abbreviations named with $\le 3$ characters] The number of \texttt{abbreviation}s with names of at most three characters, such as \texttt{es} or \texttt{CG}.
\end{description}

\paragraph{Layering.}
These statistics concern the arrangement of algorithms, sub-procedures and implementations in locales and theories.
Each theory is in one of three classes:
\begin{itemize}
\item the \emph{example theory} (\texttt{*\_Example}), the only theory in which the reference algorithms are instantiated with concrete data structures;
\item the \emph{sub-procedure theory} (\texttt{*\_Subprocedures}), which implements the sub-procedures of an algorithm;
\item an \emph{algorithm theory}, which is any other theory. \vspace*{\baselineskip}
\end{itemize}
\begin{description}
\item[Interpretations in algorithm theories, in the sub-procedure theory]\vspace*{-\baselineskip}
\item[and in the example theory] The number of \texttt{global\_interpretation} and \texttt{interpretation} commands in theories of each class, given as global/plain.
  In the reference families, interpretations occur only in the example theories.
\item[\texttt{sublocale} commands] The number of \texttt{sublocale} commands.
  The reference theories do not use them to relate algorithms.
\item[Algorithm theories importing concrete data structures] The number of algorithm theories that import an implementation by red-black trees, or a theory that builds an executable instance on one.
\item[Locales with qualified data-structure parents] The number of locale definitions that import a data-structure specification as a qualified parent (\texttt{Graph:~Pair\_Graph\_Specs}, \texttt{set\_ops:~Set2}).
  Locales that only inherit such a parent are not counted.
\item[Sub-procedure contracts in the definitions locale] Whether the specifications of the sub-procedures, which are parameters of a loop, are assumptions of the locale that defines the loop (as in BFS), or of a separate locale used only for proofs.
  This was determined by reading the source; ``--'' means that there are no sub-procedures.
\end{description}

\paragraph{Limitations.}
The statistics are computed from the source text, not from the logical content that Isabelle constructs, so some of them are approximations:
\begin{itemize}
\item the count of cartouches used as fact references;
\item the classification of a \texttt{by} at the start of a proof as beginning a line;
\item the recording of only the first method of a call with two methods.
\end{itemize}
They measure conformity of form; they do not measure the difficulty of the mathematics.
A theory without loops, such as the characterisation of augmenting paths in \texttt{Alt\_BFS\_Bipartite}, naturally has longer structured proofs than a loop theory, even when it is written in the same style.
\par}



%% file: measurements/tex_tables/dfs_square.tex
\begin{table}[H]
\centering
\scriptsize
\setlength{\tabcolsep}{3pt}
\caption{Depth-first search, Square-n family~\protect\citecode{ch9CodeSuite} ($m \approx 4n$, 10 instances per size, averaged over the medians of 3 repeats per instance).}
\label{tab:dfs-square}

\end{table}

%% file: measurements/tex_tables/dfs_long.tex
\begin{table}[H]
\centering
\scriptsize
\setlength{\tabcolsep}{3pt}
\caption{Depth-first search, Long-n family~\protect\citecode{ch9CodeSuite} ($m \approx 4n$, 10 instances per size, averaged over the medians of 3 repeats per instance).}
\label{tab:dfs-long}
%
\end{table}

%% file: measurements/tex_figures/dfs_long.tex
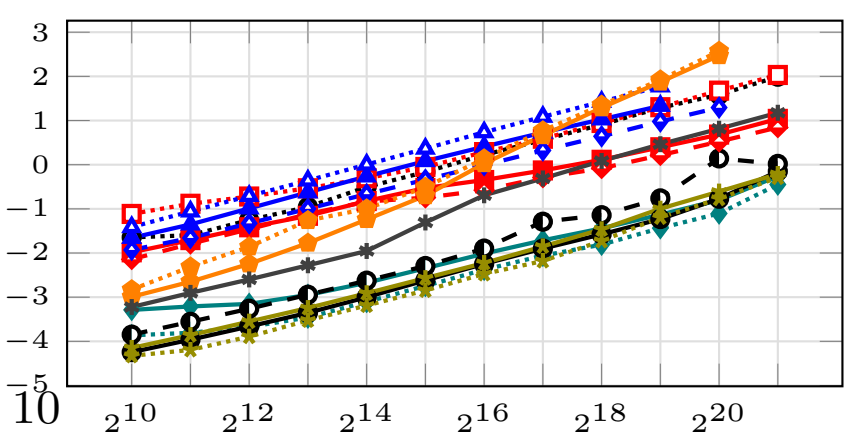
\begin{figure}[H]
\centering
\begin{tikzpicture}
\begin{groupplot}[group style={group size=1 by 1, horizontal sep=0.7cm, vertical sep=1.1cm}, xmode=log, ymode=log, log basis x=2, ytickten={-6,...,6}, log basis y=10, yticklabel={$\pgfmathparse{int(round(\tick))}\pgfmathresult$}, after end axis/.code={\node[anchor=north east, font=\footnotesize, inner sep=1pt] at (rel axis cs:0,0) {$10$};}, grid=major, grid style={gray!25}, tick label style={font=\tiny}, label style={font=\scriptsize}, every axis plot/.append style={thick, mark size=1.5pt, legend image code/.code={\draw[mark repeat=2, mark phase=2] plot coordinates {(0cm,0cm) (0.15cm,0cm) (0.3cm,0cm)};}}, width=0.55\textwidth, height=4cm, title style={font=\scriptsize, align=center}]
\nextgroupplot[title={Long-n (sparse), Tb.~\ref{tab:dfs-long}}, xtick={1024,2048,4096,8192,16384,32768,65536,131072,262144,524288,1048576,2097152}, xticklabels={$2^{10}$,,$2^{12}$,,$2^{14}$,,$2^{16}$,,$2^{18}$,,$2^{20}$,}]
\addplot[teal, mark=diamond*, mark options={solid}] coordinates {(1024,0.000515) (2048,0.000617) (4096,0.000705) (8192,0.001124) (16384,0.002119) (32768,0.004519) (65536,0.00968) (131072,0.019368) (262144,0.036148) (524288,0.073973) (1048576,0.158045) (2097152,0.549907)};
\label{leg-dfs_long:scipy_csr}
\addplot[teal, densely dotted, mark=diamond*, mark options={solid}] coordinates {(1024,0.000137) (2048,0.000156) (4096,0.000213) (8192,0.000346) (16384,0.000784) (32768,0.001798) (65536,0.004195) (131072,0.008713) (262144,0.01551) (524288,0.036604) (1048576,0.078726) (2097152,0.355943)};
\label{leg-dfs_long:scipy_prebuilt}
\addplot[black, mark=*, mark options={solid}] coordinates {(1024,5.7e-05) (2048,0.000109) (4096,0.000217) (8192,0.000442) (16384,0.000998) (32768,0.00241) (65536,0.005673) (131072,0.012625) (262144,0.026505) (524288,0.058742) (1048576,0.169354) (2097152,0.707776)};
\label{leg-dfs_long:sml_csr}
\addplot[black, dashed, mark=halfcircle*, mark options={solid, rotate=90}] coordinates {(1024,0.000142) (2048,0.000277) (4096,0.000547) (8192,0.001137) (16384,0.002368) (32768,0.005079) (65536,0.012616) (131072,0.051161) (262144,0.071914) (524288,0.172719) (1048576,1.38689) (2097152,1.04272)};
\label{leg-dfs_long:sml_lists}
\addplot[black, densely dotted, mark=*, mark options={solid, fill=white}] coordinates {(1024,0.02261) (2048,0.025287) (4096,0.055825) (8192,0.110791) (16384,0.312342) (32768,0.683522) (65536,1.66499) (131072,3.63755) (262144,8.07819) (524288,19.4538) (1048576,38.3092) (2097152,98.3038)};
\label{leg-dfs_long:sml_rbt}
\addplot[red, mark=square*, mark options={solid}] coordinates {(1024,0.01038) (2048,0.019633) (4096,0.038299) (8192,0.069624) (16384,0.150318) (32768,0.29607) (65536,0.449993) (131072,0.740744) (262144,1.29106) (524288,2.48038) (1048576,4.81104) (2097152,10.8941)};
\label{leg-dfs_long:scala_csr}
\addplot[red, dashed, mark=halfsquare*, mark options={solid}] coordinates {(1024,0.007197) (2048,0.016176) (4096,0.034741) (8192,0.076153) (16384,0.146495) (32768,0.178939) (65536,0.265034) (131072,0.51319) (262144,0.827257) (524288,1.63958) (1048576,3.33442) (2097152,6.97443)};
\label{leg-dfs_long:scala_lists}
\addplot[red, densely dotted, mark=square*, mark options={solid, fill=white}] coordinates {(1024,0.076719) (2048,0.130446) (4096,0.190274) (8192,0.292851) (16384,0.492687) (32768,0.888475) (65536,1.92952) (131072,3.87158) (262144,8.80433) (524288,20.0333) (1048576,47.5355) (2097152,107.583)};
\label{leg-dfs_long:scala_rbt}
\addplot[blue, mark=triangle*, mark options={solid}] coordinates {(1024,0.022079) (2048,0.044406) (4096,0.102471) (8192,0.236325) (16384,0.542508) (32768,1.20836) (65536,2.56202) (131072,5.34324) (262144,10.6299) (524288,21.5329)};
\label{leg-dfs_long:go_csr}
\addplot[blue, dashed, mark=halfdiamond*, mark options={solid}] coordinates {(1024,0.011741) (2048,0.02194) (4096,0.046995) (8192,0.10055) (16384,0.215524) (32768,0.45829) (65536,0.975581) (131072,2.10253) (262144,4.39269) (524288,9.5171) (1048576,19.6631)};
\label{leg-dfs_long:go_lists}
\addplot[blue, densely dotted, mark=triangle*, mark options={solid, fill=white}] coordinates {(1024,0.038742) (2048,0.087388) (4096,0.19631) (8192,0.43858) (16384,1.00876) (32768,2.31048) (65536,5.43937) (131072,12.1706) (262144,26.6455) (524288,59.3333)};
\label{leg-dfs_long:go_rbt}
\addplot[olive, mark=star, mark options={solid}] coordinates {(1024,7.1e-05) (2048,0.000138) (4096,0.00028) (8192,0.000568) (16384,0.001228) (32768,0.002705) (65536,0.005943) (131072,0.014079) (262144,0.034466) (524288,0.100842) (1048576,0.246345) (2097152,0.61616)};
\label{leg-dfs_long:lemon_csr}
\addplot[olive, densely dotted, mark=star, mark options={solid}] coordinates {(1024,4.7e-05) (2048,6.4e-05) (4096,0.000128) (8192,0.000298) (16384,0.000649) (32768,0.001422) (65536,0.003359) (131072,0.006607) (262144,0.018832) (524288,0.064078) (1048576,0.169504) (2097152,0.512505)};
\label{leg-dfs_long:lemon_linked}
\addplot[orange, mark=pentagon*, mark options={solid}] coordinates {(1024,0.001019) (2048,0.002239) (4096,0.005683) (8192,0.016801) (16384,0.057537) (32768,0.19925) (65536,1.05851) (131072,4.56434) (262144,19.4682) (524288,74.4654) (1048576,295.699)};
\label{leg-dfs_long:jgrapht_csr}
\addplot[orange, densely dotted, mark=pentagon*, mark options={solid}] coordinates {(1024,0.001482) (2048,0.004852) (4096,0.013836) (8192,0.054554) (16384,0.100344) (32768,0.311406) (65536,1.36276) (131072,5.8806) (262144,22.4282) (524288,83.6779) (1048576,372.873)};
\label{leg-dfs_long:jgrapht_hashgraph}
\addplot[darkgray, mark=asterisk, mark options={solid}] coordinates {(1024,0.000598) (2048,0.001252) (4096,0.002523) (8192,0.005279) (16384,0.011124) (32768,0.049121) (65536,0.20284) (131072,0.488224) (262144,1.21424) (524288,2.90866) (1048576,6.5286) (2097152,15.0152)};
\label{leg-dfs_long:networkx}
\end{groupplot}
\end{tikzpicture}
\caption{DFS: filled marks are exports over CSR arrays, half-filled marks those over adjacency lists, hollow marks those over red-black trees. \ref{leg-dfs_long:sml_csr}~SML CSR, \ref{leg-dfs_long:sml_lists}~SML lists, \ref{leg-dfs_long:sml_rbt}~SML RBT, \ref{leg-dfs_long:scala_csr}~Scala CSR, \ref{leg-dfs_long:scala_lists}~Scala lists, \ref{leg-dfs_long:scala_rbt}~Scala RBT, \ref{leg-dfs_long:go_csr}~Go CSR, \ref{leg-dfs_long:go_lists}~Go lists, \ref{leg-dfs_long:go_rbt}~Go RBT, \ref{leg-dfs_long:lemon_csr}~LEMON CSR, \ref{leg-dfs_long:lemon_linked}~LEMON linked, \ref{leg-dfs_long:jgrapht_csr}~JGraphT CSR, \ref{leg-dfs_long:jgrapht_hashgraph}~JGraphT hash, \ref{leg-dfs_long:scipy_csr}~SciPy CSR, \ref{leg-dfs_long:scipy_prebuilt}~SciPy prebuilt, \ref{leg-dfs_long:networkx}~NetworkX.}
\label{fig:dfs-long}
\end{figure}

%% file: measurements/tex_tables/dfs_dense.tex
\begin{table}[H]
\centering
\scriptsize
\setlength{\tabcolsep}{3pt}
\caption{Depth-first search, dense connected rand. family~\protect\citecode{ch9CodeSuite} ($m = n^2/16$, 10 instances per size, averaged over the medians of 3 repeats per instance). The instances are those of \cref{tab:bfs-dense}.}
\label{tab:dfs-dense}
%
\end{table}

%% file: measurements/tex_tables/dfs_road.tex
\begin{table}[H]
\centering
\scriptsize
\setlength{\tabcolsep}{3pt}
\caption{Depth-first search, USA-road-t family~\protect\citecode{ch9RoadNetworks} ($2.35n \le m \le 2.78n$, one network per size, 10 sources each, median of 3 repeats each).}
\label{tab:dfs-road}
%
\end{table}

%% file: measurements/tex_tables/bfs_square.tex
\begin{table}[H]
\centering
\scriptsize
\setlength{\tabcolsep}{3pt}
\caption{Breadth-first search, Square-n family~\protect\citecode{ch9CodeSuite} ($m \approx 4n$, 10 instances per size, averaged over the medians of 3 repeats per instance).}
\label{tab:bfs-square}
%
\end{table}

%% file: measurements/tex_tables/bfs_long.tex
\begin{table}[H]
\centering
\scriptsize
\setlength{\tabcolsep}{3pt}
\caption{Breadth-first search, Long-n family~\protect\citecode{ch9CodeSuite} ($m \approx 4n$, 10 instances per size, averaged over the medians of 3 repeats per instance).}
\label{tab:bfs-long}
%
\end{table}

%% file: measurements/tex_figures/bfs_long.tex
\begin{figure}[H]
\centering
\begin{tikzpicture}
\begin{groupplot}[group style={group size=1 by 1, horizontal sep=0.7cm, vertical sep=1.1cm}, xmode=log, ymode=log, log basis x=2, ytickten={-6,...,6}, log basis y=10, yticklabel={$\pgfmathparse{int(round(\tick))}\pgfmathresult$}, after end axis/.code={\node[anchor=north east, font=\footnotesize, inner sep=1pt] at (rel axis cs:0,0) {$10$};}, grid=major, grid style={gray!25}, tick label style={font=\tiny}, label style={font=\scriptsize}, every axis plot/.append style={thick, mark size=1.5pt, legend image code/.code={\draw[mark repeat=2, mark phase=2] plot coordinates {(0cm,0cm) (0.15cm,0cm) (0.3cm,0cm)};}}, width=0.55\textwidth, height=4cm, title style={font=\scriptsize, align=center}]
\nextgroupplot[title={Long-n (sparse), Tb.~\ref{tab:bfs-long}}, xtick={1024,2048,4096,8192,16384,32768,65536,131072,262144,524288,1048576,2097152}, xticklabels={$2^{10}$,,$2^{12}$,,$2^{14}$,,$2^{16}$,,$2^{18}$,,$2^{20}$,}]
\addplot[teal, mark=diamond*, mark options={solid}] coordinates {(1024,0.000414) (2048,0.000496) (4096,0.000669) (8192,0.000972) (16384,0.001759) (32768,0.003866) (65536,0.008341) (131072,0.016295) (262144,0.03063) (524288,0.058367) (1048576,0.114167) (2097152,0.3578)};
\label{leg-bfs_long:scipy_csr}
\addplot[teal, densely dotted, mark=diamond*, mark options={solid}] coordinates {(1024,0.000116) (2048,0.000137) (4096,0.000177) (8192,0.000288) (16384,0.000562) (32768,0.001209) (65536,0.002583) (131072,0.005528) (262144,0.009254) (524288,0.02051) (1048576,0.035773) (2097152,0.129284)};
\label{leg-bfs_long:scipy_prebuilt}
\addplot[black, mark=*, mark options={solid}] coordinates {(1024,5.2e-05) (2048,0.000106) (4096,0.000201) (8192,0.000398) (16384,0.000862) (32768,0.001908) (65536,0.004202) (131072,0.009863) (262144,0.020023) (524288,0.044679) (1048576,0.123269) (2097152,0.459474)};
\label{leg-bfs_long:sml_csr}
\addplot[black, densely dotted, mark=*, mark options={solid, fill=white}] coordinates {(1024,0.019283) (2048,0.063479) (4096,0.226421) (8192,0.792974) (16384,3.07226) (32768,11.907) (65536,51.3526) (131072,191.417)};
\label{leg-bfs_long:sml_rbt}
\addplot[red, mark=square*, mark options={solid}] coordinates {(1024,0.010131) (2048,0.019657) (4096,0.037473) (8192,0.071962) (16384,0.138428) (32768,0.26441) (65536,0.437181) (131072,0.64832) (262144,1.1137) (524288,2.05056) (1048576,3.93167) (2097152,8.4506)};
\label{leg-bfs_long:scala_csr}
\addplot[red, densely dotted, mark=square*, mark options={solid, fill=white}] coordinates {(1024,0.238576) (2048,0.438412) (4096,1.14383) (8192,3.47314) (16384,19.7841) (32768,66.4567) (65536,560.943)};
\label{leg-bfs_long:scala_rbt}
\addplot[blue, mark=triangle*, mark options={solid}] coordinates {(1024,0.038653) (2048,0.03347) (4096,0.076412) (8192,0.169451) (16384,0.374639) (32768,0.816783) (65536,1.74701) (131072,3.65573) (262144,7.42008) (524288,14.9271)};
\label{leg-bfs_long:go_csr}
\addplot[blue, densely dotted, mark=triangle*, mark options={solid, fill=white}] coordinates {(1024,0.104017) (2048,0.381918) (4096,1.43518) (8192,5.65823) (16384,22.0516) (32768,91.1077) (65536,549.349)};
\label{leg-bfs_long:go_rbt}
\addplot[olive, mark=star, mark options={solid}] coordinates {(1024,0.000138) (2048,0.000122) (4096,0.000247) (8192,0.000502) (16384,0.00104) (32768,0.002232) (65536,0.004791) (131072,0.012237) (262144,0.028601) (524288,0.08706) (1048576,0.167749) (2097152,0.392756)};
\label{leg-bfs_long:lemon_csr}
\addplot[olive, densely dotted, mark=star, mark options={solid}] coordinates {(1024,2.1e-05) (2048,4e-05) (4096,8.4e-05) (8192,0.00017) (16384,0.000374) (32768,0.000854) (65536,0.00172) (131072,0.004146) (262144,0.018674) (524288,0.042303) (1048576,0.081004) (2097152,0.281969)};
\label{leg-bfs_long:lemon_linked}
\addplot[orange, mark=pentagon*, mark options={solid}] coordinates {(1024,0.000753) (2048,0.001394) (4096,0.002621) (8192,0.00559) (16384,0.011507) (32768,0.022912) (65536,0.055665) (131072,0.221551) (262144,0.562108) (524288,1.37404) (1048576,3.14757) (2097152,6.89994)};
\label{leg-bfs_long:jgrapht_csr}
\addplot[orange, densely dotted, mark=pentagon*, mark options={solid}] coordinates {(1024,0.000708) (2048,0.001501) (4096,0.002839) (8192,0.007235) (16384,0.018727) (32768,0.042753) (65536,0.065271) (131072,0.123578) (262144,0.416343) (524288,0.835573) (1048576,1.66516) (2097152,3.39374)};
\label{leg-bfs_long:jgrapht_hashgraph}
\addplot[darkgray, mark=asterisk, mark options={solid}] coordinates {(1024,0.000553) (2048,0.001102) (4096,0.002214) (8192,0.004445) (16384,0.009162) (32768,0.021509) (65536,0.056279) (131072,0.138901) (262144,0.26975) (524288,0.622639) (1048576,1.3922) (2097152,3.34391)};
\label{leg-bfs_long:networkx}
\end{groupplot}
\end{tikzpicture}
\caption{BFS: filled marks are exports over CSR arrays, hollow marks those over red-black trees. \ref{leg-bfs_long:sml_csr}~SML imp., \ref{leg-bfs_long:sml_rbt}~SML func., \ref{leg-bfs_long:scala_csr}~Scala imp., \ref{leg-bfs_long:scala_rbt}~Scala func., \ref{leg-bfs_long:go_csr}~Go imp., \ref{leg-bfs_long:go_rbt}~Go func., \ref{leg-bfs_long:lemon_csr}~LEMON CSR, \ref{leg-bfs_long:lemon_linked}~LEMON linked, \ref{leg-bfs_long:jgrapht_csr}~JGraphT CSR, \ref{leg-bfs_long:jgrapht_hashgraph}~JGraphT hash, \ref{leg-bfs_long:scipy_csr}~SciPy CSR, \ref{leg-bfs_long:scipy_prebuilt}~SciPy prebuilt, \ref{leg-bfs_long:networkx}~NetworkX.}
\label{fig:bfs-long}
\end{figure}
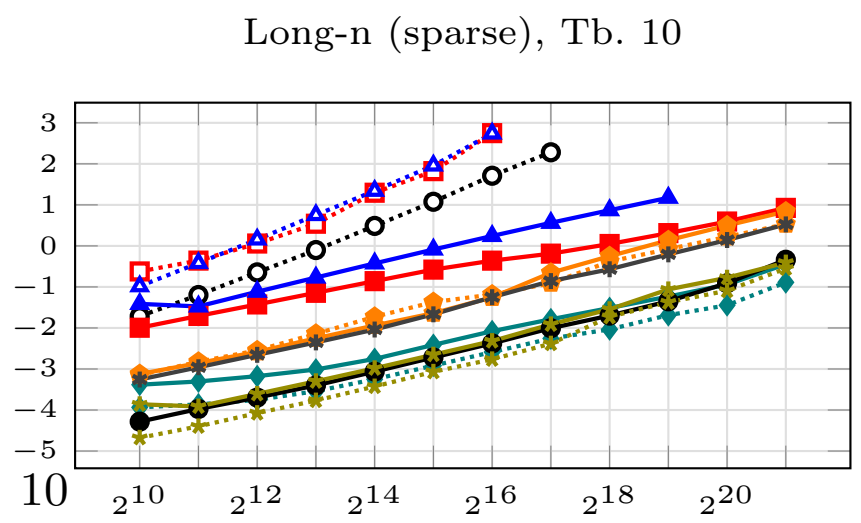

%% file: measurements/tex_tables/bfs_dense.tex
\begin{table}[H]
\centering
\scriptsize
\setlength{\tabcolsep}{3pt}
\caption{Breadth-first search, dense connected rand. family~\protect\citecode{ch9CodeSuite} ($m = n^2/16$, 10 instances per size, averaged over the medians of 3 repeats per instance).}
\label{tab:bfs-dense}
%
\end{table}

%% file: measurements/tex_tables/bfs_road.tex
\begin{table}[H]
\centering
\scriptsize
\setlength{\tabcolsep}{3pt}
\caption{Breadth-first search, USA-road-t family~\protect\citecode{ch9RoadNetworks} ($2.35n \le m \le 2.78n$, one network per size, 10 source-target pairs each, median of 3 repeats each).}
\label{tab:bfs-road}
%
\end{table}

%% file: measurements/tex_tables/dijkstra_square.tex
\begin{table}[H]
\centering
\scriptsize
\setlength{\tabcolsep}{3pt}
\caption{Dijkstra's algorithm, Square-n family~\protect\citecode{ch9CodeSuite} ($m \approx 4n$, 10 instances per size, averaged over the medians of 3 repeats per instance).}
\label{tab:dijkstra-square}
%
\end{table}

%% file: measurements/tex_tables/dijkstra_long.tex
\begin{table}[H]
\centering
\scriptsize
\setlength{\tabcolsep}{3pt}
\caption{Dijkstra's algorithm, Long-n family~\protect\citecode{ch9CodeSuite} ($m \approx 4n$, 10 instances per size, averaged over the medians of 3 repeats per instance).}
\label{tab:dijkstra-long}
%
\end{table}

%% file: measurements/tex_tables/dijkstra_road.tex
\begin{table}[H]
\centering
\scriptsize
\setlength{\tabcolsep}{3pt}
\caption{Dijkstra's algorithm, USA-road-t family~\protect\citecode{ch9RoadNetworks} ($2.35n \le m \le 2.78n$, one network per size, 10 source-target pairs each, median of 3 repeats each).}
\label{tab:dijkstra-road}
%
\end{table}

%% file: measurements/tex_tables/kruskal_square.tex
\begin{table}[H]
\centering
\scriptsize
\setlength{\tabcolsep}{3pt}
\caption{Kruskal's algorithm, Square-n family~\protect\citecode{ch9CodeSuite} ($m \approx 4n$, 10 instances per size, averaged over the medians of 3 repeats per instance).}
\label{tab:kruskal-square}
%
\end{table}

%% file: measurements/tex_tables/kruskal_long.tex
\begin{table}[H]
\centering
\scriptsize
\setlength{\tabcolsep}{3pt}
\caption{Kruskal's algorithm, Long-n family~\protect\citecode{ch9CodeSuite} ($m \approx 4n$, 10 instances per size, averaged over the medians of 3 repeats per instance).}
\label{tab:kruskal-long}
%
\end{table}

%% file: measurements/tex_tables/kruskal_dense.tex
\begin{table}[H]
\centering
\scriptsize
\setlength{\tabcolsep}{3pt}
\caption{Kruskal's algorithm, dense connected rand. family~\protect\citecode{ch9CodeSuite} ($m = n^2/16$, 10 instances per size, averaged over the medians of 3 repeats per instance). The instances are those of \cref{tab:bfs-dense}.}
\label{tab:kruskal-dense}
%
\end{table}

%% file: measurements/tex_tables/push_relabel_rmf-long.tex
\begin{table}[H]
\centering
\scriptsize
\setlength{\tabcolsep}{2pt}
\caption{Push-relabel, RMF-long family~\protect\citecode{flows2Genrmf} ($m \approx 5n$, 3 instances per size). Plots were cut off at $n=2^{23}$. The family does not meet the structural prerequisites for SML AFP~\protect\citecode{lammichPrpuMaxflowAFP}.}
\label{tab:push-relabel-rmf-long}
\resizebox{\textwidth}{!}{%
}
\end{table}

%% file: measurements/tex_tables/push_relabel_wash-rlg-long.tex
\begin{table}[H]
\centering
\scriptsize
\setlength{\tabcolsep}{2pt}
\caption{Push-relabel, WASH-RLG-long family~\protect\citecode{flows2Washington} ($m \approx 3n$, 3 instances per size). Plots were cut off at $n=2^{23}$.}
\label{tab:push-relabel-wash-rlg-long}
\resizebox{\textwidth}{!}{%
}
\end{table}

%% file: measurements/tex_tables/push_relabel_ak.tex
\begin{table}[H]
\centering
\scriptsize
\setlength{\tabcolsep}{2pt}
\caption{Push-relabel, AK family~\protect\citecode{cherkasskyGoldbergAkCode} ($m \approx 3n/2$, 1 instance per size, median of 3 repeats).}
\label{tab:push-relabel-ak}
\resizebox{\textwidth}{!}{%
}
\end{table}

%% file: measurements/tex_tables/hungarian_low-cost.tex
\begin{table}[H]
\centering
\scriptsize
\setlength{\tabcolsep}{3pt}
\caption{Hungarian method, low-cost family~\protect\citecode{dimacsAssign} ($m = n \log_2 n$, 10 instances per size, averaged over the medians of 3 repeats per instance).}
\label{tab:hungarian-low-cost}
%
\end{table}

%% file: measurements/tex_tables/hungarian_two-cost.tex
\begin{table}[H]
\centering
\scriptsize
\setlength{\tabcolsep}{3pt}
\caption{Hungarian method, two-cost family~\protect\citecode{dimacsTwoCostGen} ($m = n \log_2 n$, 10 instances per size, averaged over the medians of 3 repeats per instance).}
\label{tab:hungarian-two-cost}
%
\end{table}

%% file: measurements/tex_tables/hungarian_geometric.tex
\begin{table}[H]
\centering
\scriptsize
\setlength{\tabcolsep}{3pt}
\caption{Hungarian method, geometric family~\protect\citecode{dimacsGeometricGen} ($m = n^2/4$, 10 instances per size, averaged over the medians of 3 repeats per instance).}
\label{tab:hungarian-geometric}
%
 \\
\bottomrule
\end{tabular}
\end{table}

%% file: measurements/tex_tables/mi_low-two-cost.tex
\begin{table}[H]
\centering
\scriptsize
\setlength{\tabcolsep}{3pt}
\caption{Matroid intersection, unweighted problems on the low-cost and two-cost families~\protect\citecode{dimacsTwoCostGen}, which share their graphs ($m = n \log_2 n$, 5 instances per size, run once in each family, mean over both runs).}
\label{tab:mi-low-two-cost}
\begin{tabular}{@{}rrrrrr@{}}
\toprule
$n$ & SML & Scala & Go & \begin{tabular}[t]{@{}r@{}}Sage (our oracle)\\\citecode{sageIntersectionUnweighted}\end{tabular} & \begin{tabular}[t]{@{}r@{}}Sage (plain)\\\citecode{sageIntersectionUnweighted,sagePartitionMatroid,sageRankMatroid}\end{tabular} \\
\midrule
\multicolumn{6}{@{}l}{\emph{Matching}} \\
32 & $<$0.001 (--) & 0.038 (--) & 0.008 (--) & 0.003 (--) & 0.021 (--) \\
64 & $<$0.001 (--) & 0.068 (--) & 0.025 (--) & 0.009 (--) & 0.137 (--) \\
128 & 0.0006 (1.00) & 0.123 (206) & 0.094 (157) & 0.053 (88.8) & 1.49 (2480) \\
256 & 0.002 (1.00) & 0.229 (99.3) & 0.337 (146) & 0.337 (147) & 14.8 (6421) \\
512 & 0.009 (1.00) & 0.488 (53.6) & 1.52 (167) & 4.42 (485) & 176 (19357) \\
1024 & 0.032 (1.00) & 0.830 (26.0) & 5.70 (179) & 46.3 (1452) & TO \\
$2^{11}$ & 0.146 (1.00) & 2.27 (15.5) & 28.1 (193) & TO & TO \\
$2^{12}$ & 0.394 (1.00) & 7.29 (18.5) & 116 (294) & TO & TO \\
$2^{13}$ & 1.67 (1.00) & 29.1 (17.4) & 476 (284) & TO & TO \\
$2^{14}$ & 7.93 (1.00) & 123 (15.5) & TO & TO & TO \\
$2^{15}$ & 37.0 (1.00) & 530 (14.3) & TO & TO & TO \\
\midrule
\multicolumn{6}{@{}l}{\emph{$b$-matching}} \\
32 & $<$0.001 (--) & 0.013 (--) & 0.019 (--) & 0.011 (--) & 0.233 (--) \\
64 & $<$0.001 (--) & 0.039 (--) & 0.067 (--) & 0.076 (--) & 2.56 (--) \\
128 & 0.002 (1.00) & 0.234 (117) & 0.623 (312) & 1.63 (815) & 40.2 (20079) \\
256 & 0.018 (1.00) & 0.614 (34.9) & 4.79 (272) & 31.2 (1775) & 591 (33606) \\
512 & 0.121 (1.00) & 2.24 (18.5) & 30.1 (249) & 539 (4460) & TO \\
1024 & 0.885 (1.00) & 11.8 (13.3) & 201 (227) & TO & TO \\
$2^{11}$ & 4.83 (1.00) & 59.6 (12.4) & 1069 (221) & TO & TO \\
$2^{12}$ & 22.0 (1.00) & 267 (12.2) & TO & TO & TO \\
$2^{13}$ & 123 (1.00) & TO & TO & TO & TO \\
$2^{14}$ & 537 (1.00) & TO & TO & TO & TO \\
$2^{15}$ & TO & TO & TO & TO & TO \\
\bottomrule
\end{tabular}
\end{table}

%% file: measurements/tex_tables/mi_low-cost.tex
\begin{table}[H]
\centering
\scriptsize
\setlength{\tabcolsep}{3pt}
\caption{Matroid intersection, weighted problems on the low-cost family~\protect\citecode{dimacsAssign} ($m = n \log_2 n$, 5 instances per size, one run each). The unweighted problems are in \cref{tab:mi-low-two-cost}.}
\label{tab:mi-low-cost}
\begin{tabular}{@{}rrrrrr@{}}
\toprule
$n$ & SML & Scala & Go & \begin{tabular}[t]{@{}r@{}}Sage (our oracle)\\\citecode{sageIntersection}\end{tabular} & \begin{tabular}[t]{@{}r@{}}Sage (plain)\\\citecode{sageIntersection,sagePartitionMatroid,sageRankMatroid}\end{tabular} \\
\midrule
\multicolumn{6}{@{}l}{\emph{Weighted matching}} \\
32 & $<$0.001 (--) & 0.028 (--) & 0.067 (--) & 0.005 (--) & 0.047 (--) \\
64 & 0.001 (1.00) & 0.086 (86.4) & 0.253 (253) & 0.030 (29.6) & 0.954 (954) \\
128 & 0.004 (1.00) & 0.154 (40.6) & 0.807 (212) & 0.198 (52.2) & 19.9 (5231) \\
256 & 0.011 (1.00) & 0.300 (28.3) & 2.24 (212) & 1.49 (140) & 398 (37580) \\
512 & 0.037 (1.00) & 0.826 (22.3) & 7.38 (199) & 11.7 (317) & TO \\
1024 & 0.119 (1.00) & 2.29 (19.2) & 22.1 (186) & 103 (868) & TO \\
$2^{11}$ & 0.456 (1.00) & 8.12 (17.8) & 88.1 (193) & 868 (1904) & TO \\
$2^{12}$ & 1.69 (1.00) & 29.3 (17.4) & 349 (206) & TO & TO \\
$2^{13}$ & 6.81 (1.00) & 126 (18.5) & TO & TO & TO \\
$2^{14}$ & 31.0 (1.00) & 577 (18.6) & TO & TO & TO \\
$2^{15}$ & 157 (1.00) & TO & TO & TO & TO \\
\midrule
\multicolumn{6}{@{}l}{\emph{Weighted $b$-matching}} \\
32 & 0.001 (1.00) & 0.021 (21.2) & 0.209 (209) & 0.024 (23.8) & 1.90 (1905) \\
64 & 0.003 (1.00) & 0.060 (23.2) & 0.662 (255) & 0.166 (63.7) & 41.2 (15831) \\
128 & 0.011 (1.00) & 0.215 (19.9) & 2.74 (254) & 2.12 (197) & 1015$^{1/5}$ (93994) \\
256 & 0.044 (1.00) & 0.840 (19.2) & 10.8 (248) & 23.9 (546) & TO \\
512 & 0.193 (1.00) & 3.13 (16.2) & 44.1 (229) & 259 (1346) & TO \\
1024 & 0.997 (1.00) & 13.5 (13.6) & 192 (193) & TO & TO \\
$2^{11}$ & 4.69 (1.00) & 61.8 (13.2) & 866 (185) & TO & TO \\
$2^{12}$ & 22.7 (1.00) & 290 (12.8) & TO & TO & TO \\
$2^{13}$ & 114 (1.00) & TO & TO & TO & TO \\
$2^{14}$ & 591 (1.00) & TO & TO & TO & TO \\
$2^{15}$ & TO & TO & TO & TO & TO \\
\bottomrule
\end{tabular}
\end{table}

%% file: measurements/tex_tables/mi_two-cost.tex
\begin{table}[H]
\centering
\scriptsize
\setlength{\tabcolsep}{3pt}
\caption{Matroid intersection, weighted problems on the two-cost family~\protect\citecode{dimacsTwoCostGen} ($m = n \log_2 n$, 5 instances per size, one run each). The unweighted problems are in \cref{tab:mi-low-two-cost}.}
\label{tab:mi-two-cost}
\begin{tabular}{@{}rrrrrr@{}}
\toprule
$n$ & SML & Scala & Go & \begin{tabular}[t]{@{}r@{}}Sage (our oracle)\\\citecode{sageIntersection}\end{tabular} & \begin{tabular}[t]{@{}r@{}}Sage (plain)\\\citecode{sageIntersection,sagePartitionMatroid,sageRankMatroid}\end{tabular} \\
\midrule
\multicolumn{6}{@{}l}{\emph{Weighted matching}} \\
32 & $<$0.001 (--) & 0.006 (--) & 0.013 (--) & 0.005 (--) & 0.054 (--) \\
64 & $<$0.001 (--) & 0.023 (--) & 0.051 (--) & 0.027 (--) & 1.04 (--) \\
128 & 0.001 (1.00) & 0.077 (77.0) & 0.191 (191) & 0.179 (179) & 23.1 (23107) \\
256 & 0.005 (1.00) & 0.167 (33.4) & 0.810 (162) & 1.33 (266) & 462 (92370) \\
512 & 0.022 (1.00) & 0.488 (21.8) & 3.42 (153) & 10.4 (463) & TO \\
1024 & 0.097 (1.00) & 1.57 (16.2) & 15.0 (154) & 91.5 (942) & TO \\
$2^{11}$ & 0.391 (1.00) & 5.86 (15.0) & 65.0 (166) & 755 (1929) & TO \\
$2^{12}$ & 1.59 (1.00) & 24.0 (15.0) & 280 (175) & TO & TO \\
$2^{13}$ & 7.17 (1.00) & 106 (14.7) & 1137 (158) & TO & TO \\
$2^{14}$ & 32.6 (1.00) & 459 (14.1) & TO & TO & TO \\
$2^{15}$ & 151 (1.00) & TO & TO & TO & TO \\
\midrule
\multicolumn{6}{@{}l}{\emph{Weighted $b$-matching}} \\
32 & 0.0002 (1.00) & 0.008 (38.0) & 0.051 (257) & 0.022 (112) & 1.74 (8711) \\
64 & 0.002 (1.00) & 0.025 (15.7) & 0.235 (147) & 0.150 (93.8) & 37.0 (23125) \\
128 & 0.006 (1.00) & 0.116 (19.3) & 1.32 (220) & 2.11 (351) & 792$^{1/5}$ (131998) \\
256 & 0.030 (1.00) & 0.573 (19.2) & 7.48 (251) & 23.3 (781) & TO \\
512 & 0.165 (1.00) & 2.88 (17.5) & 39.2 (238) & 257 (1557) & TO \\
1024 & 1.07 (1.00) & 15.0 (14.0) & 209 (196) & TO & TO \\
$2^{11}$ & 5.84 (1.00) & 76.0 (13.0) & 1050 (180) & TO & TO \\
$2^{12}$ & 30.3 (1.00) & 373 (12.3) & TO & TO & TO \\
$2^{13}$ & 164 (1.00) & TO & TO & TO & TO \\
$2^{14}$ & 856 (1.00) & TO & TO & TO & TO \\
$2^{15}$ & TO & TO & TO & TO & TO \\
\bottomrule
\end{tabular}
\end{table}

%% file: measurements/tex_tables/mi_geometric.tex
\begin{table}[H]
\centering
\scriptsize
\setlength{\tabcolsep}{3pt}
\caption{Matroid intersection, geometric family~\protect\citecode{dimacsGeometricGen} ($m = n^2/4$, 5 instances per size, one run each).}
\label{tab:mi-geometric}
\begin{tabular}{@{}rrrrrr@{}}
\toprule
$n$ & SML & Scala & Go & \begin{tabular}[t]{@{}r@{}}Sage (our oracle)\\\citecode{sageIntersection,sageIntersectionUnweighted}\end{tabular} & \begin{tabular}[t]{@{}r@{}}Sage (plain)\\\citecode{sageIntersection,sageIntersectionUnweighted,sagePartitionMatroid,sageRankMatroid}\end{tabular} \\
\midrule
\multicolumn{6}{@{}l}{\emph{Matching}} \\
32 & $<$0.001 (--) & 0.007 (--) & 0.012 (--) & 0.004 (--) & 0.038 (--) \\
64 & 0.001 (1.00) & 0.043 (43.4) & 0.084 (84.0) & 0.023 (23.0) & 0.650 (650) \\
128 & 0.003 (1.00) & 0.281 (82.7) & 0.666 (196) & 0.296 (87.1) & 20.0 (5892) \\
256 & 0.029 (1.00) & 0.751 (26.3) & 5.34 (187) & 4.32 (151) & 680 (23790) \\
512 & 0.233 (1.00) & 4.56 (19.6) & 39.7 (171) & 66.3 (285) & TO \\
1024 & 1.72 (1.00) & 37.4 (21.7) & 306 (178) & TO & TO \\
$2^{11}$ & 13.1 (1.00) & OOM & TO & TO & TO \\
$2^{12}$ & 105 (1.00) & OOM & TO & TO & TO \\
\midrule
\multicolumn{6}{@{}l}{\emph{$b$-matching}} \\
32 & 0.0002 (1.00) & 0.044 (222) & 0.076 (381) & 0.050 (252) & 0.882 (4409) \\
64 & 0.003 (1.00) & 0.258 (99.3) & 0.877 (337) & 0.759 (292) & 25.2 (9702) \\
128 & 0.012 (1.00) & 0.566 (48.8) & 4.27 (368) & 12.9 (1113) & 908$^{2/5}$ (78304) \\
256 & 0.104 (1.00) & 2.44 (23.6) & 30.9 (298) & 132 (1277) & TO \\
512 & 0.374 (1.00) & 9.17 (24.5) & 94.9 (254) & 125 (333) & TO \\
1024 & 8.53 (1.00) & 176 (20.6) & TO & TO & TO \\
$2^{11}$ & 42.9 (1.00) & OOM & TO & TO & TO \\
$2^{12}$ & 402 (1.00) & OOM & TO & TO & TO \\
\midrule
\multicolumn{6}{@{}l}{\emph{Weighted matching}} \\
32 & 0.002 (1.00) & 0.135 (61.2) & 0.613 (279) & 0.009 (4.00) & 0.109 (49.5) \\
64 & 0.057 (1.00) & 1.51 (26.6) & 17.2 (302) & 0.111 (1.94) & 5.89 (103) \\
128 & 1.86 (1.00) & 40.4 (21.7) & 556 (298) & 1.90 (1.02) & 367 (197) \\
256 & 56.1 (1.00) & 1120$^{4/5}$ (20.0) & TO & 40.8 (0.73) & TO \\
512 & TO & TO & TO & 1089 (--) & TO \\
1024 & TO & TO & TO & TO & TO \\
$2^{11}$ & TO & TO & TO & TO & TO \\
$2^{12}$ & TO & TO & TO & TO & TO \\
\midrule
\multicolumn{6}{@{}l}{\emph{Weighted $b$-matching}} \\
32 & 0.004 (1.00) & 0.111 (25.3) & 1.41 (320) & 0.084 (19.0) & 9.99 (2270) \\
64 & 0.088 (1.00) & 2.04 (23.3) & 33.2 (378) & 2.14 (24.4) & 786$^{2/5}$ (8954) \\
128 & 3.93 (1.00) & 81.4 (20.7) & 372$^{1/5}$ (94.5) & 54.0 (13.7) & TO \\
256 & 154 (1.00) & TO & TO & TO & TO \\
512 & TO & TO & TO & TO & TO \\
1024 & TO & TO & TO & TO & TO \\
$2^{11}$ & TO & TO & TO & TO & TO \\
$2^{12}$ & TO & TO & TO & TO & TO \\
\bottomrule
\end{tabular}
\end{table}

%% file: measurements/tex_figures/mi_appendix.tex
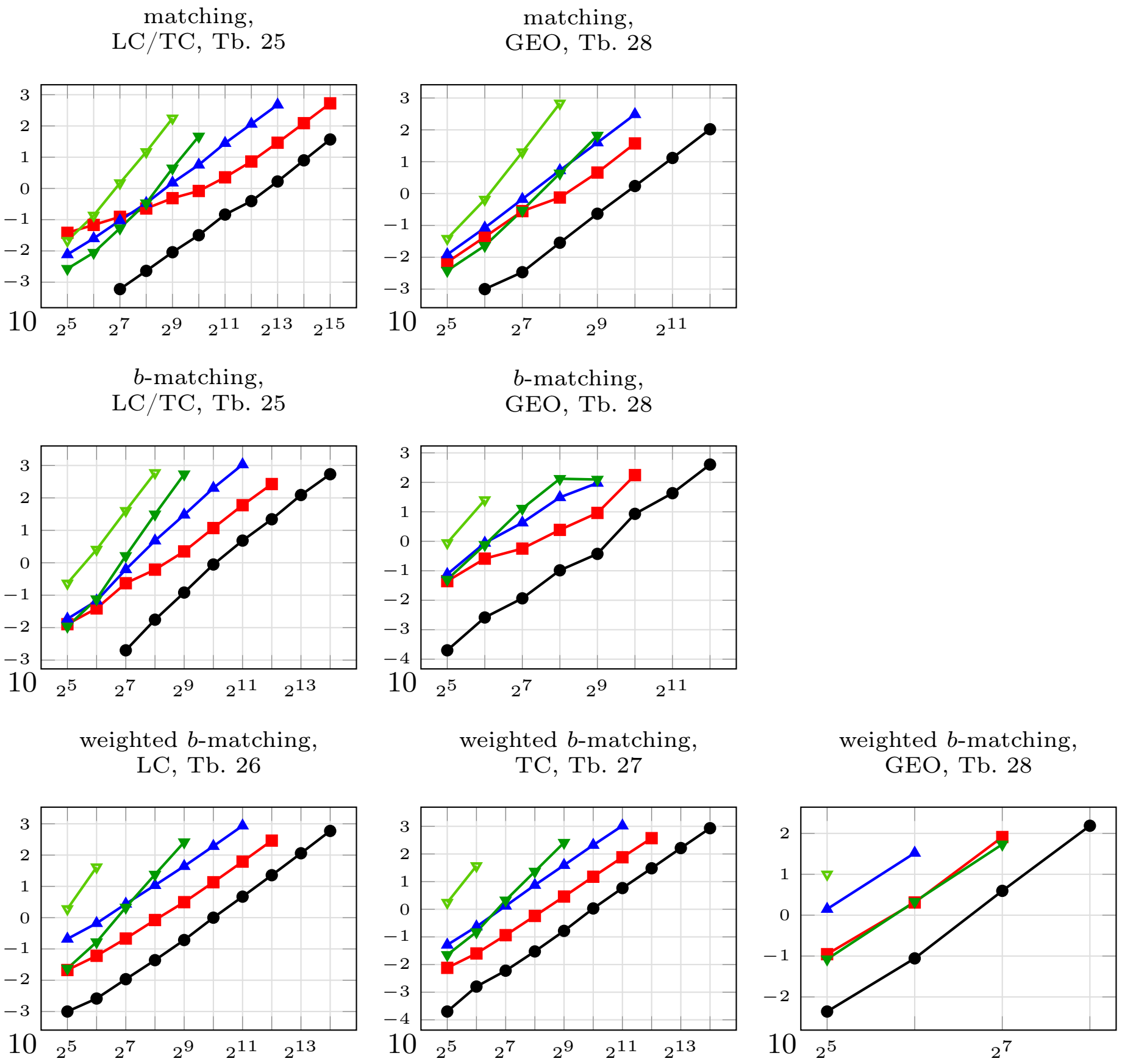
\begin{figure}[!t]
\centering
\begin{tikzpicture}
\begin{groupplot}[group style={group size=3 by 3, horizontal sep=0.7cm, vertical sep=1.5cm}, xmode=log, ymode=log, log basis x=2, ytickten={-6,...,6}, log basis y=10, yticklabel={$\pgfmathparse{int(round(\tick))}\pgfmathresult$}, after end axis/.code={\node[anchor=north east, font=\footnotesize, inner sep=1pt] at (rel axis cs:0,0) {$10$};}, grid=major, grid style={gray!25}, tick label style={font=\tiny}, label style={font=\scriptsize}, every axis plot/.append style={thick, mark size=1.5pt, legend image code/.code={\draw[mark repeat=2, mark phase=2] plot coordinates {(0cm,0cm) (0.15cm,0cm) (0.3cm,0cm)};}}, width=0.41\textwidth, height=4cm, title style={font=\scriptsize, align=center}]
\nextgroupplot[title={matching,\\ LC/TC, Tb.~\ref{tab:mi-low-two-cost}}, xtick={32,64,128,256,512,1024,2048,4096,8192,16384,32768}, xticklabels={$2^{5}$,,$2^{7}$,,$2^{9}$,,$2^{11}$,,$2^{13}$,,$2^{15}$}]
\addplot[black, mark=*, mark options={solid}] coordinates {(128,0.0006) (256,0.0023) (512,0.0091) (1024,0.0319) (2048,0.146) (4096,0.3939) (8192,1.6717) (16384,7.9309) (32768,36.9612)};
\label{leg-mi_appendix:sml}
\addplot[red, mark=square*, mark options={solid}] coordinates {(32,0.0381) (64,0.068) (128,0.1234) (256,0.2285) (512,0.488) (1024,0.8298) (2048,2.2693) (4096,7.2908) (8192,29.1024) (16384,122.699) (32768,529.77)};
\label{leg-mi_appendix:scala}
\addplot[blue, mark=triangle*, mark options={solid}] coordinates {(32,0.0077) (64,0.0251) (128,0.0941) (256,0.3368) (512,1.52) (1024,5.6987) (2048,28.1054) (4096,115.635) (8192,475.585)};
\label{leg-mi_appendix:go}
\addplot[green!60!black, mark=triangle*, mark options={solid, rotate=180}] coordinates {(32,0.0027) (64,0.0088) (128,0.0533) (256,0.3371) (512,4.4178) (1024,46.3067)};
\label{leg-mi_appendix:sage_oracle}
\addplot[green!40!lime!80!black, mark=triangle, mark options={solid, rotate=180}] coordinates {(32,0.0213) (64,0.1368) (128,1.4883) (256,14.7688) (512,176.148)};
\label{leg-mi_appendix:sage_plain}
\nextgroupplot[title={matching,\\ GEO, Tb.~\ref{tab:mi-geometric}}, xtick={32,64,128,256,512,1024,2048,4096}, xticklabels={$2^{5}$,,$2^{7}$,,$2^{9}$,,$2^{11}$,}]
\addplot[black, mark=*, mark options={solid}] coordinates {(64,0.001) (128,0.0034) (256,0.0286) (512,0.2326) (1024,1.7212) (2048,13.0984) (4096,104.6)};
\addplot[red, mark=square*, mark options={solid}] coordinates {(32,0.007) (64,0.0434) (128,0.2812) (256,0.7514) (512,4.5648) (1024,37.411)};
\addplot[blue, mark=triangle*, mark options={solid}] coordinates {(32,0.0122) (64,0.084) (128,0.6664) (256,5.3428) (512,39.69) (1024,306.419)};
\addplot[green!60!black, mark=triangle*, mark options={solid, rotate=180}] coordinates {(32,0.0038) (64,0.023) (128,0.296) (256,4.3184) (512,66.3236)};
\addplot[green!40!lime!80!black, mark=triangle, mark options={solid, rotate=180}] coordinates {(32,0.0384) (64,0.6504) (128,20.0336) (256,680.399)};
\nextgroupplot[group/empty plot, after end axis/.code={}]
\nextgroupplot[title={$b$-matching,\\ LC/TC, Tb.~\ref{tab:mi-low-two-cost}}, xtick={32,64,128,256,512,1024,2048,4096,8192,16384,32768}, xticklabels={$2^{5}$,,$2^{7}$,,$2^{9}$,,$2^{11}$,,$2^{13}$,,$2^{15}$}]
\addplot[black, mark=*, mark options={solid}] coordinates {(128,0.002) (256,0.0176) (512,0.1208) (1024,0.8848) (2048,4.8277) (4096,21.9557) (8192,122.823) (16384,537.35)};
\addplot[red, mark=square*, mark options={solid}] coordinates {(32,0.0128) (64,0.0392) (128,0.2341) (256,0.6139) (512,2.2392) (1024,11.756) (2048,59.649) (4096,267.343)};
\addplot[blue, mark=triangle*, mark options={solid}] coordinates {(32,0.0186) (64,0.0674) (128,0.623) (256,4.7898) (512,30.0511) (1024,201.104) (2048,1068.51)};
\addplot[green!60!black, mark=triangle*, mark options={solid, rotate=180}] coordinates {(32,0.0109) (64,0.0756) (128,1.6307) (256,31.2375) (512,538.775)};
\addplot[green!40!lime!80!black, mark=triangle, mark options={solid, rotate=180}] coordinates {(32,0.2325) (64,2.5636) (128,40.1576) (256,591.466)};
\nextgroupplot[title={$b$-matching,\\ GEO, Tb.~\ref{tab:mi-geometric}}, xtick={32,64,128,256,512,1024,2048,4096}, xticklabels={$2^{5}$,,$2^{7}$,,$2^{9}$,,$2^{11}$,}]
\addplot[black, mark=*, mark options={solid}] coordinates {(32,0.0002) (64,0.0026) (128,0.0116) (256,0.1036) (512,0.3742) (1024,8.5322) (2048,42.9306) (4096,402.136)};
\addplot[red, mark=square*, mark options={solid}] coordinates {(32,0.0444) (64,0.2582) (128,0.5662) (256,2.4412) (512,9.1744) (1024,176.165)};
\addplot[blue, mark=triangle*, mark options={solid}] coordinates {(32,0.0762) (64,0.877) (128,4.274) (256,30.9208) (512,94.8882)};
\addplot[green!60!black, mark=triangle*, mark options={solid, rotate=180}] coordinates {(32,0.0504) (64,0.7592) (128,12.9144) (256,132.284) (512,124.772)};
\addplot[green!40!lime!80!black, mark=triangle, mark options={solid, rotate=180}] coordinates {(32,0.8818) (64,25.2246)};
\nextgroupplot[group/empty plot, after end axis/.code={}]
\nextgroupplot[title={weighted $b$-matching,\\ LC, Tb.~\ref{tab:mi-low-cost}}, xtick={32,64,128,256,512,1024,2048,4096,8192,16384,32768}, xticklabels={$2^{5}$,,$2^{7}$,,$2^{9}$,,$2^{11}$,,$2^{13}$,,$2^{15}$}]
\addplot[black, mark=*, mark options={solid}] coordinates {(32,0.001) (64,0.0026) (128,0.0108) (256,0.0438) (512,0.1926) (1024,0.9974) (2048,4.6882) (4096,22.728) (8192,114.246) (16384,591.224)};
\addplot[red, mark=square*, mark options={solid}] coordinates {(32,0.0212) (64,0.0604) (128,0.215) (256,0.84) (512,3.1256) (1024,13.53) (2048,61.7766) (4096,289.812)};
\addplot[blue, mark=triangle*, mark options={solid}] coordinates {(32,0.209) (64,0.6618) (128,2.7446) (256,10.8436) (512,44.0712) (1024,192.475) (2048,865.915)};
\addplot[green!60!black, mark=triangle*, mark options={solid, rotate=180}] coordinates {(32,0.0238) (64,0.1656) (128,2.1226) (256,23.9152) (512,259.232)};
\addplot[green!40!lime!80!black, mark=triangle, mark options={solid, rotate=180}] coordinates {(32,1.9048) (64,41.1606)};
\nextgroupplot[title={weighted $b$-matching,\\ TC, Tb.~\ref{tab:mi-two-cost}}, xtick={32,64,128,256,512,1024,2048,4096,8192,16384,32768}, xticklabels={$2^{5}$,,$2^{7}$,,$2^{9}$,,$2^{11}$,,$2^{13}$,,$2^{15}$}]
\addplot[black, mark=*, mark options={solid}] coordinates {(32,0.0002) (64,0.0016) (128,0.006) (256,0.0298) (512,0.165) (1024,1.0702) (2048,5.8356) (4096,30.3272) (8192,163.512) (16384,856.322)};
\addplot[red, mark=square*, mark options={solid}] coordinates {(32,0.0076) (64,0.0252) (128,0.1156) (256,0.5732) (512,2.8804) (1024,15.0222) (2048,75.9716) (4096,373.217)};
\addplot[blue, mark=triangle*, mark options={solid}] coordinates {(32,0.0514) (64,0.2348) (128,1.3214) (256,7.4788) (512,39.2202) (1024,209.26) (2048,1049.62)};
\addplot[green!60!black, mark=triangle*, mark options={solid, rotate=180}] coordinates {(32,0.0224) (64,0.15) (128,2.1072) (256,23.2682) (512,256.963)};
\addplot[green!40!lime!80!black, mark=triangle, mark options={solid, rotate=180}] coordinates {(32,1.7422) (64,37.0004)};
\nextgroupplot[title={weighted $b$-matching,\\ GEO, Tb.~\ref{tab:mi-geometric}}, xtick={32,64,128,256,512,1024,2048,4096}, xticklabels={$2^{5}$,,$2^{7}$,,$2^{9}$,,$2^{11}$,}]
\addplot[black, mark=*, mark options={solid}] coordinates {(32,0.0044) (64,0.0878) (128,3.9338) (256,153.912)};
\addplot[red, mark=square*, mark options={solid}] coordinates {(32,0.1112) (64,2.0438) (128,81.4022)};
\addplot[blue, mark=triangle*, mark options={solid}] coordinates {(32,1.409) (64,33.1658)};
\addplot[green!60!black, mark=triangle*, mark options={solid, rotate=180}] coordinates {(32,0.0838) (64,2.139) (128,54.0068)};
\addplot[green!40!lime!80!black, mark=triangle, mark options={solid, rotate=180}] coordinates {(32,9.9866)};
\end{groupplot}
\end{tikzpicture}
\caption{Matroid Intersection, matching, $b$-matching and weighted $b$-matching; on matching and $b$-matching, LC and TC share their instances (LC/TC): \ref{leg-mi_appendix:sml}~SML, \ref{leg-mi_appendix:scala}~Scala, \ref{leg-mi_appendix:go}~Go, \ref{leg-mi_appendix:sage_oracle}~Sage (our oracle), \ref{leg-mi_appendix:sage_plain}~Sage (plain).}
\label{fig:mi-appendix}
\end{figure}

%% file: measurements/tex_tables/mcf_hybrid.tex
\begin{table}[H]
\centering
\scriptsize
\setlength{\tabcolsep}{3pt}
\caption{Hybrid min-cost flow solver, \texttt{gridgen\_deg}~\protect\citecode{mcfDataGridgenDeg} family generated by GRIDGEN. A density sweep induced by degree increase at the fixed node count $n = 4097$. The first column is the average in-degree $m/n$ which always equals the density.}
\label{tab:mcf-gridgen-deg}
\begin{tabular}{@{}rrlrrrrr@{\hspace{6pt}}|@{\hspace{6pt}}r@{}}
\toprule
deg & $m$ & Alg. & Oracle & Cert. & Check & Glue & Total & Total/Oracle \\
\midrule
\rowcolor{gray!25}4 & 16388 & NS & 5.70 & 0.042 & 1.59 & 0.964 & 8.30 & 1.46 \\
 &  & COS & 13.0 & 0.227 & 1.72 & 1.39 & 16.3 & 1.26 \\
 &  & CAS & 65.1 & 0.152 & 1.62 & 1.44 & 68.3 & 1.05 \\
 &  & SSP & 9.86 & 0.101 & 1.59 & 1.38 & 12.9 & 1.31 \\
\addlinespace
\rowcolor{gray!25}8 & 32776 & NS & 14.9 & 0.196 & 2.08 & 1.60 & 18.8 & 1.26 \\
 &  & COS & 27.6 & 0.439 & 2.33 & 2.51 & 32.9 & 1.19 \\
 &  & CAS & 228 & 0.385 & 2.08 & 2.56 & 233 & 1.02 \\
 &  & SSP & 544 & 0.363 & 2.10 & 2.60 & 549 & 1.01 \\
\addlinespace
\rowcolor{gray!25}16 & 65552 & NS & 23.0 & 0.321 & 3.97 & 3.05 & 30.0 & 1.32 \\
 &  & COS & 38.0 & 0.810 & 4.53 & 4.74 & 48.0 & 1.26 \\
 &  & CAS & 420 & 0.661 & 3.98 & 4.95 & 430 & 1.02 \\
 &  & SSP & 1017 & 0.694 & 3.96 & 4.34 & 1026 & 1.01 \\
\addlinespace
\rowcolor{gray!25}32 & 131104 & NS & 30.0 & 0.621 & 11.5 & 6.06 & 48.0 & 1.60 \\
 &  & COS & 60.0 & 1.46 & 12.6 & 9.38 & 83.0 & 1.39 \\
 &  & CAS & 812 & 1.12 & 11.3 & 9.93 & 835 & 1.03 \\
 &  & SSP & 1208 & 1.09 & 11.4 & 9.56 & 1230 & 1.02 \\
\addlinespace
\rowcolor{gray!25}64 & 262208 & NS & 48.0 & 1.20 & 23.2 & 11.4 & 84.0 & 1.74 \\
 &  & COS & 108 & 2.70 & 25.3 & 19.3 & 155 & 1.44 \\
 &  & CAS & 2182 & 1.93 & 23.2 & 19.9 & 2227 & 1.02 \\
 &  & SSP & 1964 & 1.94 & 23.1 & 21.0 & 2010 & 1.02 \\
\addlinespace
\rowcolor{gray!25}128 & 524416 & NS & 74.0 & 2.21 & 45.4 & 24.1 & 146 & 1.97 \\
 &  & COS & 204 & 4.47 & 49.8 & 40.4 & 299 & 1.46 \\
 &  & CAS & 6048 & 3.47 & 45.6 & 42.9 & 6140 & 1.01 \\
 &  & SSP & 2947 & 3.15 & 45.5 & 42.3 & 3038 & 1.03 \\
\addlinespace
\rowcolor{gray!25}256 & 1048832 & NS & 107 & 3.93 & 60.6 & 80.9 & 252 & 2.37 \\
 &  & COS & 459 & 8.47 & 70.3 & 127 & 665 & 1.45 \\
 &  & CAS & 16650 & 6.45 & 60.6 & 136 & 16853 & 1.01 \\
 &  & SSP & 4574 & 6.48 & 60.6 & 139 & 4780 & 1.04 \\
\addlinespace
\rowcolor{gray!25}512 & 2097664 & NS & 174 & 8.59 & 181 & 109 & 472 & 2.71 \\
 &  & COS & 1149 & 16.5 & 199 & 221 & 1585 & 1.38 \\
 &  & CAS & 115550 & 13.5 & 180 & 237 & 115980 & 1.00 \\
 &  & SSP & 6478 & 13.8 & 183 & 245 & 6920 & 1.07 \\
\addlinespace
\rowcolor{gray!25}1024 & 4195328 & NS & 316 & 16.6 & 365 & 233 & 930 & 2.95 \\
 &  & COS & 2507 & 32.6 & 401 & 481 & 3421 & 1.36 \\
 &  & CAS & 501030 & 27.6 & 361 & 511 & 501930 & 1.00 \\
 &  & SSP & 12384 & 25.3 & 361 & 511 & 13281 & 1.07 \\
\addlinespace
\rowcolor{gray!25}$2^{11}$ & 8390656 & NS & 546 & 29.1 & 728 & 438 & 1741 & 3.19 \\
 &  & COS & 5715 & 56.1 & 797 & 977 & 7545 & 1.32 \\
 &  & CAS & 1424300 & 60.5 & 734 & 1005 & 1426100 & 1.00 \\
 &  & SSP & 24552 & 52.0 & 724 & 1041 & 26369 & 1.07 \\
\addlinespace
\rowcolor{gray!25}$2^{12}$ & 16781312 & NS & 957 & 59.2 & 1471 & 889 & 3377 & 3.53 \\
 &  & COS & 11638 & 127 & 1614 & 1988 & 15367 & 1.32 \\
 &  & CAS & TO & TO & TO & TO & TO & TO \\
 &  & SSP & 43839 & 107 & 1455 & 2138 & 47539 & 1.08 \\
\bottomrule
\end{tabular}
\end{table}
\begin{table}[H]
\centering
\scriptsize
\setlength{\tabcolsep}{3pt}
\caption{Hybrid min-cost flow solver, \texttt{netgen\_deg}~\protect\citecode{mcfDataNetgenDeg} family generated by NETGEN~\protect\cite{klingmanNETGENProgramGenerating1974}. A density sweep by increasing degrees at the fixed node count $n = 4096$. The first column is the average in-degree $m/n$ which is equal to the out-degree and the density.}
\label{tab:mcf-netgen-deg}
\begin{tabular}{@{}rrlrrrrr@{\hspace{6pt}}|@{\hspace{6pt}}r@{}}
\toprule
deg & $m$ & Alg. & Oracle & Cert. & Check & Glue & Total & Total/Oracle \\
\midrule
\rowcolor{gray!25}2 & 8192 & NS & 7.79 & 0.021 & 0.649 & 0.697 & 9.15 & 1.18 \\
 &  & COS & 20.7 & 0.119 & 0.696 & 0.946 & 22.5 & 1.08 \\
 &  & CAS & 103 & 0.081 & 0.674 & 0.965 & 105 & 1.02 \\
 &  & SSP & 123 & 0.084 & 0.675 & 0.961 & 125 & 1.01 \\
\addlinespace
\rowcolor{gray!25}4 & 16384 & NS & 10.9 & 0.048 & 1.61 & 1.00 & 13.5 & 1.24 \\
 &  & COS & 19.3 & 0.244 & 1.72 & 1.42 & 22.7 & 1.18 \\
 &  & CAS & 286 & 0.198 & 1.62 & 1.47 & 290 & 1.01 \\
 &  & SSP & 212 & 0.195 & 1.63 & 1.46 & 215 & 1.01 \\
\addlinespace
\rowcolor{gray!25}8 & 32768 & NS & 16.8 & 0.198 & 2.10 & 2.43 & 21.6 & 1.28 \\
 &  & COS & 26.8 & 0.434 & 2.36 & 3.45 & 33.0 & 1.23 \\
 &  & CAS & 972 & 0.377 & 2.09 & 3.46 & 978 & 1.01 \\
 &  & SSP & 537 & 0.370 & 2.12 & 3.46 & 543 & 1.01 \\
\addlinespace
\rowcolor{gray!25}16 & 65536 & NS & 24.0 & 0.330 & 4.07 & 3.06 & 31.0 & 1.31 \\
 &  & COS & 38.0 & 0.803 & 4.62 & 4.80 & 48.0 & 1.27 \\
 &  & CAS & 3843 & 0.802 & 4.06 & 4.14 & 3852 & 1.00 \\
 &  & SSP & 911 & 0.720 & 4.06 & 4.76 & 921 & 1.01 \\
\addlinespace
\rowcolor{gray!25}32 & 131072 & NS & 36.0 & 0.628 & 11.5 & 6.03 & 54.0 & 1.50 \\
 &  & COS & 56.0 & 1.45 & 12.7 & 9.38 & 79.0 & 1.42 \\
 &  & CAS & 8854 & 1.26 & 11.5 & 10.2 & 8877 & 1.00 \\
 &  & SSP & 1265 & 1.12 & 11.5 & 9.39 & 1287 & 1.02 \\
\addlinespace
\rowcolor{gray!25}64 & 262144 & NS & 52.0 & 1.21 & 22.7 & 11.7 & 88.0 & 1.68 \\
 &  & COS & 94.0 & 2.67 & 25.1 & 19.3 & 141 & 1.50 \\
 &  & CAS & 23755 & 2.40 & 22.8 & 18.8 & 23799 & 1.00 \\
 &  & SSP & 1910 & 1.94 & 22.8 & 19.3 & 1954 & 1.02 \\
\addlinespace
\rowcolor{gray!25}128 & 524288 & NS & 84.0 & 2.29 & 45.3 & 26.6 & 158 & 1.88 \\
 &  & COS & 191 & 4.46 & 49.6 & 38.4 & 283 & 1.49 \\
 &  & CAS & 75297 & 4.21 & 45.5 & 41.3 & 75388 & 1.00 \\
 &  & SSP & 3335 & 3.43 & 45.2 & 40.4 & 3424 & 1.03 \\
\addlinespace
\rowcolor{gray!25}256 & 1048576 & NS & 123 & 4.20 & 89.8 & 56.9 & 274 & 2.23 \\
 &  & COS & 409 & 8.52 & 99.4 & 83.8 & 601 & 1.47 \\
 &  & CAS & 281740 & 8.03 & 91.1 & 90.9 & 281930 & 1.00 \\
 &  & SSP & 4564 & 6.09 & 89.8 & 84.1 & 4744 & 1.04 \\
\addlinespace
\rowcolor{gray!25}512 & 2097152 & NS & 190 & 8.44 & 181 & 117 & 495 & 2.61 \\
 &  & COS & 1058 & 16.0 & 197 & 182 & 1453 & 1.37 \\
 &  & CAS & 924820 & 15.7 & 182 & 192 & 925210 & 1.00 \\
 &  & SSP & 7854 & 12.7 & 179 & 186 & 8231 & 1.05 \\
\addlinespace
\rowcolor{gray!25}1024 & 4194304 & NS & 339 & 15.8 & 358 & 243 & 956 & 2.82 \\
 &  & COS & 2309 & 32.1 & 396 & 408 & 3146 & 1.36 \\
 &  & CAS & TO & TO & TO & TO & TO & TO \\
 &  & SSP & 13945 & 26.2 & 357 & 390 & 14718 & 1.05 \\
\addlinespace
\rowcolor{gray!25}$2^{11}$ & 8388608 & NS & 575 & 29.6 & 716 & 512 & 1833 & 3.19 \\
 &  & COS & 5100 & 56.4 & 797 & 828 & 6782 & 1.33 \\
 &  & CAS & TO & TO & TO & TO & TO & TO \\
 &  & SSP & 26008 & 48.8 & 709 & 807 & 27572 & 1.06 \\
\addlinespace
\rowcolor{gray!25}$2^{12}$ & 16252992 & NS & 916 & 53.5 & 1489 & 919 & 3378 & 3.69 \\
 &  & COS & 10320 & 110 & 1638 & 1501 & 13569 & 1.31 \\
 &  & CAS & TO & TO & TO & TO & TO & TO \\
 &  & SSP & 51289 & 115 & 1522 & 1585 & 54511 & 1.06 \\
\bottomrule
\end{tabular}
\end{table}
\begin{table}[H]
\centering
\scriptsize
\setlength{\tabcolsep}{3pt}
\caption{Hybrid min-cost flow solver, \texttt{netgen\_lo\_sr}~\protect\citecode{mcfDataNetgenLoSr} family generated by NETGEN~\protect\cite{klingmanNETGENProgramGenerating1974}. A node sweep at the density $m = n^{1.5}$.}
\label{tab:mcf-netgen-lo-sr}
\begin{tabular}{@{}rrlrrrrr@{\hspace{6pt}}|@{\hspace{6pt}}r@{}}
\toprule
$n$ & $m$ & Alg. & Oracle & Cert. & Check & Glue & Total & Total/Oracle \\
\midrule
\rowcolor{gray!25}256 & 4096 & NS & 0.266 & 0.010 & 0.279 & 0.287 & 0.842 & 3.17 \\
 &  & COS & 0.936 & 0.022 & 0.324 & 0.393 & 1.68 & 1.79 \\
 &  & CAS & 0.635 & 0.017 & 0.279 & 0.405 & 1.34 & 2.10 \\
 &  & SSP & 0.792 & 0.017 & 0.284 & 0.398 & 1.49 & 1.88 \\
\addlinespace
\rowcolor{gray!25}512 & 11585 & NS & 0.696 & 0.015 & 1.04 & 0.576 & 2.33 & 3.34 \\
 &  & COS & 2.60 & 0.069 & 1.16 & 0.893 & 4.72 & 1.82 \\
 &  & CAS & 2.09 & 0.025 & 1.04 & 0.903 & 4.06 & 1.94 \\
 &  & SSP & 3.11 & 0.025 & 1.03 & 0.895 & 5.06 & 1.63 \\
\addlinespace
\rowcolor{gray!25}1024 & 32768 & NS & 1.87 & 0.163 & 2.84 & 1.50 & 6.37 & 3.40 \\
 &  & COS & 7.13 & 0.337 & 3.14 & 2.35 & 13.0 & 1.82 \\
 &  & CAS & 7.01 & 0.244 & 2.86 & 2.42 & 12.5 & 1.79 \\
 &  & SSP & 10.3 & 0.248 & 2.82 & 2.37 & 15.7 & 1.53 \\
\addlinespace
\rowcolor{gray!25}$2^{11}$ & 92682 & NS & 6.21 & 0.352 & 7.85 & 3.84 & 18.2 & 2.94 \\
 &  & COS & 20.4 & 0.900 & 8.70 & 6.43 & 36.4 & 1.79 \\
 &  & CAS & 27.7 & 0.702 & 7.89 & 6.71 & 43.0 & 1.55 \\
 &  & SSP & 40.0 & 0.715 & 7.95 & 6.68 & 55.3 & 1.38 \\
\addlinespace
\rowcolor{gray!25}$2^{12}$ & 262144 & NS & 17.8 & 1.01 & 22.5 & 11.6 & 52.9 & 2.96 \\
 &  & COS & 62.5 & 2.65 & 25.0 & 18.8 & 109 & 1.74 \\
 &  & CAS & 86.5 & 1.94 & 22.6 & 19.1 & 130 & 1.50 \\
 &  & SSP & 145 & 1.92 & 22.5 & 19.0 & 188 & 1.30 \\
\addlinespace
\rowcolor{gray!25}$2^{13}$ & 741455 & NS & 58.7 & 3.06 & 43.1 & 34.7 & 140 & 2.38 \\
 &  & COS & 203 & 6.08 & 50.2 & 57.3 & 317 & 1.56 \\
 &  & CAS & 278 & 4.54 & 43.1 & 59.5 & 386 & 1.39 \\
 &  & SSP & 562 & 4.82 & 43.2 & 60.6 & 671 & 1.19 \\
\addlinespace
\rowcolor{gray!25}$2^{14}$ & 2097152 & NS & 203 & 8.52 & 123 & 117 & 450 & 2.22 \\
 &  & COS & 822 & 16.7 & 142 & 187 & 1168 & 1.42 \\
 &  & CAS & 1081 & 11.4 & 124 & 194 & 1411 & 1.30 \\
 &  & SSP & 2104 & 12.9 & 122 & 198 & 2437 & 1.16 \\
\addlinespace
\rowcolor{gray!25}$2^{15}$ & 5931642 & NS & 754 & 20.8 & 354 & 298 & 1427 & 1.89 \\
 &  & COS & 2832 & 44.3 & 409 & 566 & 3851 & 1.36 \\
 &  & CAS & 3539 & 30.9 & 353 & 597 & 4520 & 1.28 \\
 &  & SSP & 7417 & 33.6 & 353 & 594 & 8397 & 1.13 \\
\addlinespace
\rowcolor{gray!25}$2^{16}$ & 16777216 & NS & 2822 & 54.0 & 1102 & 1102 & 5080 & 1.80 \\
 &  & COS & 9482 & 114 & 1267 & 1731 & 12594 & 1.33 \\
 &  & CAS & 12679 & 104 & 1106 & 1886 & 15775 & 1.24 \\
 &  & SSP & 27502 & 109 & 1106 & 1899 & 30616 & 1.11 \\
\bottomrule
\end{tabular}
\end{table}
\begin{table}[H]
\centering
\scriptsize
\setlength{\tabcolsep}{3pt}
\caption{Hybrid min-cost flow solver, \texttt{road\_paths}~\protect\citecode{mcfDataRoad} family on the road networks of~\protect\cite{demetrescuShortestPathProblem2009}. US state road network (real data). The first column lists the instances, ordered by increasing number of nodes $n$.}
\label{tab:mcf-road-paths}
\begin{tabular}{@{}lrrrlrrrrr@{\hspace{6pt}}|@{\hspace{6pt}}r@{}}
\toprule
Instance & $n$ & $m$ & $m/n$ & Alg. & Oracle & Cert. & Check & Glue & Total & Total/Oracle \\
\midrule
01\_DC & 9559 & 29768 & 3.11 & NS & 15.2 & 0.165 & 2.09 & 1.85 & 19.3 & 1.27 \\
 &  &  &  & COS & 31.0 & 0.415 & 2.24 & 2.54 & 36.2 & 1.17 \\
 &  &  &  & CAS & 4.75 & 0.164 & 2.08 & 2.66 & 9.65 & 2.03 \\
\rowcolor{gray!25} &  &  &  & SSP & 4.73 & 0.174 & 2.08 & 2.69 & 9.67 & 2.05 \\
\addlinespace
02\_DE & 49109 & 120576 & 2.46 & NS & 143 & 1.23 & 9.11 & 8.05 & 162 & 1.13 \\
 &  &  &  & COS & 239 & 2.25 & 9.27 & 10.7 & 261 & 1.09 \\
\rowcolor{gray!25} &  &  &  & CAS & 23.2 & 0.634 & 8.63 & 11.8 & 44.3 & 1.91 \\
 &  &  &  & SSP & 23.2 & 0.617 & 8.83 & 11.6 & 44.3 & 1.91 \\
\addlinespace
03\_NH & 116920 & 265402 & 2.27 & NS & 678 & 2.28 & 19.2 & 17.9 & 718 & 1.06 \\
 &  &  &  & COS & 845 & 4.13 & 20.6 & 24.1 & 894 & 1.06 \\
\rowcolor{gray!25} &  &  &  & CAS & 76.7 & 1.47 & 19.3 & 27.3 & 125 & 1.63 \\
 &  &  &  & SSP & 76.9 & 1.45 & 19.2 & 26.9 & 124 & 1.62 \\
\addlinespace
04\_NV & 261155 & 620924 & 2.38 & NS & 2854 & 4.53 & 45.6 & 40.9 & 2945 & 1.03 \\
 &  &  &  & COS & 3339 & 8.91 & 49.0 & 58.0 & 3455 & 1.03 \\
 &  &  &  & CAS & 280 & 2.98 & 46.0 & 63.6 & 393 & 1.40 \\
\rowcolor{gray!25} &  &  &  & SSP & 279 & 2.92 & 45.3 & 63.5 & 391 & 1.40 \\
\addlinespace
05\_WI & 519157 & 1266876 & 2.44 & NS & 9150 & 10.7 & 96.2 & 87.1 & 9344 & 1.02 \\
 &  &  &  & COS & 8859 & 19.8 & 103 & 122 & 9104 & 1.03 \\
 &  &  &  & CAS & 1121 & 6.27 & 96.7 & 144 & 1368 & 1.22 \\
\rowcolor{gray!25} &  &  &  & SSP & 1116 & 5.74 & 96.6 & 142 & 1360 & 1.22 \\
\addlinespace
06\_FL & 1048506 & 2653624 & 2.53 & NS & TO & TO & TO & TO & TO & TO \\
 &  &  &  & COS & TO & TO & TO & TO & TO & TO \\
\rowcolor{gray!25} &  &  &  & CAS & 1842 & 12.5 & 211 & 309 & 2375 & 1.29 \\
 &  &  &  & SSP & 1859 & 11.4 & 212 & 302 & 2384 & 1.28 \\
\addlinespace
07\_TX & 2073870 & 5156088 & 2.49 & NS & TO & TO & TO & TO & TO & TO \\
 &  &  &  & COS & TO & TO & TO & TO & TO & TO \\
 &  &  &  & CAS & 6688 & 21.4 & 427 & 590 & 7727 & 1.16 \\
\rowcolor{gray!25} &  &  &  & SSP & 6665 & 24.2 & 425 & 609 & 7723 & 1.16 \\
\bottomrule
\end{tabular}
\end{table}
\begin{table}[H]
\centering
\scriptsize
\setlength{\tabcolsep}{3pt}
\caption{Hybrid min-cost flow solver, \texttt{goto\_8}~\protect\citecode{mcfDataGoto8} family generated by GOTO~\protect\cite{goldbergImplementingScalingPushRelabel1993}. A node sweep at the fixed density $m = 8n$.}
\label{tab:mcf-goto-8}
\begin{tabular}{@{}rrlrrrrr@{\hspace{6pt}}|@{\hspace{6pt}}r@{}}
\toprule
$n$ & $m$ & Alg. & Oracle & Cert. & Check & Glue & Total & Total/Oracle \\
\midrule
\rowcolor{gray!25}256 & $2^{11}$ & NS & 0.638 & 0.011 & 0.175 & 0.211 & 1.03 & 1.62 \\
 &  & COS & 2.81 & 0.024 & 0.186 & 0.260 & 3.28 & 1.17 \\
 &  & CAS & 5.39 & 0.025 & 0.177 & 0.264 & 5.86 & 1.09 \\
 &  & SSP & 32.9 & 0.045 & 0.170 & 0.276 & 33.4 & 1.01 \\
\addlinespace
\rowcolor{gray!25}$2^{11}$ & $2^{14}$ & NS & 23.0 & 0.060 & 1.57 & 0.842 & 26.0 & 1.11 \\
 &  & COS & 40.0 & 0.260 & 1.62 & 1.27 & 43.0 & 1.08 \\
 &  & CAS & 132 & 0.205 & 1.57 & 1.29 & 135 & 1.02 \\
 &  & SSP & 3672 & 0.220 & 1.54 & 1.24 & 3675 & 1.00 \\
\addlinespace
\rowcolor{gray!25}$2^{12}$ & $2^{15}$ & NS & 108 & 0.293 & 2.26 & 2.49 & 113 & 1.05 \\
 &  & COS & 139 & 0.598 & 2.34 & 3.34 & 145 & 1.04 \\
 &  & CAS & 2054 & 0.362 & 2.25 & 3.39 & 2060 & 1.00 \\
 &  & SSP & 26607 & 0.380 & 2.25 & 3.37 & 26613 & 1.00 \\
\addlinespace
$2^{13}$ & $2^{16}$ & NS & 318 & 0.505 & 4.29 & 3.13 & 326 & 1.02 \\
\rowcolor{gray!25} &  & COS & 247 & 1.01 & 4.52 & 4.87 & 258 & 1.04 \\
 &  & CAS & 1253 & 0.619 & 4.26 & 5.12 & 1263 & 1.01 \\
 &  & SSP & 69189 & 0.622 & 4.27 & 5.11 & 69199 & 1.00 \\
\addlinespace
$2^{14}$ & $2^{17}$ & NS & 1271 & 0.823 & 8.44 & 5.74 & 1286 & 1.01 \\
\rowcolor{gray!25} &  & COS & 655 & 1.74 & 9.01 & 9.79 & 676 & 1.03 \\
 &  & CAS & 3523 & 1.12 & 8.53 & 10.3 & 3543 & 1.01 \\
 &  & SSP & 274510 & 1.32 & 8.48 & 10.2 & 274530 & 1.00 \\
\addlinespace
$2^{15}$ & $2^{18}$ & NS & 31593 & 1.69 & 17.2 & 12.1 & 31624 & 1.00 \\
\rowcolor{gray!25} &  & COS & 2342 & 3.04 & 18.7 & 22.2 & 2386 & 1.02 \\
 &  & CAS & 185270 & 2.29 & 17.2 & 20.5 & 185310 & 1.00 \\
 &  & SSP & 1270300 & 2.30 & 17.2 & 80.5 & 1270400 & 1.00 \\
\addlinespace
$2^{16}$ & $2^{19}$ & NS & 35666 & 3.08 & 34.1 & 25.8 & 35729 & 1.00 \\
\rowcolor{gray!25} &  & COS & 5140 & 5.88 & 36.3 & 42.9 & 5225 & 1.02 \\
 &  & CAS & 34066 & 4.14 & 34.1 & 44.8 & 34149 & 1.00 \\
 &  & SSP & 3148600 & 4.13 & 34.1 & 61.8 & 3148700 & 1.00 \\
\bottomrule
\end{tabular}
\end{table}
\begin{table}[H]
\centering
\scriptsize
\setlength{\tabcolsep}{3pt}
\caption{Hybrid min-cost flow solver, \texttt{goto\_8} family (continued from \cref{tab:mcf-goto-8}).}
\begin{tabular}{@{}rrlrrrrr@{\hspace{6pt}}|@{\hspace{6pt}}r@{}}
\toprule
$n$ & $m$ & Alg. & Oracle & Cert. & Check & Glue & Total & Total/Oracle \\
\midrule
$2^{17}$ & $2^{20}$ & NS & 152080 & 6.73 & 68.7 & 54.5 & 152210 & 1.00 \\
\rowcolor{gray!25} &  & COS & 19734 & 12.9 & 73.1 & 92.0 & 19912 & 1.01 \\
 &  & CAS & 105990 & 8.02 & 68.9 & 93.1 & 106160 & 1.00 \\
 &  & SSP & TO & TO & TO & TO & TO & TO \\
\addlinespace
$2^{18}$ & $2^{21}$ & NS & 3203400 & 12.9 & 140 & 147 & 3203700 & 1.00 \\
\rowcolor{gray!25} &  & COS & 41971 & 25.8 & 155 & 211 & 42363 & 1.01 \\
 &  & CAS & TO & TO & TO & TO & TO & TO \\
 &  & SSP & TO & TO & TO & TO & TO & TO \\
\addlinespace
$2^{19}$ & $2^{22}$ & NS & 2413900 & 24.9 & 290 & 285 & 2414500 & 1.00 \\
\rowcolor{gray!25} &  & COS & 270890 & 43.7 & 308 & 388 & 271630 & 1.00 \\
 &  & CAS & TO & TO & TO & TO & TO & TO \\
 &  & SSP & TO & TO & TO & TO & TO & TO \\
\addlinespace
$2^{20}$ & $2^{23}$ & NS & TO & TO & TO & TO & TO & TO \\
\rowcolor{gray!25} &  & COS & 1009100 & 85.7 & 640 & 874 & 1010700 & 1.00 \\
 &  & CAS & TO & TO & TO & TO & TO & TO \\
 &  & SSP & TO & TO & TO & TO & TO & TO \\
\addlinespace
$2^{21}$ & $2^{24}$ & NS & TO & TO & TO & TO & TO & TO \\
\rowcolor{gray!25} &  & COS & 656740 & 150 & 1489 & 1931 & 660310 & 1.00 \\
 &  & CAS & TO & TO & TO & TO & TO & TO \\
 &  & SSP & TO & TO & TO & TO & TO & TO \\
\addlinespace
$2^{22}$ & $2^{25}$ & NS & TO & TO & TO & TO & TO & TO \\
 &  & COS & TO & TO & TO & TO & TO & TO \\
 &  & CAS & TO & TO & TO & TO & TO & TO \\
 &  & SSP & TO & TO & TO & TO & TO & TO \\
\bottomrule
\end{tabular}
\end{table}

%% file: measurements/tex_tables/hybrid_low-cost.tex
\begin{table}[H]
\centering
\scriptsize
\setlength{\tabcolsep}{3pt}
\caption{Hybrid bipartite matching solver with CSA~\protect\citecode{goldbergCSACode} as oracle, low-cost family~\protect\citecode{dimacsAssign} (10 instances per size, averaged over the medians of 3 repeats per instance). A node sweep at the density $m = n \log_2 n$.}
\label{tab:hybrid-low-cost}
\begin{tabular}{@{}rrrrrrr@{\hspace{6pt}}|@{\hspace{6pt}}r@{}}
\toprule
$n$ & $m$ & Oracle & Cert. & Check & Glue & Total & Total/Oracle \\
\midrule
$2^{11}$ & 22528 & 1.29 & 0.116 & 0.042 & 0.774 & 2.22 & 1.72 \\
$2^{12}$ & 49152 & 2.90 & 0.247 & 0.092 & 1.64 & 4.88 & 1.68 \\
$2^{13}$ & 106496 & 7.12 & 0.555 & 0.197 & 3.63 & 11.5 & 1.61 \\
$2^{14}$ & 229376 & 16.3 & 1.35 & 0.437 & 8.25 & 26.3 & 1.62 \\
$2^{15}$ & 491520 & 41.8 & 4.96 & 1.50 & 20.8 & 69.1 & 1.65 \\
$2^{16}$ & 1048576 & 99.4 & 14.0 & 4.08 & 45.4 & 163 & 1.64 \\
$2^{17}$ & 2228224 & 284 & 37.3 & 9.75 & 96.9 & 427 & 1.51 \\
$2^{18}$ & 4718592 & 877 & 122 & 22.9 & 212 & 1234 & 1.41 \\
$2^{19}$ & 9961472 & 2193 & 227 & 45.0 & 424 & 2890 & 1.32 \\
$2^{20}$ & 20971520 & 6237 & 545 & 98.9 & 967 & 7849 & 1.26 \\
\bottomrule
\end{tabular}
\end{table}

%% file: measurements/tex_tables/hybrid_two-cost.tex
\begin{table}[H]
\centering
\scriptsize
\setlength{\tabcolsep}{3pt}
\caption{Hybrid bipartite matching solver with CSA~\protect\citecode{goldbergCSACode} as oracle, two-cost family~\protect\citecode{dimacsTwoCostGen} (10 instances per size, averaged over the medians of 3 repeats per instance). A node sweep at the density $m = n \log_2 n$.}
\label{tab:hybrid-two-cost}
\begin{tabular}{@{}rrrrrrr@{\hspace{6pt}}|@{\hspace{6pt}}r@{}}
\toprule
$n$ & $m$ & Oracle & Cert. & Check & Glue & Total & Total/Oracle \\
\midrule
$2^{11}$ & 22528 & 4.89 & 0.116 & 0.041 & 0.745 & 5.79 & 1.18 \\
$2^{12}$ & 49152 & 12.2 & 0.248 & 0.092 & 1.63 & 14.2 & 1.16 \\
$2^{13}$ & 106496 & 27.5 & 0.583 & 0.198 & 3.65 & 31.9 & 1.16 \\
$2^{14}$ & 229376 & 65.6 & 1.73 & 0.471 & 8.15 & 76.0 & 1.16 \\
$2^{15}$ & 491520 & 150 & 5.23 & 1.49 & 20.3 & 177 & 1.18 \\
$2^{16}$ & 1048576 & 356 & 14.9 & 3.96 & 43.4 & 418 & 1.18 \\
$2^{17}$ & 2228224 & 1438 & 37.0 & 9.47 & 94.3 & 1579 & 1.10 \\
$2^{18}$ & 4718592 & 4362 & 114 & 22.9 & 200 & 4699 & 1.08 \\
$2^{19}$ & 9961472 & 10805 & 264 & 49.7 & 431 & 11550 & 1.07 \\
$2^{20}$ & 20971520 & 31019 & 552 & 102 & 956 & 32630 & 1.05 \\
\bottomrule
\end{tabular}
\end{table}

%% file: measurements/tex_tables/hybrid_geometric.tex
\begin{table}[H]
\centering
\scriptsize
\setlength{\tabcolsep}{3pt}
\caption{Hybrid bipartite matching solver with CSA~\protect\citecode{goldbergCSACode} as oracle, geometric family~\protect\citecode{dimacsGeometricGen} (10 instances per size, averaged over the medians of 3 repeats per instance). A node sweep on complete bipartite graphs, $m = n^2/4$.}
\label{tab:hybrid-geometric}
\begin{tabular}{@{}rrrrrrr@{\hspace{6pt}}|@{\hspace{6pt}}r@{}}
\toprule
$n$ & $m$ & Oracle & Cert. & Check & Glue & Total & Total/Oracle \\
\midrule
256 & $2^{14}$ & 0.545 & 0.061 & 0.023 & 0.577 & 1.21 & 2.21 \\
512 & $2^{16}$ & 2.09 & 0.410 & 0.090 & 2.31 & 4.90 & 2.35 \\
1024 & $2^{18}$ & 8.63 & 0.749 & 0.352 & 11.2 & 21.0 & 2.43 \\
$2^{11}$ & $2^{20}$ & 44.2 & 4.53 & 1.83 & 53.9 & 104 & 2.36 \\
$2^{12}$ & $2^{22}$ & 258 & 45.0 & 6.97 & 241 & 551 & 2.13 \\
$2^{13}$ & $2^{24}$ & 1348 & 171 & 34.6 & 1101 & 2654 & 1.97 \\
$2^{14}$ & $2^{26}$ & 6990 & 641 & 193 & 4000 & 11824 & 1.69 \\
\bottomrule
\end{tabular}
\end{table}